\documentclass[12pt,letterpaper]{article}

\pdfoutput=1

\usepackage{epsfig}
\usepackage{amsmath}
\usepackage{amssymb}
\usepackage{amsthm}
\usepackage{colonequals}
\usepackage{indentfirst}
\usepackage{xspace}
\usepackage{multirow}
\usepackage{hyperref}
\usepackage{xcolor}
\definecolor{darkred}{rgb}{0.5,0.15,0.15}
\hypersetup{colorlinks=true,urlcolor=darkred,linkcolor=darkred,citecolor=darkred}
\usepackage{verbatim}
\usepackage[letterpaper,margin=1in,headheight=15pt]{geometry}
\usepackage{mathpazo}
\usepackage{authblk}
\usepackage{tikz-cd}
\usepackage{capt-of}
\usepackage{stmaryrd}
\usepackage[missing=]{gitinfo2}

\usepackage{tikz}
\usetikzlibrary{calc}   
\usetikzlibrary{topaths}
\usetikzlibrary{decorations}
\usetikzlibrary{decorations.pathmorphing}

\numberwithin{equation}{section}

\newcommand{\eo}{{extended operator}}
\renewcommand{\o}{{operator}}
\newcommand{\op}{{\mathcal{O}}}

\newcommand{\be}{\begin{equation}}
\newcommand{\ee}{\end{equation}}
\newcommand{\bp}{\begin{pmatrix}}
\newcommand{\ep}{\end{pmatrix}}
\newcommand{\bsp}{\left(\begin{smallmatrix}}
\newcommand{\esp}{\end{smallmatrix}\right)}

\newcommand{\fg}{{\mathfrak g}}
\newcommand{\fh}{{\mathfrak h}}

\newcommand{\CA}{{\mathcal A}}
\newcommand{\CX}{{\mathcal X}}
\newcommand{\CL}{{\mathcal L}}
\newcommand{\CN}{{\mathcal N}}
\newcommand{\CO}{{\mathcal O}}
\newcommand{\CY}{{\mathcal Y}}
\newcommand{\pd}{\partial}
\newcommand{\ol}{\overline}
\newcommand{\ot}{\otimes}

\newcommand{\cA}{\ensuremath{\mathcal A}}
\newcommand{\cB}{\ensuremath{\mathcal B}}
\newcommand{\cC}{\ensuremath{\mathcal C}}
\newcommand{\cN}{\ensuremath{\mathcal N}}
\newcommand{\cM}{\ensuremath{\mathcal M}}
\newcommand{\cD}{\ensuremath{\mathcal D}}
\newcommand{\cL}{\ensuremath{\mathcal L}}

\newcommand{\cO}{\ensuremath{\mathcal O}}
\newcommand{\cP}{\ensuremath{\mathcal P}}

\newcommand{\cX}{\ensuremath{\mathcal X}}

\newcommand{\R}{\ensuremath{\mathbb R}}
\newcommand{\C}{\ensuremath{\mathbb C}}

\newcommand{\Z}{\ensuremath{\mathbb Z}}

\newcommand{\N}{{\mathcal N}}

\newcommand{\hk}{hyperk\"ahler\xspace}

\newcommand{\de}{\mathrm{d}}

\newcommand{\norm}[1]{\lVert#1\rVert}

\newcommand{\eps}{\epsilon}

\newcommand{\ti}[1]{\textit{#1}}

\newcommand{\cclass}[1]{{{\llbracket#1\rrbracket}}}
\newcommand{\pprod}{{*}}
\newcommand{\sprod}[2]{{\{#1,#2\}}}

\newcommand{\ops}{{\mathrm {Op}}}

\newcommand{\Spin}{{\mathrm {Spin}}}

\DeclareMathOperator{\img}{im}
\newcommand{\Gv}{G^\vee}

\newcommand{\Ahat}{\wh{A}_{G}}
\newcommand{\Bhatv}{\wh{B}_{G^{\vee}}}

\newcommand{\wh}{\widehat}

\newtheorem{dummy}{dummy}[section]              

\newtheorem{theorem}[dummy]{Theorem}

\newtheorem{proposition}[dummy]{Proposition}

\newtheorem{claim}[dummy]{Claim}

\newcommand{\insfigscaled}[3]{

\medskip
\noindent
\begin{minipage}{\linewidth}

\makebox[\linewidth]{\includegraphics[keepaspectratio=true,scale=#2]{#1.pdf}}

\captionof{figure}{#3}

\label{fig:#1}
\end{minipage}
\medskip

}

\begin{document}

\title{Secondary products in supersymmetric field theory}
\author[1,2]{Christopher Beem}
\author[3]{David Ben-Zvi}
\author[4]{Mathew Bullimore}
\author[5]{Tudor~Dimofte}
\author[3]{Andrew Neitzke}
\affil[1]{\small Mathematical Institute, University of Oxford, Woodstock Road, Oxford OX2 6GG, UK}
\affil[2]{St. John's College, University of Oxford, St. Giles Road, Oxford OX1 3JP, UK}
\affil[3]{Dept. of Mathematics, University of Texas, Austin, TX 78712, USA}
\affil[4]{Dept. of Mathematical Sciences, Durham University, Durham DH1 3LE, UK}
\affil[5]{Dept. of Mathematics and QMAP, UC Davis, One Shields Ave, Davis, CA 95616, USA}

\date{{{\tiny \color{gray} \tt \gitAuthorIsoDate}}
{{\tiny \color{gray} \tt \gitAbbrevHash}}}

\maketitle

{\abstract{The product of local operators in a topological quantum field theory in dimension greater than one is commutative, as is more generally the product of extended operators of codimension greater than one. In theories of cohomological type these commutative products are accompanied by secondary operations, which capture linking or braiding of operators, and behave as (graded) Poisson brackets with respect to the primary product. We describe the mathematical structures involved and illustrate this general phenomenon in a range of physical examples arising from supersymmetric field theories in spacetime dimension two, three, and four. In the Rozansky-Witten twist of three-dimensional $\cN=4$ theories, this gives an intrinsic realization of the holomorphic symplectic structure of the moduli space of vacua. We further give a simple mathematical derivation of the assertion that introducing an $\Omega$-background precisely deformation quantizes this structure. We then study the secondary product structure of extended operators, which subsumes that of local operators but is often much richer. We calculate interesting cases of secondary brackets of line operators in Rozansky-Witten theories and in four-dimensional $\cN=4$ super Yang-Mills theories, measuring the noncommutativity of the spherical category in the geometric Langlands program.}}

\vfill\eject

\tableofcontents

\setcounter{page}{1}

\section{Introduction}

Mathematical perspectives on topological quantum field theory (TQFT) have evolved signficantly since their initial axiomatization by Atiyah \cite{Atiyah:1989vu}, inspired by Segal's approach to conformal field theory (CFT) \cite{Segal:1987sk}. Atiyah defined a $d$-dimensional TQFT as a functor from the cobordism category of $d$-manifolds to the category of vector spaces, multiplicative under disjoint unions. The contemporary view of TQFT extends this structure in two directions. First, it takes into account homological structures that express the local constancy of the theory over spaces of bordisms of manifolds, thereby capturing aspects of the topology of these spaces. Such structures are ubiquitous in ``Witten-type'' or ``cohomological'' TQFTs obtained via topological twist from supersymmetric QFTs \cite{Witten:1988ze, Witten:1988xj}, and also in the setting of two-dimensional topological conformal field theories (TCFTs) \cite{Getzler:1994yd,Segal:99,Costello:2004ei}. Additionally, one may consider ``extended'' TQFTs, which express the locality of field theories in the language of higher categories of manifolds with corners, and capture additional physical entities such as extended operators and defects. The cornerstone of this edifice is the Cobordism Hypothesis \cite{Baez:1995xq,Lurie:2009keu,Ayala:2017wcr} (see \cite{Freed:2012hx} for an elementary review), which gives a powerful ``generators and relations'' description of fully extended cohomological TQFTs.

Our aim in this paper is to extract a key structure that emerges naturally from this mathematical formalism and understand it concretely in several familiar physical cohomological TQFTs. This exercise turns out to have practical benefits. From a physical point of view, it will illuminate certain fundamental structures in TQFTs (and indeed sometimes in the underlying non-topological QFTs) that seem to have been previously underappreciated or even unnoticed. In particular we arrive at a deeper understanding of the role of the Poisson bracket in three-dimensional TQFTs and the formal mechanism of its quantization by the $\Omega$-background. Mathematically, we gain access to a variety of rich examples coming from physics; we uncover some new features of well-known mathematical structures (such as the canonical deformation quantization of $E_d$-algebras by rotation equivariance); and we acquire an explicit understanding of certain phenomena whose previous characterization was more formal (\emph{e.g.}, the noncommutativity of $E_d$-categories, in particular the spherical category of the Geometric Langlands program). 

The structure we aim to address is the existence of \emph{higher products} in TQFT operator algebras.
We can briefly illustrate what we mean by this, in the simplest case of local operators.
  Recall that the local operators in a TQFT always form an algebra, with a primary product that we denote by `$*$' coming from the collisions
\begin{equation}
\label{intro-p1}
(\CO_1 * \CO_2)(y) = \lim_{x\to y} \CO_1(x)\CO_2(y)~.
\end{equation}
Topological invariance ensures that the limit in \eqref{intro-p1} is non-singular, and moreover that it does not depend on the manner in which the operators are brought together. Thus in dimension $d=1$ the product is associative, while in dimension $d\geqslant 2$, where operators can be moved around each other, the product is commutative.

Now suppose that a TQFT is of cohomological type, such as a twist of a supersymmetric theory. Then topological invariance only holds in the cohomology of a supercharge~$Q$. In particular, only the cohomology class of a collision product such as \eqref{intro-p1} is guaranteed to be well defined and independent of the way the limit is taken. Working on the ``chain level,'' \emph{i.e.} working with $Q$-closed operators themselves, expected properties such as commutativity may fail. This allows for the existence of secondary operations, akin to Massey products.

For cohomology classes of local operators, the most important secondary operation turns out to be a Lie bracket of degree $1-d$, which acts as a derivation with respect to the primary product.
This secondary product has a surprisingly simple and concrete physical definition in terms of \emph{topological descent}.
Topological descent was introduced in \cite{Witten:1988ze} (and further expanded upon in \cite{Moore:1997pc}) as a way to produce extended operators from local ones in cohomological TQFT. Recall that the $k$-th descendant $\CO^{(k)}$ of a $Q$-closed operator $\CO$ is a $k$-form on spacetime whose integral on any $k$-dimensional cycle is again $Q$-closed. The secondary product $\{\CO_1,\CO_2\}$ of two $Q$-closed operators (representing cohomology classes) may then be constructed by integrating the $(d-1)$-th descendent of $\CO_1$ around a small $S^{d-1}$ surrounding $\CO_2$~,
\begin{equation}
\label{intro-p2} 
\{\CO_1,\CO_2\} := \oint_{S^{d-1}_y}\CO_1^{(d-1)}(x)\,\CO_2(y)~.
\end{equation}
This is again a $Q$-closed local operator, and represents a well-defined cohomology class in the topological operator algebra.

\subsection{\texorpdfstring{$E_d$}{E(d)} and shifted Poisson algebras}
\label{sec:intro-Ed}

We take a moment to discuss the fundamental mathematical structures that give rise to secondary operations.
In the modern mathematical formulation of cohomological TQFT~\cite{Lurie:2009keu} --- as in the earlier formulation of TCFT \cite{Getzler:1994yd,Segal:99,Costello:2004ei} --- there are not only operations corresponding to individual bordisms, such as \eqref{intro-p1}, but there is in addition a family of homotopies identifying the operations as the bordism varies continuously over a moduli space. In particular, this means that the products of $n$ operators are encoded by the topology of the configuration space $\cC_n(\R^d)$ of $n$ points in $\R^d$ --- \emph{i.e.}, by the various ways that these operators can move around each other.

It is convenient to excise small discs around the operator insertion points, resulting in the homotopy equivalent space of embeddings of $n$ disjoint $d$-discs inside $\R^d$, or equivalently inside a sufficiently large disc. This leads to one of the fundamental algebraic notions of homotopical algebra, that of an $E_d$, or $d$-disc, \emph{algebra} --- an algebra over the operad of little $d$-discs. The algebra is endowed with multilinear operations parametrized by configurations of $d$-discs inside a large disc, and compositions governed by the combinatorics of embedding such configurations into still larger discs. If in addition we allow operations corresponding to rotations of the discs, the resulting structure is called an oriented $d$-disc algebra. (For some initial references, see the original sources~\cite{May:1972,Cohen:1976}, the recent review~\cite{Sinha} and the modern treatment in~\cite{HA}.%
\footnote{The $E_d$ terminology is the standard one in the math literature. We prefer the disc algebra terminology, as in the papers of Ayala-Francis, \emph{e.g.}~\cite{ayalafrancis,ayalafrancisPoincare} which, besides sounding less technical, matches the TQFT setting better: a framed TQFT gives rise to a framed disc algebra, which is an ordinary or unframed $E_d$ algebra, while an oriented TQFT gives rise to an oriented disc algebra, which is confusingly a framed $E_d$ algebra.})

Disc algebra structures make sense in a variety of algebraic contexts. These include on the level of cohomology (\emph{i.e.}, of graded vector spaces), on the chain level (\emph{i.e.}, on chain complexes), 
and on the level of categories or higher categories. Physically, these different situations arise when we consider, respectively, the $Q$-cohomology of local operators, the space of physical local operators considered as a chain complex up to quasi-isomorphism, and categories of extended operators (see Section~\ref{intro extended ops}).

At the level of cohomology, an $E_d$ or framed disc algebra becomes very simple: it is a graded variant of a Poisson algebra, known as a $P_d$-algebra or $d$-braid algebra (see the reviews~\cite{Sinha} and~\cite{Cattaneo:2006}). This means that in addition to the primary product, (cohomology classes of) local operators in a $d$-dimensional TQFT carry a \emph{secondary product}, a Lie bracket $\{\CO_1,\CO_2\}$ of degree $1-d$, which acts as a derivation with respect to the primary product. 
In terms of configuration space, the secondary product is associated to the top homology class of $\cC_2(\R^d)\simeq S^{d-1}$. Unraveling the mathematical formulation leads to concrete formulas for the secondary product involving topological descent, such \eqref{intro-p2}. We expand on the definition \eqref{intro-p2} and its relation to configuration space in Sections \ref{sec:algebras} and \ref{sec:top-2}.

If we forget the $\Z$-grading, we find a dichotomy between the case of $d$ odd, where a $P_d$ structure is a conventional Poisson structure, and the case of $d$ even, where a $P_d$ structure becomes an odd (fermionic) Poisson structure, better known as a Gerstenhaber structure.
Similarly, an oriented disc algebra (allowing rotations of the discs)  in odd dimensions is described by a Poisson algebra with an action of an additional exterior algebra (the homology of the orthogonal group)~\cite{WahlSalvatore}, while in even dimensions we find the so-called Batalin-Vilkovisky (BV) algebras (with an extra exterior algebra action for $d\geq 4$).

While we work primarily at the level of cohomology in this paper, the chain-level disc algebra structure on the space of physical local operators carries much richer information and is essential for many applications.%
\footnote{When working on the chain level, we will implicitly be using the machinery of $\infty$-categories, which is a formal language to manage structures defined up to coherent homotopies. In particular this means we can freely transport structures between quasi-isomorphic chain complexes. Thus, for example, the distinction between spaces of little discs in a $d$-disc or of points in $\R^d$ is suppressed, as is that between associative and $A_\infty$ or Lie and $L_\infty$ algebras.}
At the level of chains, disc structures do not boil down to separate primary (commutative) and secondary (Lie bracket) operations. However, one can extract from any chain-level disc algebra its ``Lie part", which forms (up to a degree shift) a homotopy Lie algebra, \emph{i.e.} $L_\infty$-algebra. The nontriviality of this $L_\infty$ structure forms an obstruction to the honest chain-level commutativity of the operator product. Just as in the more familiar case of $A_\infty$ algebras, the chain level $L_\infty$ structure can be detected on the level of cohomology using an infinite sequence of higher bracket operations (Massey products) $L_3,L_4,\dots$, which extend the
bracket operation $L_2$, see \emph{e.g.}~\cite{Lada:1992wc,Kontsevich-notes,Gaiotto:2015aoa, Gaiotto:2015zna}.

\subsubsection{Relation with shifted Poisson geometry and factorization algebras}
\label{subsubsec:Poisson_and_factorization}

Disc algebras are at the center of two of the most active areas of current research in physics-inspired geometry.

Shifted Poisson (or $P_d$) algebras form the local building blocks in the theory of shifted symplectic, and more generally shifted Poisson, geometry~\cite{PTVV,CPTVV} (see also the surveys~\cite{PantevVezzosi,Safronov}). This theoretical framework provides a powerful and general algebro-geometric setting for the AKSZ-BV construction of d-dimensional field theories~\cite{AKSZ} (in the more differential geometric setting of the Poisson sigma model and its generalizations see in particular the review~\cite{Cattaneo:2006}, the papers~\cite{Kotov:2010wr,Fiorenza:2011jr}, the survey page~\cite{nlab:AKSZ} and references therein). In particular, one can describe geometric objects with the property that spaces of maps from $d$-dimensional manifolds into them are locally (derived) critical loci of action functionals. These theories in turn provide the starting point for the perturbative construction of topological field theories by a process of deformation quantization. In particular~\cite{CPTVV,PantevVezzosi} discuss the construction of disc algebra structures from shifted Poisson spaces that are expected to match those found on local and line operators.

In the powerful approach to perturbative quantum field theory developed by Costello and Gwilliam~\cite{CG}, the observables carry the structure of {\em factorization algebras}. Factorization algebras first arose in the setting of two-dimensional chiral CFT in the work of Beilinson and Drinfeld~\cite{BD} as a geometric formulation of the theory of vertex algebras, \emph{i.e.}, of the meromorphic operator product expansion. An important perspective on disc algebras is as the topological (and thus simplest) instances of factorization algebras --- namely, by a theorem of Lurie, $E_d$ (\emph{i.e.} framed $d$-disc) algebras are identified with {\em locally constant} factorization algebras, \emph{i.e.}, the structure carried by observables of TQFTs~\cite{HA}.
This was discussed specifically in the context of topologically twisted supersymmetric theories and their holomorphically twisted cousins in the lectures \cite{CostelloScheimbauer}.
 The recent paper~\cite{ElliottSafronov} produces $E_d$-algebras from the factorization algebras of topologically twisted supersymmetric theories in the formalism of~\cite{CG} by analyzing the subtle distinction between cohomological trivialization of the stress tensor, the infinitesimal (or ``de Rham'') form of topological invariance, and topological invariance (local constancy) in the stronger (``Betti") sense. 

Note that constructing a factorization algebras of observables
requires extra structure in a field theory, \emph{e.g.}, a Lagrangian formulation. It would be interesting to measure the precise distance between the $E_d$-algebras of local operators in TQFT and the factorization algebras of observables in a perturbative TQFT, built by way of this general formalism.

\subsection{Two-dimensional theories}
\label{sec:intro-2d}

In dimension $d=2$, the Gerstenhaber and BV algebra structures on the cohomology of local operators is well known. In the context of the BRST cohomology of topological conformal field theories this structure was constructed by Lian and Zuckerman in the early 1990's \cite{Lian:1992mn} (see also \cite{Penkava:1992sh}). The relevance of operads in topological field theory was discovered by Kontsevich (see in particular~\cite{Kontsevich93}), and Getzler proved that the operad controlling the structure of operators in oriented two-dimensional TQFT is identified with that of BV algebras~\cite{Getzler:1994yd}. There has also been extensive work lifting this structure to the chain level. The underlying $L_\infty$-algebra is identified with the fundamental homotopy-Lie-algebra structure of string field theory \cite{Witten:1992yj}. 

The Gerstenhaber and BV algebra structure on local operators plays an important role also in the study of mirror symmetry. In the B-model with a K\"ahler target $\CX$, the local operators are given by the Dolbeault cohomology of polyvector fields and the secondary product is induced by the Schouten-Nijenhuis bracket. (We rederive this in detail in Section~\ref{sec:b-model}.) Thus the bracket is interesting for example for $\CX=\C$ or $\CX={\mathbb P}^n$. However, for $\CX$ compact Calabi-Yau, Hodge theory combines with the theory of BV algebras to prove vanishing of the bracket on Dolbeault cohomology, and to deduce the Tian-Todorov unobstructedness of deformations (see~\cite{KKP08,KKP14} for the modern perspective). Dually, in the A-model, a nonvanishing bracket requires a noncompact target and twist fields, described mathematically via symplectic cohomology \cite{Seidel}, \emph{cf.} \cite{Ganatra, GPS}.
In the case of A-twisted Landau-Ginzburg models, the bracket (in fact, the entire $L_\infty$ structure on local operators) was recently studied in \cite{Gaiotto:2015aoa, Gaiotto:2015zna}.

In a 2d TQFT, the local operators can also be identified with the Hochschild cohomology of the category of boundary conditions (or D-branes) \cite{Kontsevich:1994dn, Costello:2004ei,KontsevichSoibelman:2009, Kapustin:2004df}. In this guise the secondary product matches the original appearance of Gerstenhaber algebras \cite{Gerst} as the structure carried by the Hochschild cohomology of associative algebras, and more generally the Hochschild cohomology of categories (while Hochschild cohomology of Calabi-Yau categories carries a BV algebra structure). The chain-level lift of the Gerstenhaber bracket to an $E_2$ structure on Hochschild cochains is a fundamental result in homotopical algebra, known as the Deligne conjecture~\cite{Tamarkin:98, McClureSmith:99, KontsevichSoibelman:2000, BergerFresse:2001}, see also~\cite{HA}.

\subsection{Three-dimensional theories and the \texorpdfstring{$\Omega$}{Omega} background}
\label{subsec:intro_three_dimensional}

We spend a large part of this paper studying the secondary bracket in the case ${d=3}$. We find a rich set of examples --- new, to the best of our knowledge, to the physics literature --- with concrete applications.

In odd dimensions, the secondary product defines an even, \emph{i.e.} bosonic, Poisson bracket on the algebra of local operators.
We devote Section~\ref{sec:3dn=4} to the Rozansky-Witten twist of three-dimensional $\CN=4$ theories, where we find that the descent operation \eqref{intro-p2} captures the \emph{geometric} Poisson bracket on a holomorphic symplectic target space. 
Applying this result to physical 3d $\CN=4$ gauge theories, we quickly deduce that the Poisson bracket on the Higgs and Coulomb branch chiral rings is intrinsically topological, and thus not renormalized.
We also show that for sigma-models with compact targets, such as those originally studied by Rozansky and Witten \cite{Rozansky:1996bq}, the secondary bracket on topological local operators vanishes, just as it does in the 2d B-model on compact CY manifolds.

Secondary products in higher dimensions turns out to provide a useful perspective on $\Omega$-backgrounds \cite{Moore:1997dj, Moore:1998et, Nekrasov:2002qd,Nekrasov:2003rj}. In the physics literature, it has been argued from a variety of angles that an $\Omega$-background can give rise to quantization of operator algebras, \emph{e.g.} \cite{Nekrasov:2009rc, Shadchin:2006yz, Dimofte:2010tz, Yagi2014, Dimofte:2009bv, Alday:2009fs, Drukker:2009id, Teschner:2010je, Gaiotto:2010be, Nekrasov:2010ka}. From a TQFT perspective, turning on an $\Omega$-background  amounts to working equivariantly with respect to rotations about one or more axes in $d$-dimensional spacetime. Such rotations of spacetime induce an action on the configuration spaces that control products in a $d$-disc algebra. One then expects the $\Omega$-background to lead to deformations of disc algebras whose products are controlled by the \emph{equivariant homology} of configuration space\footnote{Note that the action of the orthogonal group, and hence the corresponding equivariant deformations, are part of the structure of an oriented disc algebra.}.

In the case of $d=3$, we can make this idea quite concrete. The configuration space $\cC_2(\R^3)\simeq S^2$ has two homology classes: the point class $[p]$, inducing the primary product of local operators, and the fundamental class $[S^2]$, inducing the Poisson bracket. After turning on equivariance with respect to rotations about an axis, localization gives us an identity
\begin{equation}
\epsilon [S^2] = [N] - [S]  \qquad \text{in~~~} H^{U(1)}_\bullet(S^2)~,
\end{equation}
where $\epsilon$ is the equivariant parameter and $[N]$, $[S]$ are the equivariant cohomology classes of the fixed points at the North and South poles. Translating this identity to products in the operator algebra, we find
\begin{equation}
\epsilon \{\CO_1,\CO_2\} = \CO_1 * \CO_2 - \CO_2 * \CO_1~,
\end{equation}
with the RHS encoding the difference of primary products taken in opposite orders along the fixed axis of rotations --- in other words, a commutator. It follows that 
the $\Omega$-deformation produces a canonical ``deformation quantization" of topological local operators with their secondary bracket.%
\footnote{The general formalism does not however guarantee {\em flatness} of the deformation. For example, the space of states in the B-model is deformed by the $\Omega$-background from Dolbeault to de Rham cohomology, and thus jumps radically for noncompact targets. It would be interesting to find physical mechanisms that do ensure flatness.} %

We discuss this topological approach to quantization further in Section~\ref{sec:omega}. In the special case of 3d Rozansky-Witten theory, it offers a precise topological explanation of a physical result of Yagi~\cite{Yagi2014} (also related to deformation quantization on canonical coisotropic branes in 2d A-models \cite{Kapustin:2001ij, Kapustin:2006pk, Gukov:2008ve, Nekrasov:2010ka}).

More generally, turning on equivariance around a single axis in $d$ dimensions deforms an $E_d$ algebra to an $E_{d-2}$ algebra. This deformation is defined and studied in \cite{BZNeitzke}.
In three dimensions, the equivariant form of the $E_3$ operad is identified with a graded version of the so-called $BD_1$ operad, which controls deformation quantizations. In the case of $d=2$, one recovers Getzler's theorem~\cite{Getzler:1994yd}.

\subsection{Extended operators}
\label{intro extended ops}

In the final sections of the paper we begin an investigation of the rich structures arising from higher products involving extended operators. $k$-dimensional extended operators have a primary product in which the extended operators are aligned in parallel and brought together in the transverse dimensions. Moreover, as with local operators, the extended operators can be moved around each other in the transverse $d-k$ directions, resulting in additional operations controlled by the topology of the configuration spaces of points (or little discs) in $\R^{d-k}$, \emph{i.e.}, a $(d-k)$-disc structure.

Extended operators in isolation are already more complicated entities than local operators. Topological line operators naturally form a category, in which individual line operators are objects and the morphisms between two lines are given by the topological interfaces between them. The associative composition of morphisms is given by the collision of interfaces. Likewise, $k$-dimensional extended operators have the structure of a $k$-category, with higher morphisms given by interfaces between interfaces (see, \emph{e.g.}, \cite{Kapustin:2010ta} for a physical exposition of this mathematical structure). Extended operators with $k$-dimensional support in a $d$-dimensional TQFT thus have the structure of a $(d-k)$-disc $k$-category.

In this paper we will only take the first steps in exploring this structure, restricting our attention to line operators --- thus, to ordinary categories, or ``one-categories.''
In general terms, the structures associated to line operators should be as follows:
\begin{itemize}
\item In $d=2$, line operators form self-interfaces of the theory itself. The 1-disc structure is simply the associative composition of interfaces, or at the chain-level, the homotopy-associative (\emph{i.e.}, $E_1=A_\infty$) lift of this composition~\cite{Gaiotto:2015aoa, Gaiotto:2015zna}. Here the transverse configuration space does not give rise to any additional structures. 

\item In $d=3$, we encounter in this fashion the familiar braiding of line operators, such as the braiding of Wilson lines in Chern-Simons theory. Indeed an $E_2$ structure on an abelian category is identified with the more familiar notion of braided tensor category (see \emph{e.g.} Example 5.1.2.4 in~\cite{HA}). As we review, in the setting of Rozansky-Witten theory line operators are given by the derived category of coherent sheaves, where this braided structure is described somewhat indirectly~\cite{Rozansky:1996bq, RobertsWillerton,Kapustin:2008sc} (see also~\cite{BFN} where the local version is described as a Drinfeld center).

\item In $d=4$, we see nothing interesting on the level of abelian categories, since $E_3$ structures on abelian categories reduce to symmetric monoidal (\emph{i.e.}, commutative tensor) structures. However, on the derived level the $E_3$ structure can be nondegenerate (shifted symplectic).
\end{itemize}
We thus set as a goal to understand concretely the facets of the disc algebra structure on derived categories of line operators, in particular in Rozansky-Witten theory ($d=3$) and Geometric Langlands (GL) twisted $\cN=4$ super Yang-Mills theories ($d=4$).

First, by considering local operators as self-interfaces of the ``trivial line" one can recover the disc structure on local operators. This is in fact precisely the information captured by the perturbative construction of factorization algebras of observables as in~\cite{CG}, and is utilized to great effect in~\cite{CostelloFrancis} and~\cite{Costello:13} to recover the braided categories of representations of quantum (loop) groups in perturbative Chern-Simons theory and 4d gauge theories.  This approach suffices to describe all line operators in the case of RW theory on an affine target, but not in general. In 4d GL-twisted Yang-Mills, it gives no information at all, because the fermionic Poisson bracket vanishes on the completely bosonic ring of topological local operators. 

We thus probe the next level of structure carried by topological line operators 
 by defining a secondary product between local operators and line operators. (The mathematical formalism underlying this construction is developed in~\cite{AyalaFrancisTanaka}, though we are not aware of any literature from either the mathematical or physical tradition where this structure is fully expressed or explored in examples.) The secondary bracket is defined much as for pairs of local operators: we integrate the descendant of a local operator on a small sphere linking a line operator. This defines an action of local operators as self-interfaces of any line operator which, as we explain, are \emph{central}: they commute with all other interfaces between lines. We also introduce briefly in Section~\ref{further OPEs} a further level of structure, defining a new line operator as the secondary product of a pair of line operators.

In Section~\ref{sec:extended RW example} we calculate this secondary product of local and line operators in Rozansky-Witten theory by first showing how to describe secondary products in 3d as primary products in the reduction of the theory on a circle. We then interpret the secondary product geometrically as describing the flow of the line operator (coherent sheaf) along a Hamiltonian vector field defined by the local operator.

Finally, in Section~\ref{sec:4d} we investigate the secondary product of local and line operators in four dimensions, in the gauge theoretic setting for the Geometric Langlands Program introduced in~\cite{Kapustin:2006pk}. We consider the GL twist of $\cN=4$ super-Yang-Mills with gauge group $G$ with the canonical values $\Psi=0$ of the twisting parameter (the ``$\Ahat$ model'' in the terminology of~\cite{Witten:2009mh}) and its $S$-dual description, GL-twisted $\cN=4$ with gauge group $\Gv$ and $\Psi=\infty$ (the ``$\Bhatv$ model''). This theory carries topological local operators given by invariant polynomials of an adjoint-valued scalar field (the equivariant cohomology of a point). There are also topological line operators, in particular the topological Wilson line operators in $\Bhatv$ and topological 't Hooft line operators in $\Ahat$, both labelled by representations of the dual group $G^\vee$.

We find that the secondary product of local operators with these line operators is highly nontrivial: it defines an action of a $\text{rank}(G)$-dimensional abelian Lie algebra (a principal nilpotent centralizer) by central self-interfaces on the line operators, see Theorem~\ref{infi Ngo redux}.  We refine and reinterpret in this way the construction of Witten~\cite{Witten:2009mh}, who found (by studying a particular three-dimensional configuration) the effect of this action on the underlying vector space of a corresponding $\Gv$ representation, thereby giving a physical interpretation to a construction of Ginzburg~\cite{Ginzburg1995}. In particular this measures explicitly (we believe for the first time) the noncommutativity of the 3-disc structure on the category of line operators, also known as the spherical (or derived Satake) category, one of the central objects in the geometric Langlands program\footnote{The existence of a 3-disc structure on the spherical category was first observed by Lurie in 2005. It is mentioned in~\cite{LurieICM,Toenbranes,ArinkinGaitsgory} -- a related construction appears in~\cite{BFN} -- and the factorization homology of this 3-disc structure is described in~\cite{darioTFT}.}. Our action is an infinitesimal version of the \emph{Ng\^o action} studied in~\cite{BZG}, where a group of central symmetries of the spherical category was constructed.

\subsection{Further directions}
We conclude the introduction by briefly mentioning three important problems in which we expect the higher product structure on operators to play a central role: deformations, higher-form symmetries, and holomorphic twists.

There is a strongly expected relation between the homotopy Lie algebra structure of local operators 
and the deformation theory of the TQFT,
though we are not aware of a precise general formulation in the literature beyond $d=2$. 
On the one hand, it is well-known that one can use descendants of local operators to deform a TQFT.
On the other hand, there is a long-standing philosophy associated with Deligne, Drinfeld, Feigin, and 
Kontsevich (see for example~\cite{KontsevichSoibelmanBook, Kontsevich-notes}) that formal deformation problems are associated with dg or homotopy Lie algebras, as the spaces of solutions of the associated Maurer-Cartan equations. In the setting of derived algebraic geometry, this philosophy becomes the general Koszul duality equivalence~\cite{LurieICM} between Lie algebras and formal moduli problems.

We expect that the $L_\infty$ algebra given by the chain-level bracket of local operators controls a space of deformations of the corresponding TQFT, in any dimension.%
\footnote{It's important to note that the entire disc algebra structure (not just the $L_\infty$ part) is important for the deformation problem --- $d$-disc algebras~\cite{LurieICM,ayalafrancisPoincare} define enhanced ``slightly noncommutative" formal moduli spaces, whose rings of functions are themselves $d$-disc algebras.} %
(For discussions of the relation of $L_\infty$-algebras and the space of deformations a quantum field theory see~\cite{Park:2003kn,SoibelmanCFT,Gaiotto:2015aoa, Gaiotto:2015zna}.) In particular, this is well known in two dimensions, where Hochschild cohomology controls the deformations of dg categories of boundary conditions and hence their associated TQFTs. 
In higher dimensions, however, we expect to need extended operators to describe all deformations of a TQFT. For example, in Rozansky-Witten theory local operators only control the {\em exact} deformations of the underlying holomorphic symplectic manifold, while line operators can be used to produce deformations that vary the class of the holomorphic symplectic form. 

The $(d-k)$-disc structure of $k$-dimensional extended operators also plays a key
role in the theory of generalized global symmetries of quantum field theories~\cite{Gaiotto:2014kfa}, which we mention very brielfy. Namely, in TQFT, one can define the notion of a $(p-1)$-form symmetry as the data of an $E_{p}$-space $G$ and an $E_{p}$-map from $G$ to the $E_{p}$-algebra of codimension-$p$ extended operators.

For example, for $p=1$, $E_1$-spaces are (homotopical versions of) monoids. We thus find a monoid acting by ordinary ($0$-form) symmetries of a TQFT, via a homomorphism to the monoid (in fact $E_1$ $(d-1)$-category) of self-interfaces 
of the theory. We will encounter a particular example of this when considering flavor symmetries in Rozansky-Witten theory, in Section \ref{sec:3d-sym}.
For $p=2$, an $E_2$-space is a suitable homotopic version of a commutative monoid, and we can ask for such an object --- rather than just a commutative group as in~\cite{Gaiotto:2014kfa} --- to act by 1-form symmetries of a TQFT, via codimension-2 extended operators.

One other interesting future direction would be to investigate higher products in holomorphically (rather than topologically) twisted supersymmetric theories, such as the twists of 4d $\CN=1$ and $\CN=2$ theories introduced by \cite{Johansen:1994aw, Kapustin:2006hi}. Holomorphic twists were studied in the factorization algebra formalism of \cite{CG} in \emph{e.g.} \cite{Costello:2011np, Costello:13, Elliott:2015rja}. They form an important bridge between full, physical SUSY QFTs and topologically twisted ones, and they admit higher products closely analogous to those of TQFTs, whose general structure was briefly outlined in \cite{CostelloScheimbauer}. A particularly interesting higher product in (hybrid) holomorphically-topologically twisted 4d $\CN=1$ gauge theory was computed in \cite{Costello:13}, where it played a central role in defining the coproduct for a Yangian algebra. 
There seem to be many other interesting examples to uncover; however, in this paper we will focus on the --- already rich --- case of purely topological twists instead.

\subsection{Acknowledgments} 
The preliminary ideas that led to this project congealed during seminars and discussions at the 2015 BIRS program Geometric Unification from Six-Dimensional Physics, attended by all five authors.
Further developments occurred while MB, TD, AN, and DBZ participated in the Aspen Center for Physics program on Boundaries and Defects in QFT, supported by National Science Foundation grant PHY-1066293.  We would like to express our gratitude to BIRS and ACP for their hospitality. 
 
DBZ would like to thank Sam Gunningham and Pavel Safronov for numerous helpful discussions, in particular in relation to~\cite{BZG,pavel}, and John Francis for teaching him about disc algebras ever since~\cite{BFN}. DBZ would like to acknowledge the National Science Foundation for its support through individual grant DMS-1705110. Parts of DBZ's research were supported by a Membership at the Institute for Advanced Study (as part of the Program on Locally Symmetric Spaces) and a Visiting Fellowship at All Souls College Oxford.

MB gratefully acknowledges support from ERC STG grant 306260 and the Mathematical Institute at the University of Oxford where part of this work was completed. 

TD would like to thank Kevin Costello, Justin Hilburn, and Philsang Yoo for many insightful discussions. He would also like to thank the students in his UC Davis graduate topics course (MAT 280: \emph{QFT and Representation Theory}) for helping work through examples of secondary products in 2d and 3d, and asking illuminating questions. TD is grateful to St. John's College, Oxford for its hospitality via a Visiting Fellowship.
TD's research is supported in part by NSF CAREER grant DMS-1753077.

AN thanks the National Science Foundation for its support through
individual grants DMS-1151693 and DMS-1711692, and the
Simons Foundation for its support through a 
Simons Fellowship in Mathematics.

We would also like to thank Kevin Costello and Greg Moore for helpful feedback in the final preparation of this manuscript.

\section{Algebras of topological operators}
\label{sec:algebras}

We now go back to basics. Our goal is to define physically, from first principles, the primary and secondary products on cohomology classes of local operators in a TQFT of cohomological type, and to understand their interplay with configuration spaces.

In this section we recall the structure the primary product and its properties. We will establish our conventions and assumptions for describing algebraic structures in cohomological TQFTs and their cousins. Though the constructions are standard, we make an effort to treat clearly the underlying chain-level structures that will be responsible for the higher operations described in the following section. In this paper we will only be interested in local structures in spacetime, so we will work exclusively in flat $d$-dimensional Euclidean space $\R^d$.

\subsection{The topological sector}

\label{subsec:topological_sector}

The basic symmetry structure underlying a cohomological TQFT is the twisted super-Poincar\'e algebra generated by charges $\{P_\mu, Q, Q_\mu\}$, with $P_{\mu}$ being the generator of spacetime translations in the $x^\mu$ direction, and odd supercharges $Q$ and $Q_\mu$ obeying
\begin{equation}  \label{eq:QQ-P}
Q^2 = [Q_\mu, Q_{\nu}] = [Q_\mu, P_\nu] = [Q, P_\nu] = 0~, \quad 	[Q, Q_\mu] = iP_\mu~.
\end{equation}
Throughout this paper we will use the graded commutator $[a,b] := ab - (-1)^{F(a) F(b)} ba$, where $F$ is the $\Z/2\Z$ fermion number. In situations where fermion number can be lifted to a $\Z$ grading, we assume that $F(Q)=1$ and $F(Q_\mu)=-1$.

Standard examples of this structure arise from topological twisting of supersymmetric field theories. In such cases, $Q$ (resp. $Q_\mu$) is not a scalar (resp. vector) under the ordinary Euclidean rotation group $\Spin(d)_E$. Rather it is scalar (resp. vector) under an ``improved'' rotation group $\Spin(d)' \subset G_R \times \Spin(d)_E$, where $G_R$ is the $R$-symmetry group.%
\footnote{The existence of a (non-anomalous) improved rotation group enhances a $d$-disc structure to a framed $d$-disc structure, as explained in Section \ref{sec:intro-Ed} of the Introduction. For most of our purposes in this paper, it will not matter if such an improved rotation group can actually be defined, though it will be present in our examples.}

Let $\ops_\delta$ denote the vector space of states associated to a sphere of radius $\delta$. In a conformal field theory, the state-operator correspondence gives a basis for $\ops_\delta$ consisting of local operators inserted at the center of the sphere. In a more general theory, $\ops_\delta$ might be larger: it is equivalent to the vector space of all operators, local or otherwise, with support strictly inside the ball of radius $\delta$, including multiple-point insertions (Figure \ref{fig:OpsB}). 
We will nevertheless abuse language and call an element of $\ops_\delta$ a ``local operator.'' The infinitesimal symmetries $P_\mu$, $Q$, and $Q_\mu$ act on the space $\ops_\delta$. In particular, we can consider  $\op \in \ops_\delta$ with
\begin{equation}
	Q (\op) = 0.
\end{equation}
We call these \ti{topological operators}.

\insfigscaled{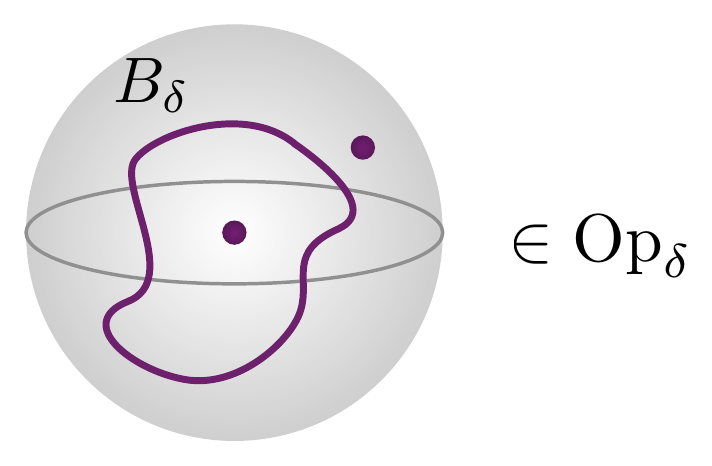}{0.65}{An illustration of an element of $\ops_\delta$, coming from a multi-point insertion of ordinary local operators and a compact loop operator, all supported in the ball $B_\delta$.}

We consider topological operators modulo $Q$-exact operators, \emph{i.e.}, the cohomology of $Q$ acting on $\ops_\delta$, 
\begin{equation}
\label{eq:define_op_cohomology}
\cA_\delta = \frac{\ker Q}{\img Q}~.
\end{equation}
Passing to this quotient is automatic in QFT: once we restrict our attention exclusively to $Q$-closed operators, any correlation function involving a $Q$-exact operator will vanish. Thus, after restricting to $Q$-closed operators, $Q$-exact operators are equivalent to zero.
Note also that for any $\delta > \delta'$, the path integral over the annulus $B_\delta \setminus B_{\delta'}$ induces a canonical map $\cA_\delta \to \cA_{\delta'}$. We assume that this map is actually an isomorphism; thus, once we pass to $Q$-cohomology, the size of the ball we consider is irrelevant. With this in mind, from now on we suppress the label $\delta$ and just call the $Q$-cohomology $\cA$.

\subsection{The topological algebra}
\label{sec:top-algebra}

Let us fix a large ball $B_1 \subset \R^d$ of size (say) $1$, and a pair of points $(x_1, x_2) \in B_1^2$ with $x_1 \neq x_2$. Given operators $\op_1, \op_2 \in \ops_\delta$ with $\delta\ll |x_1-x_2|$, we can construct an element of $\ops_1$ by inserting balls $B_\delta^{(1)},B_\delta^{(2)}$ containing $\op_1, \op_2$ (respectively) inside $B_1$. This defines a map of vector spaces
\begin{equation} \label{*Ops}
\pprod_{x_1,x_2}:  \begin{array}{ccc}	 \ops_{\delta} \otimes \ops_\delta &\to& \ops_1 \\[.2cm]
       \op_1\,,\;\;\op_2 & \mapsto & \op_1(x_1) \op_2(x_2)\,. \end{array}
\end{equation}
(When $\op_1,\op_2$ are ordinary local operators supported at isolated points, $\op_1(x_1) \op_2(x_2)$ is an ordinary two-point insertion in the path integral.)
The map $\pprod_{x_1,x_2}$ depends in a nontrivial way on the precise insertion
points $x_1$, $x_2$.

Upon passing to $Q$-cohomology classes, we obtain a much simpler structure. As the $Q$-cohomology of $\ops_\delta$ is independent of the radius $\delta$, \eqref{*Ops} induces a product operation
\begin{equation}
\pprod_{x_1,x_2}: \cA \otimes \cA \to \cA~,
\end{equation}
namely 
\begin{equation}
\label{eq:pprod-x1x2}
\cclass{\op_1} \pprod_{x_1,x_2} \cclass{\op_2} = \cclass{\op_1(x_1) \op_2(x_2)}~,
\end{equation}
where $\cclass{\cdot}$ denotes a $Q$-cohomology class. The product $\pprod_{x_1,x_2}$ is invariant under continuous deformations of $(x_1,x_2)$ as long as $x_1 \neq x_2$; this follows from the $Q$-exactness of translations as given in \eqref{eq:QQ-P}. Indeed, for an infinitesimal variation of $x_2$ we have
\begin{align}
	\partial_{x_2^\mu} \cclass{\op_1(x_1) \op_2(x_2)} &= \cclass{\op_1(x_1) \partial_{\mu} \op_2(x_2) } \\
	&= \cclass{\op_1(x_1) Q Q_\mu (\op_2(x_2))} \\
	&= (-1)^{F_1} \cclass{Q \left(\op_1(x_1) Q_\mu (\op_2(x_2)) \right)} \\
	&= 0\,,
\end{align}
and similarly for variations of $x_1$.

Said otherwise, if we define the \ti{configuration space}
\begin{equation}
		\cC_2(B) = \{ (x_1, x_2) \in B^2 \, \vert \, x_1 \neq x_2 \}
\end{equation}
then $\pprod_{x_1,x_2}$ depends only on the \ti{connected component}
of $\cC_2(B)$ in which $(x_1, x_2)$ lies.

\subsubsection{Topological algebra in dimension \texorpdfstring{$d \geqslant 2$}{d >= 2}}
\label{sec:products-dge2}

The simplest case is when the spacetime dimension $d$ at least two. In that case $\cC_2(B)$ is homotopic to $S^{d-1}$, and has only one connected component (we can interpolate from any $(x_1,x_2)$ to any other $(x'_1,x'_2)$ while keeping the points distinct.) Therefore, there is just a single product $\pprod$.
\insfigscaled{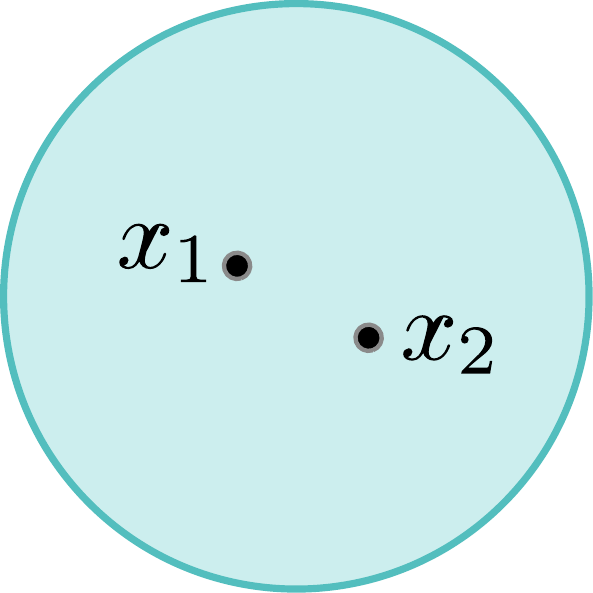}{0.6}{A point of the connected space $\cC_2(B)$ for $d = 2$.}
\noindent Moreover, $\pprod$ is \ti{graded-commutative}: to see this, pick any $x_1 \neq x_2$ and note
\begin{equation} 
\label{gr-comm}
	\cclass{\op_1} \pprod \cclass{\op_2} = \cclass{\op_1} \pprod_{x_1,x_2} \cclass{\op_2}  = (-1)^{F_1 F_2} \cclass{\op_2} \pprod_{x_2,x_1} \cclass{\op_1} = (-1)^{F_1 F_2} \cclass{\op_2} \pprod \cclass{\op_1}\,.
\end{equation}

\subsubsection{Topological algebra in dimension \texorpdfstring{$d=1$}{d=1}}
\label{sec:products-1d}

In one dimension the story is slightly different because there is not enough room to move $x_1$ and $x_2$ past one another without a collision. Said otherwise, $\cC_2(B)$ has two connected components,
\begin{equation}
\cC_{2,a}(B) = \{(x_1,x_2) \in B^2: x_1 < x_2\}~, \quad
\cC_{2,b}(B) = \{(x_1,x_2) \in B^2: x_1 > x_2\}~.	
\end{equation}
\insfigscaled{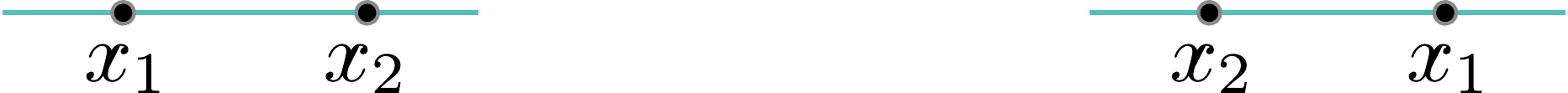}{0.6}{Left: a point of the component $\cC_{2,a}(B)$. Right: a point of the component $\cC_{2,b}(B)$.}
\noindent Consequently there are two product operations, $\pprod_a$ and $\pprod_b$, on $\cA$.
These two products are related by swapping the arguments; indeed, choosing any $x_1 < x_2$, 
\begin{equation}
	\cclass{\op_1} \pprod_a \cclass{\op_2} = \cclass{\op_1} \pprod_{x_1,x_2} \cclass{\op_2}  = (-1)^{F_1 F_2} \cclass{\op_2} \pprod_{x_2,x_1} \cclass{\op_1} = (-1)^{F_1 F_2} \cclass{\op_2} \pprod_b \cclass{\op_1}\,.
\end{equation}
Thus we can again restrict our attention to the single product $\pprod := \pprod_a$ without losing any information. Unlike the $d \geqslant 2$ case, though, here $\pprod$ need not be graded-commutative. (When $d \geqslant 2$ the product $\pprod$ is graded-commutative because we could continuously exchange $x_1$ and $x_2$; in $d=1$ there is not enough room to do this, so there is no reason for $\pprod$ to be graded-commutative.)

\subsubsection{Associativity}
\label{subsubsec:associativity}

The final elementary point is that the product $\pprod$ is \ti{associative} in any dimension. To see this we consider a class
\begin{equation}
\cclass{\op_1(x_1) \op_2(x_2) \op_3(x_3)}~,
\end{equation}
corresponding to the insertion of three small balls (containing $\op_1,\op_2,\op_3$, respectively) inside a large ball $B$.
For simplicity of exposition suppose the dimension is $d \geqslant 2$. Then by a continuous deformation we can arrange that $x_1$ and $x_2$ lie in a ball $B' \subset B$ with $x_3 \notin B'$, as shown in the top row of Fig.~\ref{fig:triple-product}. Next we replace the two operators $\op_1(x_1)$ and $\op_2(x_2)$ by a single operator $\op_{12}(x_{12})$, where $\op_{12}$ lies in the class $\cclass{\op_1 \pprod \op_2}$; after this replacement we have operators $\op_{12}(x_{12})$ 
and $\op_3(x_3)$ on the ball $B$; this configuration of operators
is in the class $\cclass{(\op_1 \pprod \op_2) \pprod \op_3}$,
so we get
\begin{equation} \label{eq:assoc-1}
	\cclass{\op_1(x_1) \op_2(x_2) \op_3(x_3)} = \cclass{(\op_1 \pprod \op_2) \pprod \op_3}\,.
\end{equation}
By instead bringing $x_2$ and $x_3$ together, as shown in the bottom
row of Figure \ref{fig:triple-product}, we get
\begin{equation} \label{eq:assoc-2}
	\cclass{\op_1(x_1) \op_2(x_2) \op_3(x_3)} = \cclass{\op_1 \pprod (\op_2 \pprod \op_3)}\,.
\end{equation}
Combining \eqref{eq:assoc-1} and \eqref{eq:assoc-2}
gives the desired associativity,
\begin{equation}
	\cclass{(\op_1 \pprod \op_2) \pprod \op_3} = \cclass{\op_1 \pprod (\op_2 \pprod \op_3)}\,.
\end{equation}
The key topological fact we used in this argument is that
the configuration space $\cC_3(B)$ is connected:
this allows us to interpolate between the configuration
with $x_1$ and $x_2$ close together and the configuration
with $x_2$ and $x_3$ close together.

\insfigscaled{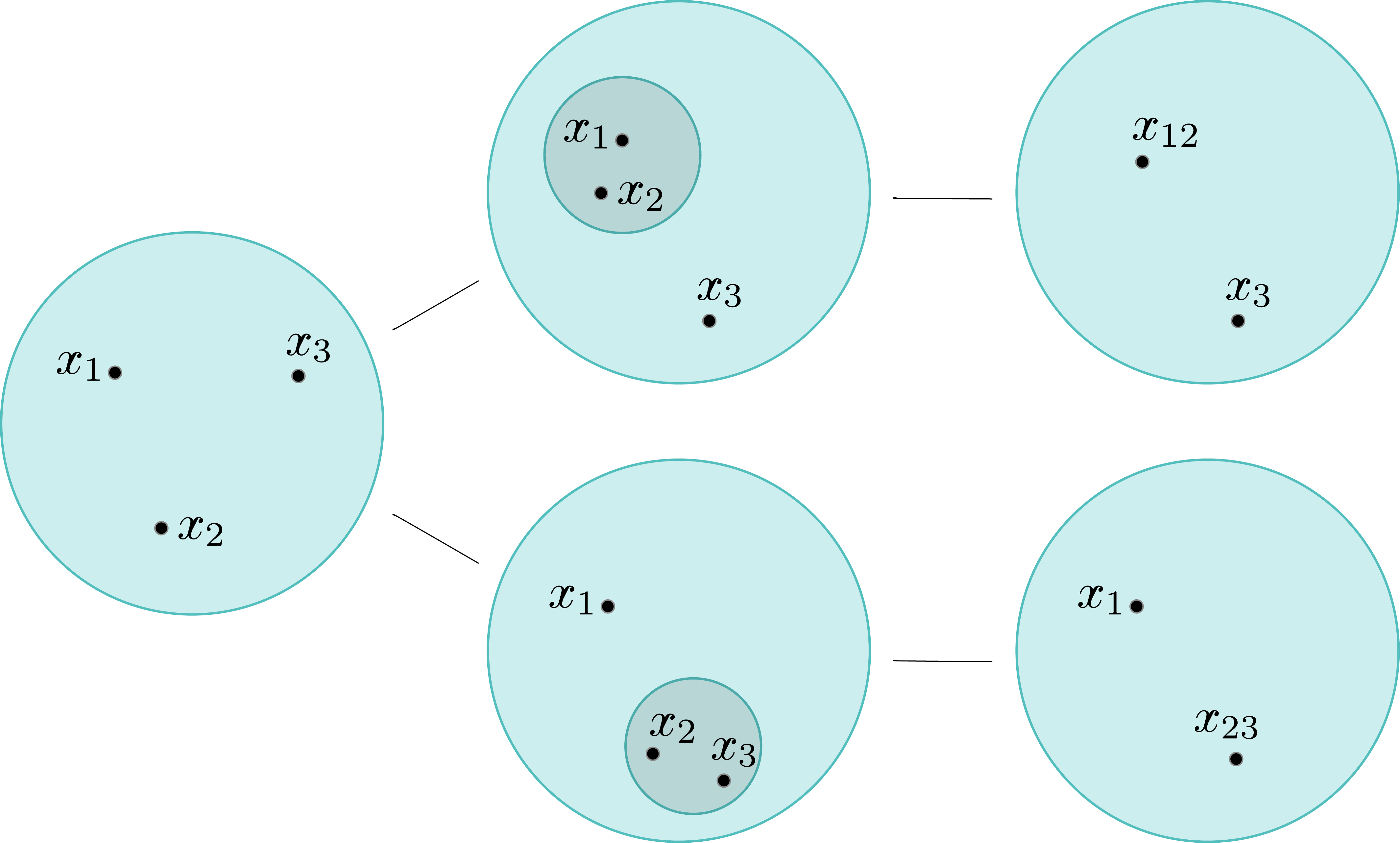}{0.25}{Connectedness of $\cC_3(B)$ leads to associativity of $*$.}

A similar argument establishes the associativity in dimension $d=1$.
In this case $\cC_3(B)$ has six components, corresponding
to the six possible orderings of the $x_i$.

\section{The Poisson structure on topological operators}
\label{sec:top-2}

In this section we develop the second, less standard way to multiply operators in topological theories: the ``secondary product.''
The secondary product promotes the associative graded-commutative algebra $\cA$ to a (super) Poisson algebra when $d$ is odd and a Gerstenhaber algebra when $d$ is even. In many examples arising from twisted supersymmetric theories, the $\Z/2\Z$-grading of $\cA$ has a refinement to a $\Z$-grading that is identified with an $R$-charge in the supersymmetric theory. In such examples, we can say more uniformly that $\cA$ inherits a {\em $P_d$-algebra structure} -- a graded Poisson bracket of degree $1-d$.

\subsection{Topological descent} 
\label{sec:descent}

To formulate the secondary product, we need to review the notion of \ti{topological descent} introduced in \cite{Witten:1988ze} and elaborated in \cite{Moore:1997pc}. Given a topological observable $\op(x)$ (corresponding concretely to an operator supported in a ball centered at $x$), one defines an associated one-form observable according to
\begin{equation} \label{eq:phi1}
	\op^{(1)}(x) = \op_\mu(x) \de x^\mu\,, \qquad \op_\mu(x) = Q_\mu (\op(x))~,
\end{equation}
a two-form observable as
 \begin{equation} \label{eq:phi2}
	\op^{(2)}(x) = \op_{\mu \nu}(x) \de x^\mu \wedge \de x^\nu\,, \qquad \op_{\mu \nu}(x) = \frac12 Q_\mu Q_\nu (\op(x))~,
\end{equation}
and more generally for any positive integer $k$,
\begin{equation}
	\op^{(k)}(x) = \frac{1}{k!} (Q_{\mu_1} \cdots Q_{\mu_k} \op)(x) \de x^{\mu_1} \wedge \cdots \wedge \de x^{\mu_k}~.
\end{equation}
The $\op^{(k)}$ are not topological operators in their own right, but they are topological ``up to total derivatives,'' in the sense that
\begin{equation} \label{eq:qphi-1}
	Q (\op^{(1)}(x)) = Q\left(Q_\mu (\op(x))\right) \de x^\mu = iP_\mu (\op(x)) \de x^\mu = \de \op(x)~,
\end{equation}
and more generally by an analogous computation,
\begin{equation} \label{eq:Q-d}
	Q (\op^{(k)}(x)) = \de \op^{(k-1)}(x)~.
\end{equation}
For later convenience we will introduce the \ti{total descendant}
\begin{equation}
	\op^{*} = \sum_{k=0}^d \op^{(k)}\,,
\end{equation}
in terms of which \eqref{eq:Q-d} becomes simply
\begin{equation}
	Q \op^* = \de \op^*~.
\end{equation}

Now, for any $k$-chain $\gamma \subset B$ we define
a new extended operator living along $\gamma$,
\begin{equation}
	\op(\gamma) = \int_\gamma \op^{(k)}\,.
\end{equation}
Since $\op^{(k)}$ is topological up to total derivatives,
$\op(\gamma)$ is topological up to boundary terms: indeed,
using Stokes's theorem and \eqref{eq:Q-d} we get
\begin{equation}
	Q (\op(\gamma)) = \op(\partial \gamma)\,.
\end{equation}
In particular, when $\gamma$ is a $k$-\ti{cycle}, we have
\begin{equation}
	Q (\op(\gamma)) = 0\,,
\end{equation}
so in this case $\op(\gamma)$ is an extended topological operator.
Moreover, by another application of 
Stokes's theorem and \eqref{eq:Q-d} we see that
the homology class $\cclass{\op(\gamma)} \in \cA$ depends only
on the homology class of $\gamma$.

This last observation might at first seem discouraging.
In prior work on topological field theory,
the extended operators $\op(\gamma)$ are usually wrapped around 
homologically nontrivial cycles in spacetime.\footnote{For example,
the TQFT interpretation of the 
Donaldson invariants of a $4$-manifold $X$, given in \cite{Witten:1988ze}, 
involves operators $\op(\gamma)$ wrapped
around cycles $\gamma \subset X$.}
When spacetime is $\R^d$, it looks like the classes
$\cclass{\op(\gamma)}$ cannot give us anything new: 
indeed the only homologically nontrivial 
compact cycles are $0$-cycles, for 
which $\op(\gamma)$ reduces to the
original topological operator $\op(x)$.
Fortunately, there is another possibility.

\subsection{The secondary product}
\label{sec:secondary}

\insfigscaled{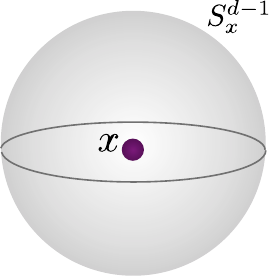}{1.0}{Construction of the 
product operator \eqref{eq:descent-product-op}: the local operator $\op_1$
is placed at $x$, and the $(d-1)$-form descendant of the local operator 
$\op_2$ is integrated over a surrounding sphere. The large ball $B$ is not shown explicitly.}

Suppose that $\op_1$ and $\op_2$ are two topological operators.
We insert $\op_2$ at a point $x \in B$. Then we apply descent to $\op_1$,
obtaining the $(d-1)$-form operator $\op_1^{(d-1)}$,
and integrate it over a sphere $S^{d-1}_x$ centered at $x$.\footnote{This
operation depends on an orientation of $S^{d-1}$; 
here and below we always choose the standard orientation, induced from the ambient spacetime.}
In other words, we consider the product operator
\begin{equation} \label{eq:descent-product-op}
	\op_1(S^{d-1}_x) \op_2(x)\,.
\end{equation}
Note that this is again a \emph{local} operator: it is supported inside a single, sufficiently large ball.
Moreover, as the product of two topological operators, $\eqref{eq:descent-product-op}$
is itself topological, so we may consider its class\footnote{Since $S^{d-1}_x$ is the boundary of a ball $B^d_x$, we might try to 
use the equation
$\op_1(S^{d-1}_x) = Q (\op_1(B^d_x))$ to show that
\eqref{eq:descent-product-op} is $Q$-exact, but 
there is a potential
obstruction: we would need to write
\begin{equation}
\op_1(S^{d-1}_x) \op_2(x) = Q (\op_1(B^d_x)) \op_2(x)  = Q (\op_1(B^d_x) \op_2(x))
\end{equation}
and this last operator is ill-defined thanks to the
colliding-point singularity at $x$.
Thus we cannot conclude that \eqref{eq:descent-product-op} is $Q$-exact,
and indeed we will see in examples below that it may not be.}
\begin{equation}
\cclass{\op_1(S^{d-1}_x) \op_2(x)} \in \cA\,.
\end{equation}
Given two spheres
$S^{d-1}_x$, $S'^{d-1}_x$ of different radii, the chain $S^{d-1}_x - S'^{d-1}_x$ is
the boundary of an annulus that does not intersect $x$,
so Stokes' theorem can be safely applied to show
that $\cclass{\op_1(S^{d-1}_x) \op_2(x)}$ is independent of the radius of $S^{d-1}_x$.
Thus we may define a new product $\sprod{\cdot}{\cdot}$ on $\cA$ by
\begin{equation} \label{eq:secondary-product}
	\sprod{\cclass{\op_1}}{\cclass{\op_2}} = \cclass{\op_1(S^{d-1}_x) \op_2(x)}.
\end{equation}
As we will show in the next few sections, 
the two products $\pprod$ and $\{\cdot,\cdot\}$ 
make $\cA$ into a \ti{$\Z/2$-graded Poisson algebra},
with bracket of parity opposite to that of $d$, or (in the $\Z$-graded setting) degree $1-d$.

\subsubsection{Descent on configuration space} \label{sec:descent-configuration-space}

In order to derive some basic properties of secondary products,
it will be advantageous to switch to
a more sophisticated point of view on descent, \emph{cf}. \cite[Sec 1.3]{CG}. Namely, instead of applying descent to the individual $\op_i$ 
to get form-valued operators $\op_i^{(k)}$
on $\R^d$, we can apply it directly to the product $\op_1(x_1) \op_2(x_2)$, 
defining form-valued operators ${(\op_1 \boxtimes \op_2)^{(k)}}$
on the configuration space $\cC_2(B)$:

\begin{equation} \label{eq:product-descendant}
	(\op_1 \boxtimes \op_2)^{(k)}(x_1,x_2) = \sum_{n=0}^k (-1)^{(k-n) F_1} \op_1^{(n)}(x_1) \wedge \op_2^{(k-n)}(x_2)\,.
\end{equation}
To be concrete, in coordinates we have \emph{e.g.} 
\begin{equation}
	(\op_1 \boxtimes \op_2)^{(1)} = \op_{1;\mu}(x_1) \op_2(x_2) \de x_1^\mu + (-1)^{F_1} \op_1(x_1) \op_{2;\mu}(x_2) \de x_2^\mu\,.
\end{equation}
The forms $(\op_1 \boxtimes \op_2)^{(k)}$ have been engineered to obey the key condition
(\emph{cf.} \eqref{eq:Q-d})
\begin{equation} \label{eq:Q-d-ext}
  Q((\op_1 \boxtimes \op_2)^{(k+1)}) = \de (\op_1 \boxtimes \op_2)^{(k)}\,.
\end{equation}
More compactly, 
we can rewrite \eqref{eq:product-descendant} in terms of the total descendant as
\begin{equation} \label{eq:product-total-descendant}
	(\op_1 \boxtimes \op_2)^{*} = \op_1^{*} \wedge \sigma^{F_1} \op_2^{*}\,,
\end{equation}
where $\sigma$ acts as $(-1)^k$ on the degree $k$ part.
Then \eqref{eq:Q-d-ext} is
\begin{equation} \label{eq:Q-d-total}
	Q(\op_1 \boxtimes \op_2)^{*} = \de (\op_1 \boxtimes \op_2)^{*}\,.
\end{equation}

Now given any $k$-cycle $\Gamma$ on $\cC_2(B)$ we can define a 
new
extended operator,
\begin{equation} \label{eq:cs-extended-op}
(\op_1 \boxtimes \op_2)(\Gamma) = \int_\Gamma (\op_1 \boxtimes \op_2)^{(k)}\,.
\end{equation}
By applying Stokes's theorem and \eqref{eq:Q-d-ext} we find that 
$(\op_1 \boxtimes \op_2)(\Gamma)$ is topological,
and $\cclass{(\op_1 \boxtimes \op_2)(\Gamma)}$ depends only on $\cclass{\op_1}$,
$\cclass{\op_2}$ and the homology class $\cclass{\Gamma} \in H_k(\cC_2(B), \Z)$.
Thus for each class $P \in H_k(\cC_2(B), \Z)$ we obtain a bilinear operation,
which we denote
\begin{equation}
	\star_P: \cA \otimes \cA \to \cA\,.
\end{equation}
The operation $\star_P$ depends linearly on the class $P$.

In a similar way, given $n$ topological operators we can
build a form-valued operator on the configuration space $\cC_n(B)$
of $n$ points.
Integrating these forms against homology classes $P \in H_\bullet(\cC_n(B), \Z)$ gives multilinear operations on $\cA$,
\begin{equation}
	\star_P: \cA^{\otimes n} \to \cA\,.
\end{equation}
Below we will only need explicitly the case $n=3$,
for which the relevant form is
\begin{equation}
	(\op_1 \boxtimes \op_2 \boxtimes \op_3)^* = \op_1^* \wedge \sigma^{F_1} \op_2^* \wedge \sigma^{F_1 + F_2} \op_3^*.
\end{equation}

\subsubsection{Symmetry of the secondary product} \label{sec:symmetry}

In this section we prove that $\sprod{\cdot}{\cdot}$ has the
symmetry property
\begin{equation} \label{eq:sprod-symmetry}
	\sprod{\cclass{\op_2}}{\cclass{\op_1}} = (-1)^{F_1 F_2 + d} \sprod{\cclass{\op_1}}{\cclass{\op_2}}.
\end{equation}
We begin by showing that
\begin{equation} \label{eq:symmetry-rewritten}
	\cclass{\op_1} \star_{\cclass{\Gamma_a}} \cclass{\op_2} = (-1)^{d} \cclass{\op_1} \star_{\cclass{\Gamma_b}} \cclass{\op_2},
\end{equation}
where the two $(d-1)$-cycles in $\cC_2(B)$ are defined as
\begin{align}
\Gamma_a &= S^{d-1}_{x_2} \times \{x_2\}, \\
\Gamma_b &= \{x_1\} \times S^{d-1}_{x_1}.
\end{align}
To relate these two cycles, note that the configuration space 
$\cC_2(B)$ is homotopy equivalent to $S^{d-1}$, via the map
\begin{equation} \label{eq:hom-equiv-2}
	(x_1,x_2) \mapsto \frac{x_1-x_2}{\norm{x_1 - x_2}}\,.
\end{equation}
It follows that its homology is (for $d \geqslant 2$)
\begin{equation}
	H_k(\cC_2(B), \Z) = \begin{cases} \Z & \text{ for } k = 0\,, \\
\Z & \text{ for } k = d-1\,, \\
0 & \text{ otherwise}\,.
\end{cases}
\end{equation}
Moreover, for any $(d-1)$-cycle $\Gamma$, the homology class
$\cclass{\Gamma} \in \Z$ is the topological degree of the map
$\Gamma \to S^{d-1}$ given by
\eqref{eq:hom-equiv-2}. 
Since $\Gamma_b$ is obtained by composing $\Gamma_a$ with the antipodal map $A:S^{d-1}\to S^{d-1}$, whose degree is $\text{deg}(A)=(-1)^d$, we have 
\begin{equation} \label{eq:gammaa-b-relation}
	\cclass{\Gamma_a} = (-1)^d \cclass{\Gamma_b}\,.
\end{equation}
The relation \eqref{eq:gammaa-b-relation} immediately implies \eqref{eq:symmetry-rewritten}.
Now we can use \eqref{eq:symmetry-rewritten} to determine the symmetry of the secondary product:\footnote{In passing from \eqref{eq:tricky-sign-before} to \eqref{eq:tricky-sign-after}
we need to take account of the tricky sign $\sigma^{F_1}$ in \eqref{eq:product-total-descendant}.
This is the only place in this paper where this sign plays a role.}
\begin{align}
	\sprod{\cclass{\op_2}}{\cclass{\op_1}} &= \cclass{\op_2(S^{d-1}_x) \op_1(x)} \notag  \\
	&= \cclass{\op_2} \star_{\cclass{\Gamma_a}} \cclass{\op_1} \notag \\
	&= (-1)^d \cclass{\op_2} \star_{\cclass{\Gamma_b}} \cclass{\op_1} \label{eq:tricky-sign-before}  \\
	&= (-1)^{d+F_2(d-1)} \cclass{\op_2(x) \op_1(S^{d-1}_x)} \label{eq:tricky-sign-after} \\
	&= (-1)^{d+F_2(d-1)+F_2(F_1+d-1)} \cclass{\op_1(S^{d-1}_x) \op_2(x)} \notag \\
	&= (-1)^{F_1 F_2+d} \sprod{\cclass{\op_1}}{\cclass{\op_2}}\,.
\end{align}
This is the desired symmetry property \eqref{eq:sprod-symmetry}.

Incidentally, our computation also
shows that \ti{any} operation $\star_{\cclass{\Gamma}}$ built from a $(d-1)$-cycle
$\Gamma \subset \cC_2(B)$ is an integer multiple of $\sprod{\cdot}{\cdot}$.
For example, we could apply descent to \ti{both}
operators $\op_1$ and $\op_2$, then integrate them around
cycles $\gamma_1$, $\gamma_2$ in $\R^d$, with $\dim \gamma_1 + \dim \gamma_2 = d-1$;
\emph{e.g.} in $d=3$ we could integrate
$\op_1^{(1)}$ and $\op_2^{(1)}$ around two circles making up 
a Hopf link (Figure \ref{fig:Hopf}).
What we have just shown is that the resulting class
is $\ell \sprod{\cclass{\op_1}}{\cclass{\op_2}}$ where $\ell$ is the linking number
between $\gamma_1$ and $\gamma_2$. 

\insfigscaled{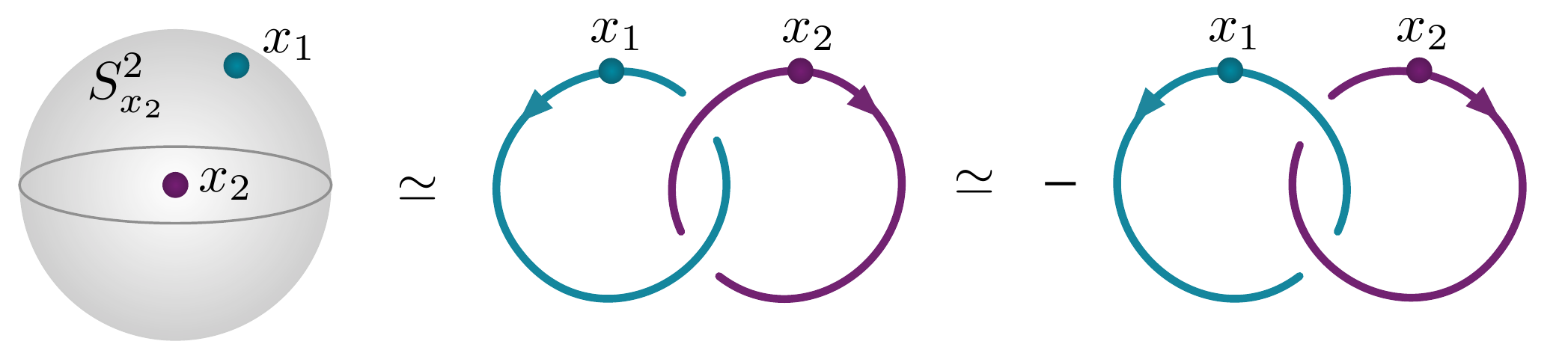}{0.7}{In $d=3$, the secondary product can (equivalently) be defined by integrating first descendants around a Hopf link.}

\subsubsection{The derivation property}

Another key property of a Poisson bracket is that it is a \ti{derivation}
of the algebra structure:
\begin{equation} \label{eq:poisson-derivation}
  \sprod{\cclass{\op_1}}{\cclass{\op_2} \pprod \cclass{\op_3}} = \sprod{\cclass{\op_1}}{\cclass{\op_2}} \pprod \cclass{\op_3} + (-1)^{(F_1+d-1)F_2} \cclass{\op_2} \pprod \sprod{\cclass{\op_1}}{\cclass{\op_3}}\,.
\end{equation}
We can prove this using the same strategy we used in Section~\ref{sec:symmetry}:
we identify the three terms as operations
\begin{equation}
	\star_{\cclass{\Gamma}}: \cA \otimes \cA \otimes \cA \to \cA
\end{equation}
coming from cycles $\Gamma$ on $\cC_3(B)$.

\insfigscaled{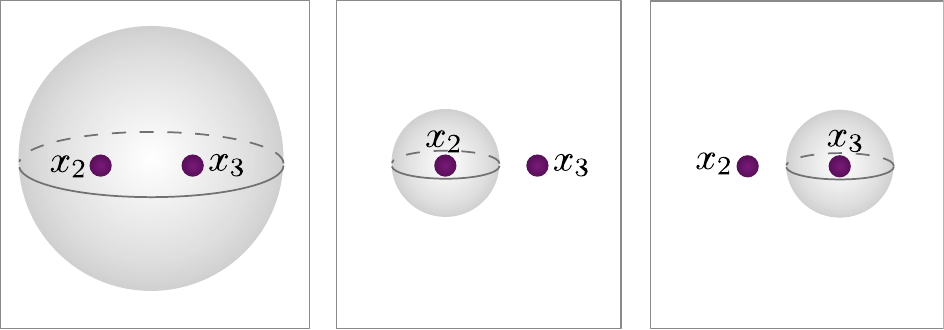}{1.0}{The three cycles $\Gamma$, $\Gamma'$ and $\Gamma''$ in $\cC_3(B)$
used to prove the derivation property of the secondary product.}

The left side of \eqref{eq:poisson-derivation} 
corresponds to the cycle
\begin{equation}
	\Gamma = S^{d-1}_{x_2,x_3} \times \{x_2\} \times \{x_3\}
\end{equation}
where $S^{d-1}_{x_2, x_3}$ is a sphere enclosing both $x_2$ and $x_3$.
The two terms on the right are
\begin{align}
	\Gamma' &= S^{d-1}_{x_2} \times \{x_2\} \times \{x_3\}\,, \\
	\Gamma'' &= S^{d-1}_{x_3} \times \{x_2\} \times \{x_3\}\,.
\end{align}
Since the sphere enclosing both $x_2$ and $x_3$ is homologous to a sum of spheres enclosing $x_2$ and $x_3$ separately, we have
\begin{equation}
	\cclass{\Gamma} = \cclass{\Gamma'} + \cclass{\Gamma''}\,.
\end{equation}
Now we use this as follows:
\begin{align}
	\sprod{\cclass{\op_1}}{\cclass{\op_2} \pprod \cclass{\op_3}} &= \cclass{\op_1(S^{d-1}_{x_2,x_3}) \op_2(x_2) \op_3(x_3)} \notag \\
	&= \star_{\cclass{\Gamma}}(\cclass{\op_1}, \cclass{\op_2}, \cclass{\op_3}) \notag \\
	&= \star_{\cclass{\Gamma'}}(\cclass{\op_1}, \cclass{\op_2}, \cclass{\op_3}) + \star_{\cclass{\Gamma''}}(\cclass{\op_1}, \cclass{\op_2}, \cclass{\op_3}) \notag \\
	&= \cclass{\op_1(S^{d-1}_{x_2}) \op_2(x_2) \op_3(x_3)} + \cclass{\op_1(S^{d-1}_{x_3}) \op_2(x_2) \op_3(x_3)} \notag \\
	&= \cclass{\op_1(S^{d-1}_{x_2}) \op_2(x_2) \op_3(x_3)} + (-1)^{(F_1+d-1)F_2} \cclass{\op_2(x_2) \op_1(S^{d-1}_{x_3})  \op_3(x_3)} \notag \\
	&= \sprod{\cclass{\op_1} \pprod \cclass{\op_2}}{\cclass{\op_3}} + (-1)^{(F_1+d-1)F_2} \sprod{\cclass{\op_2}}{\cclass{\op_1 \pprod \op_3}}\,,
\end{align}
as needed to prove \eqref{eq:poisson-derivation}.

Note that the cycles $\Gamma$, $\Gamma'$ and $\Gamma''$
lie in a subspace of $\cC_3(B)$ where $x_2$ and $x_3$ are fixed 
and only $x_1$ varies. 
Thus, for this argument it was not really necessary to use the language
of descent on configuration spaces: we could have gotten by with
ordinary descent applied to $\op_1(x_1)$ alone.

\subsubsection{The Jacobi identity}

The last property we need to check is the \ti{Jacobi identity},
\begin{align} \label{eq:jacobi}
	&\sprod{\cclass{\op_1}}{\sprod{\cclass{\op_2}}{\cclass{\op_3}}} -
(-1)^{(F_1+d-1)(F_2+d-1)} \sprod{\cclass{\op_2}}{\sprod{\cclass{\op_1}}{\cclass{\op_3}}}
	\\ & \hspace{2.5in} =\, (-1)^{(d-1)(F_1+d-1)} \sprod{\sprod{\cclass{\op_1}}{\cclass{\op_2}}}{\cclass{\op_3}}\,. \notag
\end{align}
\insfigscaled{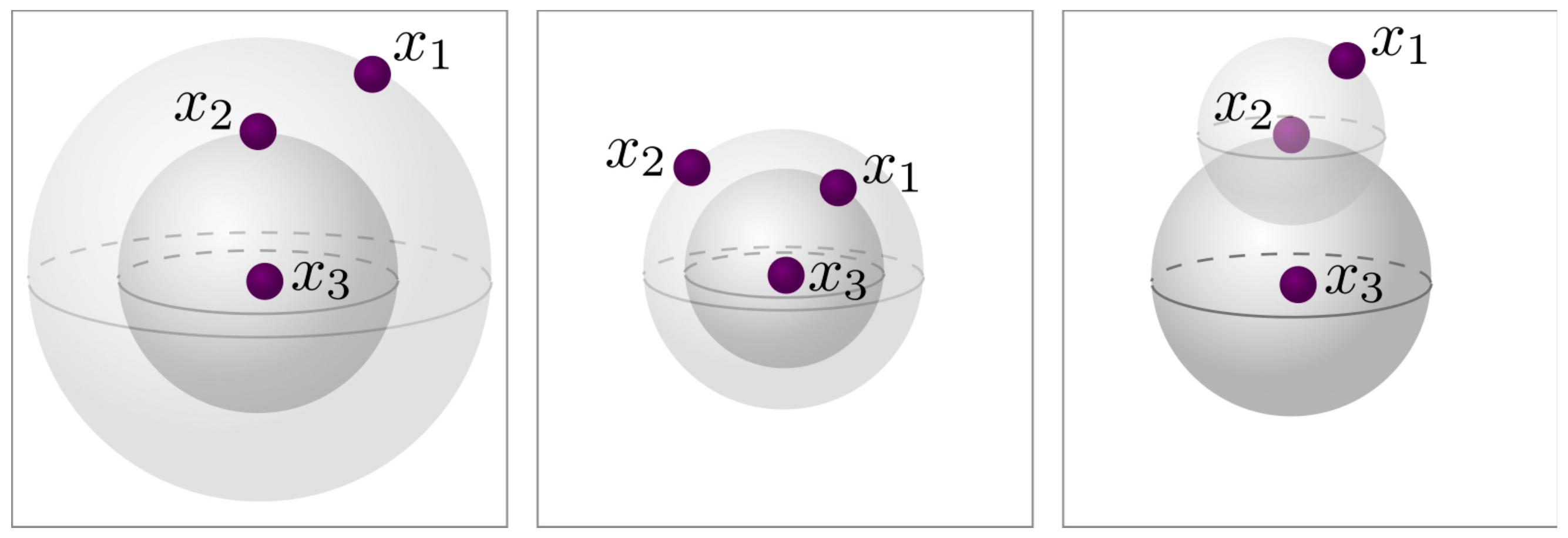}{0.4}{The cycles $\Gamma$, $\Gamma'$, $\Gamma''$ in
$\cC_3(B)$.}
The first term on the left of \eqref{eq:jacobi} is $\star_{\cclass{\Gamma}}(\cclass{\op_1},\cclass{\op_2},\cclass{\op_3})$ where
\begin{equation}
	\Gamma = S^{d-1,big}_{x_3} \times S^{d-1,small}_{x_3} \times \{x_3\}\,.
\end{equation}
The second term in \eqref{eq:jacobi} is the same with $x_2$ and $x_1$ reversed.
For convenience we may rescale distances so that $x_2$ goes around
the same sphere $S^{{d-1},small}_{x_3}$ in the second term as it did in the first term; 
then the second term is
$\star_{\cclass{\Gamma'}}(\cclass{\op_1},\cclass{\op_2},\cclass{\op_3})$ where
\begin{equation}
	\Gamma' = S^{d-1,tiny}_{x_3} \times S^{d-1,small}_{x_3} \times \{x_3\}\,.
\end{equation} 
The difference of these cycles is
\begin{equation}
  \Gamma - \Gamma' = (S^{d-1,big}_{x_3} - S^{d-1,tiny}_{x_3}) \times S^{d-1,small}_{x_3} \times \{x_3\}\,.
\end{equation}
For each fixed $x_2 \in S^{d-1,big}_{x_3}$, the chain
$(S^{d-1,big}_{x_2} - S^{d-1,tiny}_{x_2})$ is homologous
in $\R^d - \{x_2,x_3\}$ to a small sphere $S^{d-1}_{x_2}$.
Thus we have
\begin{equation} \label{eq:cycle-rel-jacobi}
	\cclass{\Gamma} - \cclass{\Gamma'} = \cclass{\Gamma''}
\end{equation}
where $\Gamma''$ is shown in \autoref{fig:jacobi}. The right side of \eqref{eq:jacobi}
is $\star_{\cclass{\Gamma''}}(\cclass{\op_1},\cclass{\op_2},\cclass{\op_3})$. Thus the relation \eqref{eq:cycle-rel-jacobi}
gives the desired \eqref{eq:jacobi}.

\subsection{No new higher operations}

As we have explained, any class in $H_\bullet(\cC_n(B), \Z)$ induces an $n$-ary operation on the $Q$-cohomology
$\cA$. In particular the two binary operations $\pprod$ and $\sprod{\cdot}{\cdot}$ are
induced by the two nontrivial homology classes in $H_\bullet(\cC_2(B), \Z)$.
One might wonder whether the higher $n$-ary operations coming from
$H_\bullet(\cC_n(B),\Z)$ bring anything new. The 
simple answer is no: the only $n$-ary operations we get from $H_\bullet(\cC_n(B),\Z)$ come from iterated
compositions of $\pprod$ and $\sprod{\cdot}{\cdot}$.
This follows from the fact
that, for $d>1$, \ti{the homology of the little $d$-discs operad is the degree $d-1$ Poisson operad}
(see \cite{Sinha} for a useful review.)

A slightly more refined answer is that nontrivial higher $n$-ary operations \emph{do} exist, corresponding to the higher $L_\infty$ operations discussed briefly in Section \ref{sec:intro-Ed} of the Introduction. However, these higher $n$-ary operations come from open chains rather than cycles in configuration space $\cC_n(B)$, and so generically map an $n$-tuple of $Q$-closed operators to an arbitrary element of $\ops_\delta$, rather than to another $Q$-closed operator. Thus, the higher $n$-ary operations are not defined on the entire $Q$-cohomology $\cA$. The only nontrivial operations guaranteed to exist on all of $\cA$ --- coming from \emph{cycles} in configuration space --- are the primary product and the Lie bracket.

\section{Example: the 2d B-model}
\label{sec:b-model}

As our first example of this formalism, we will review the construction of the secondary product in perhaps the simplest nontrivial setting: the B-twist of a two-dimensional $\cN=(2,2)$ sigma-model, \emph{a.k.a.} the B-model. It is well known \cite{Witten:1988xj, Witten:1991zz} (\emph{cf.} also \cite{Hori:2003ic}) that the B-model with K\"ahler target $\cX$ has a topological algebra of local operators that is isomorphic to the Dolbeault cohomology of polyvector fields on $\cX$,
\begin{equation}
\cA \cong H^\bullet_{\bar\partial}\big(\Omega^{0,\bullet}(\cX)\otimes \Lambda^\bullet(T^{1,0}\cX)\big)~. 
\end{equation}
In other words, $\CA$ is the $\bar\partial$-cohomology of $(0,q)$ forms valued in arbitrary exterior powers $\Lambda^p(T^{1,0} \CX)$ of the holomorphic tangent bundle.
This algebra is $\Z$-graded, with $k$-th graded component $\cA^{(k)} = \oplus_{p+q=k}H^q_{\bar\partial}\big(\Omega^{0,\bullet}(\cX)\otimes \Lambda^p(T^{1,0}\cX)\big)$. The chiral ring of the underlying (untwisted) sigma model consists of holomorphic functions on $\CX$ \cite{Lerche:1989uy}, and sits inside the topological algebra as the $0$-graded component
\begin{equation}
\C[\CX] = \CA^{(0)} \subset \CA~.
\end{equation}
Due to the absence of instanton corrrections in the B-model, the primary product on $\CA$ coincides with the ordinary wedge product of polyvector fields,
\begin{equation}
\label{b-prod}
\cclass{\CO_1}*\cclass{\CO_2} = \cclass{\CO_1\wedge \CO_2}~.
\end{equation}

It is also well known that there exists an odd (degree $-1$) Poisson bracket on $\CA$ that gives $\CA$ the structure of a graded Lie algebra, in a manner compatible with the primary product. Altogether, this endows $\CA$ with the structure of a Gerstenhaber algebra.
In terms of the geometry of the target, the Poisson bracket coincides with the Schouten-Nijenhuis (SN) bracket of polyvector fields, which extends the geometric Lie bracket of ordinary vector fields. 

We will verify in this section that the SN bracket coincides with the secondary product that arises from topological descent. To keep things simple, we begin by considering the theory with target $\C^n$, \emph{i.e.} the theory of $n$ free chiral multiplets. We will then see how to generalize the discussion to allow for more interesting K\"ahler targets.%
\footnote{It may appear unconventional for us to allow arbitrary K\"ahler target in the B-model. Indeed, in order to define the B-model on arbitrary \emph{curved} 2d spacetimes, the target must actually be Calabi-Yau, which ensures the existence of a non-anomalous axial $U(1)_A$ $R$-symmetry that can be used to twist the theory. Because we are only addressing local properties of the algebra of operators in flat spacetime, the $\CN=(2,2)$ supersymmetry algebra contains both the ``topological'' supercharge $Q$ and the vector $Q_\mu$ whenever the target is K\"ahler, regardless of whether the axial $R$-symmetry is present.}

\subsection{\texorpdfstring{$(2,2)$}{(2,2)} superalgebra}
\label{sec:b-basics}

We first recall the general structure of $\CN=(2,2)$ supersymmetry in flat two-dimensional Euclidean space, which we take to have complex coordinates $z,\bar z$. In the absence of central charges, the supersymmetry algebra is generated by $Q_\pm$ and $\ol Q_\pm$ with (anti)commutation relations
\begin{equation}
 \label{b-algebra} 
[Q_+,\ol Q_+] = 2iP_{\bar z}\,,\qquad [Q_-,\ol Q_-]=2iP_{z}~.
\end{equation}
The linear combination
\begin{equation}
Q \colonequals \ol Q_++\ol Q_- ~,
\end{equation}
is nilpotent, and we treate it as the ``scalar'' supercharge. We can further define the vector supercharge
\begin{equation}
Q_\mu = \bp Q_z \\ Q_{\bar z} \ep \colonequals \frac 1 2 \bp Q_- \\  Q_+\ep ~,
\end{equation}
such that the superalgebra takes the general form given in Eqn. \eqref{eq:QQ-P}.

With respect to the $\text{Spin}(2)\cong U(1)_E$ Lorentz group, the $U(1)_A$ axial $R$-symmetry (when present), and the $U(1)_V$ vector $R$-symmetry, the supercharges (and coordinates) transform with the following weights:
\begin{equation}
\label{b-AV} 
\begin{array}{c|cccc|cc}
& Q_+ & \ol Q_+ & Q_- & \ol Q_- & z & \bar z \\\hline
U(1)_E  & 1 &  1 & - 1 & - 1 & 2 & -2 \\[.2cm]
U(1)_A & -1 & 1 & 1 & -1 & 0 & 0 \\[.2cm]
U(1)_V & -1 & 1 & -1 & 1 & 0 & 0 
\end{array}
\end{equation}
If $U(1)_A$ is a symmetry, one can consider an improved Lorentz group, defined as the anti-diagonal of $U(1)_E\times U(1)_A$. With respect to the improved Lorentz group, $Q$ and $Q_\mu$ really do transform as a scalar and a vector. However, even if $U(1)_A$ is anomalous, we still have $Q^2=0$ and $[Q,Q_\mu]=iP_\mu$, which is good enough for our purposes. In any K\"ahler sigma-model (whose target is locally parameterized by chiral multiplets), $U(1)_V$ remains unbroken and defines a $\Z$-valued fermion-number grading, under which $Q$ and $Q_\mu$ have charges $+1$ and $-1$, respectively.

\subsection{Free chiral: target space \texorpdfstring{$\C^n$}{Cn}}
\label{sec:freechiral}

For our example we take the theory of $n$ free chiral multiplet, consisting of a complex scalar fields $\phi^i$ and complex left- and right-handed fermions $\psi_\pm^i$, $\bar\psi_{i\pm}$. The action,
\begin{equation}\label{b-Caction} 
S = \int d^2z\, \Big[ \pd_z\phi^i\pd_{\bar z}\bar\phi_i - \tfrac12 \bar\psi_{i+}\pd_z\psi_+^i - \tfrac12\bar\psi_{i-}\pd_{\bar z}\psi_-^i\Big]~,
\end{equation}
is invariant under supersymmetry transformations generated by $Q_\pm$ and $\ol Q_\pm$ that act on the fields according to
\begin{equation}
\begin{array}{c|cccc}
& Q_+ & \ol Q_+ & Q_- & \ol Q_- \\\hline
\phi^i & \psi_+^i &0&\psi_-^i &0 \\
\psi_+^i & 0 &2\pd_{\bar z}\phi^i  & 0 & 0 \\
\psi_-^i & 0 &0&0 &2\pd_z\phi^i 
\end{array}\;, \qquad 
\begin{array}{c|cccc}
& Q_+ & \ol Q_+ & Q_- & \ol Q_- \\\hline
\bar\phi_i &0 & \bar\psi_{i+} &0&\bar\psi_{i-} \\
\bar\psi_{i+} &2\pd_{\bar z}\bar\phi_i & 0&0& 0 \\
\bar\psi_{i-} &0& 0&2\pd_z\bar\phi_i & 0 \\
\end{array}~.
\end{equation}
The superalgebra relations \eqref{b-algebra} are realized modulo the equations of motion.

It is convenient to reparameterize the fermions according to their transformations under the improved Lorentz group,
\begin{equation}
\label{def-etaxi} 
\eta_i = \bar\psi_{i+}+\bar\psi_{i-}\,,\qquad \xi_i =-i(\bar\psi_{i+}-\bar\psi_{i-})~;\qquad \chi^i = \tfrac12\big(\psi_-^idz+\psi_+^id\bar z)~.
\end{equation}
Now $\eta_i$ and $\xi_i$ are scalars, while $\chi^i$ are one-forms. In terms of the relabelled fields, the action \eqref{b-Caction} takes the form
\begin{equation}
\label{b-xiaction} 
S = \int  \Big[ d\phi^i\wedge *d\bar\phi_i+  \xi_i d\chi^i - \eta_i d *\chi^i\Big]~.
\end{equation}
The supersymmetry transformations relevant for the descent procedure are given by
\begin{equation}
\label{b-SUSY}
\begin{split}
Q(\phi^i) &=  Q(\xi_i) = Q(\eta_i) = 0~,\\[.2cm]
Q(\bar\phi_i) &= \eta_i\,,\quad Q(\chi^i) = d\phi^i~, \\[.2cm]
\mathbb Q(\phi^i) &= \chi^i\,,\quad \mathbb Q(\xi_i) = -(*d\bar\phi_i)\,,\\[.2cm]
\mathbb Q(\eta_i) &= d\bar\phi_i\,,\quad \mathbb Q(\bar\phi_i) = \mathbb Q(\chi^i) = 0\,,
\end{split}
\end{equation}
where we have defined the one-form supercharge $\mathbb Q := Q_\mu dx^\mu = Q_zdz+Q_{\bar z}d\bar z$.

\subsubsection{Local operators}
\label{sec:b-poly}

We will restrict our attention to local operators that are represented by polynomial functions of the fields. This will suffice for illustrating the main features in the computation of the secondary product. 
Our analysis extends in a straightforward way to analytic functions.
(In general, other sorts of operator might be considered as well.)

The local operators corresponding to polyvector fields come from inserting copies of the fields $\phi^i,\bar\phi_i,\eta_i,\xi^i$ simultaneously at distinct points in a ball $B$. 
We may represent a multi-insertion as a monomial in $\phi,\bar\phi,\eta,\xi$, as long as we remember that insertion points are distinct; for example
\be \label{b-poly}  (\phi^i)^2\eta_j\bar\phi_k \qquad\text{means} \qquad  \phi^i(z_1,\bar z_1)\phi^i(z_2,\bar z_2)\eta_j(z_3,\bar z_3)\bar\phi_k(z_4,\bar z_4) \ee
at some distinct $z_1,z_2,z_3,z_4$. 
As usual, after passing to cohomology, the precise choice of insertion points becomes irrelevant.
The topological supercharge $Q$ acts on these operators by extending the elementary transformations $Q(\phi^i)=Q(\eta_i)=Q(\xi_i)=0$ and $Q(\bar\phi_i) = \eta_i$ by linearity and a graded Leibniz rule.
Upon identifying $\eta_i$ and $\xi_i$ with anti-holomorphic differentials and holomorphic vector fields on the target $\CX=\C^n$,
\be \eta_i \leftrightarrow \mathbf{d}\bar\phi_i\,,\qquad \xi_i \leftrightarrow \frac{\pd}{\pd\phi^i}\,, \ee
we find that these operators generate the Dolbeault complex $\C[\phi,\bar\phi,\eta,\xi]\simeq  \Omega^{0,\bullet}\C^n\otimes \Lambda^\bullet T^{1,0}\C^n$, 
with $Q$ acting as the Dolbeault differential.\footnote{More precisely, the given operators topologically generate the real analytic model of the Dolbeault complex. We will also abuse the notation $\C[\phi,\bar\phi...]$ to mean analytic functions (rather than polynomials) in $\phi$ and $\bar\phi$, \emph{i.e.}, real analytic functions.} 

The $Q$-cohomology here is extremely simple. Since the $\eta_i$ are exact, we find a topological algebra
\begin{equation}
\CA \simeq \C[\phi,\xi] = H^\bullet_{\bar\pd}( \Omega^{0,\bullet}\C^n\otimes \Lambda^\bullet T^{1,0}\C^n)~,
\end{equation}
consisting of holomorphic functions $f(\phi)$ and holomorphic vector fields $g^i(\phi)\xi_i$.  To simplify notation, we will henceforth suppress the brackets $\cclass{...}$ that indicate cohomology classes.

The algebra $\CA$ is graded by the fermion number coming from $U(1)_V$, which acts on the fields in a chiral multiplet with charges
\be \label{V-C} \begin{array}{c|cccc}
& \phi &\bar\phi & \eta & \xi \\\hline
U(1)_V & 0 & 0 & 1 & 1
\end{array}\;.\ee
Moreover, as emphasized in \eqref{b-prod}, the primary product $*$ on $\CA$ coincides with the ordinary product of polyvector fields. 
Thus $\CA\simeq\C[\phi,\xi]$ as a graded-commutative ring.

\subsubsection{Secondary product}
\label{sec:b-C-secondary}

We now turn to the secondary product of elements in $\CA$. We start with the bracket $\{\xi_i,\phi^i\}$. For this case, the definition \eqref{eq:secondary-product} says
\begin{align} 
\{\xi_i,\phi^j\} &= \cclass{\xi_i(S_{w,\bar w}^1)*\phi^j(w,\bar w)} \notag  \\[5pt]
 &\!\overset{\text{def}}{=} \Big\llbracket \oint_{S_{w,\bar w}^1} \xi_i^{(1)} \;\phi^j(w,\bar w) \Big\rrbracket  \notag \\
 &\hspace{-.3cm}\overset{\text{Stokes}}{=} \Big\llbracket \int_{D^2_{w,\bar w}} d\xi_i^{(1)} \;\phi^j(w,\bar w) \Big\rrbracket\,. \label{xp-bracket}
\end{align}
where $D^2_{w,\bar w}$ is a disc centered around the insertion point of $\phi$ and the circle $S^1_{w,\bar w}$ is its boundary. The first descendant of $\xi_i$ is computed as follows,
\begin{equation}
\xi_i^{(1)} = \mathbb Q(\xi_i) = -( * d\bar\phi_i)~.
\end{equation}
We can then observe that the two-form $d\xi_i^{(1)}$ is proportional to the equation-of-motion for~$\phi^i$,
\begin{equation}
d\xi_i^{(1)} =  d*d\bar\phi_i = \frac{\delta S}{\delta \phi^i}~.
\end{equation}
A standard manipulation of the Euclidean path integral shows that the equation of motion operator $\frac{\delta S}{\delta \phi}(z,\bar z)$ is zero up to contact terms. In particular, in any correlation function the product of operators $\frac{\delta S}{\delta \phi}(z,\bar z) \phi(w,\bar w)$ (when kept separate from any other operators) is equivalent to the insertion of a delta-function two-form ${\delta^{(2)}(z-w,\bar z-\bar w)}$. This is just integration by parts,
\begin{align}
\int D\phi D\bar\phi(\ldots)  &\frac{\delta S}{\delta \phi}(z,\bar z) \phi(w,\bar w)\, e^{-S}\notag\\
 	&=  \int D\phi D\bar\phi(\ldots) \bigg(-\frac{\delta}{\delta\phi(z,\bar z)} \big[ \phi(w,\bar w) e^{-S}\big] +\delta^{(2)}(z- w,\bar z-\bar w) e^{-S}\bigg)\\
 	&= \int  D\phi D\bar\phi(\ldots) \delta^{(2)}(z- w,\bar z-\bar w) e^{-S}~.\notag
\end{align}
Making this replacement in \eqref{xp-bracket} gives us a simple expression for the secondary product,
\begin{equation}
\label{bracket-xiphi}
\{\xi_i,\phi^j\} = \delta_i{}^j~.
\end{equation}
We could also compute the secondary product by performing descent on $\phi^j$. From \eqref{b-SUSY} we see that the relevant descendent is
\begin{equation}
{\phi^j}^{(1)} = \mathbb Q( \phi^j)=\chi^j~,
\end{equation}
which again is related to an equation of motion,
\begin{equation}
d{\phi^j}^{(1)} = d\chi^j = \frac{\delta S}{\delta \xi_j}~.
\end{equation}
The operator $\frac{\delta S}{\delta \xi_j}(z,\bar z) \xi_j(w,\bar w)$ is again equivalent to a delta-function $\delta^{(2)}(z-w,\bar z-\bar w)$, giving a second derivation of the secondary product,
\begin{equation}
\label{bracket-phixi}
\{\phi^j,\xi_i\} = \Big\llbracket \oint_{S_{w,\bar w}^1} (\phi^j)^{(1)} \;\xi_i(w,\bar w) \Big\rrbracket = \Big\llbracket \int_{D^2_{w,\bar w}} d(\phi^j)^{(1)} \;\xi_i(w,\bar w)\Big\rrbracket = \delta_i{}^j\,. 
\end{equation}
Similar manipulations show that$\{\phi^i,\phi^j\}=\{\xi_i,\xi_j\}\equiv 0$, as there is no contact term between $\frac{\delta S}{\delta \xi}$ and $\phi$, and between $\frac{\delta S}{\delta \phi}$ and $\xi$.

We observe that this calculation directly verifies the relation $\{\xi_i,\phi^j\}=\{\phi^j,\xi_i\}$, which is a special case of the symmetry relation \eqref{eq:sprod-symmetry} with $F(\xi)=1$, $F(\phi)=0$, and $d=2$.  Since the algebra $\CA$ is generated by $\phi$ and $\xi$, the secondary product of arbitrary elements of $\CA$ may now be obtained from the brackets of $\phi$ and $\xi$ together with the general ``derivation'' property \eqref{eq:poisson-derivation}. Furthermore, the Jacobi identity \eqref{eq:jacobi} is guaranteed.

We can now compare the secondary product with the Schouten-Nijenhuis (SN) bracket of polyvector fields on $\C$. The SN bracket is uniquely specified by its action on generators of the ring of (polynomial, or more generally, analytic) polyvector fields,
\begin{equation}
\label{SN-C} 
\{\xi_i,\phi^j\}_{\rm SN} = \delta_i{}^j = -\{\phi^j,\xi_i\}_{\rm SN}\,,\qquad \{\phi^i,\phi^j\}_{\rm SN}=\{\xi_i,\xi_j\}_{\rm SN} \equiv 0~,
\end{equation}
together with the fact that $\{\;,\;\}_{\rm SN}$ is (graded)symmetric, is a (graded) derivation in each argument, and satisfies the Jacobi identity. Namely, acting on arbitrary polyvector fields,
\begin{equation}
\label{SN-props}
\begin{array}{r@{\;}l} 
\{a,b\}_{\rm SN} &= -(-1)^{(F(a)-1)(F(b)-1)}\{b,a\}_{\rm SN}\,, \\[.2cm]
\{a,bc\}_{\rm SN} &= \{a,b\}_{\rm SN}\,c + (-1)^{(F(a)-1)F(b)}b\{a,c\}_{\rm SN}\,, \\[.2cm]
\{a,\{b,c\}_{\rm SN}\}_{\rm SN} &= \{\{a,b\}_{\rm SN},c\}_{\rm SN} + (-1)^{(F(a)-1)(F(b)-1)}\{b,\{a,c\}_{\rm SN}\}_{\rm SN}\,.
\end{array}
\end{equation}
The SN bracket and its various properties agrees perfectly with the secondary product, subject to the identification
\begin{equation}
\label{SN-id}
\{a,b\}_{\rm SN} =(-1)^{F(a)-1} \{a,b\}~. 
\end{equation}

\subsection{General K\"ahler target}
\label{sec:b-general}

In the B-model with a general K\"ahler target $\CX$, we expect to be able to compute the secondary product \emph{locally} on $\CX$, where it essentially reduces to the free-field computation of Section \ref{sec:freechiral}. There are two important features of the B-model that justify such an analysis.

\begin{itemize}
\item First, in the presence of any collection of $Q$-closed operators, the path integral in the B-model (meaning, the path integral of an underlying 2d $\CN=(2,2)$ theory) localizes on constant maps. What does this buy us?

In higher spacetime dimension $(d>2)$, we could evaluate correlation functions in the presence of fixed vacua, \emph{i.e.} fixed values $\phi=\phi_0$ of the bosonic fields near spacetime infinity. Then, given localization of the path integral on constant maps, we would see directly that the specialization of a correlation function to any $\phi_0$ vacuum only depends on the neighborhood of $\phi_0$ in $\CX$. In particular, primary and secondary products in the topological algebra $\CA$ must admit a consistent specialization to any $\phi_0\in \CX$, which depends only on the neighborhood of $\phi_0$. Thus they can be computed locally.

In $d=2$, a slightly different argument must be made, because a 2d quantum sigma-model does not have distinct vacua labelled by individual points $\phi_0$ of the target.  Instead, in the B-model, we can introduce Dirichlet boundary conditions that are labelled by points $\phi_0\in \CX$. In other words, we may consider the theory on $\R\times \R_+$, with a boundary condition $\cB_{\phi_0}$ at the origin of $\R_+$ that forces the bosonic fields $\phi$ to take the fixed value $\phi_0$.  (In the category of boundary conditions $D^b\text{Coh}(\CX)$, $\cB_{\phi_0}$ corresponds to a skyscraper sheaf supported at $\phi_0$.) In the presence of such a boundary condition, the path integral will again behave the way we want: localization on constant maps implies that correlation functions will only depend on a neighborhood of $\phi_0\in \CX$. In turn, this implies that primary and secondary products in the algebra $\CA$ admit a consistent local computation.

\item Second, deformations of the target-space metric are $Q$-exact, as long as they preserve the complex structure. (One usually says that the B-model only depends on the complex structure of $\CX$.) In any local patch of $\CX$, say an open neighborhood of any $\phi_0\in\CX$, we can deform the metric to be flat; then the patch becomes isomorphic to an open subset of flat $\C^n$, $n=\dim_\C\CX$. Therefore, the local analysis of primary and secondary products boils down to a computation in the theory of free chiral multiplets.

\end{itemize}

Let's now be more explicit. The topological algebra $\CA$ of local operators in the B-model with general target $\CX$ is usually identified as the Dolbeault cohomology 
\be \CA \simeq H_{\ol \pd}^\bullet\big( \Omega^{0,\bullet}\CX\otimes \Lambda^\bullet T^{1,0}\CX \big)\,. \label{A-polyvec}\ee
Locally, an element of $\CA$ may be represented as a function of the chiral multiplet fields $\phi^i,\bar\phi_i,\eta_i,\xi_i$, just as in Section~\ref{sec:b-poly}.  We identify the $\xi_i$ with a basis of holomorphic vector fields and the $\eta_i$ with a basis of anti-holomorphic 1-forms.%
\footnote{For a free theory, this identification is somewhat ad-hoc. However, in the presence of a non-trivial target-space metric, a more careful analysis shows that the linear combinations of fermions $\xi_i$, $\eta_i$ that are $Q$-closed transform unambiguously as holomorphic vector fields and anti-holomorphic 1-forms, as indicated.} %
In contrast to Section~\ref{sec:b-poly}, however, the $\bar\phi$ and $\eta$ dependence in local operators need not always be exact (due to the global structure of $\CX$). Indeed, for general $\CX$, the higher Dolbeault cohomology \eqref{A-polyvec} is nontrivial.

Given two operators $\CO_1=f_1(\phi,\bar\phi,\eta,\xi)$, $\CO_2=f_2(\phi,\bar\phi,\eta,\xi)$ that are both represented as \emph{polynomial} (or more generally, analytic) functions in a local $\C^n$ patch of $\CX$, the computation of the secondary product becomes relatively simple. We may factor the operators as
\be \CO_1 = \sum_i g_{1,i}(\phi,\xi)h_{1,i}(\bar\phi,\eta)\,,\qquad 
 \CO_2 = \sum_i g_{2,i}(\phi,\xi)h_{2,i}(\bar\phi,\eta)\,, \ee
where $g_{1,i},g_{2,i}$ represent holomorphic polyvector fields, and $h_{1,i},h_{2,i}$ are purely antiholomorphic $(0,*)$ forms. An extension of the descent analysis from Section \ref{sec:b-C-secondary} then shows that $h_{1,i}(\bar\phi,\eta)$ and $h_{2,i}(\bar\phi,\eta)$ are in the kernel of the secondary product. (In particular, correlation functions involving $\bar\phi$ and $\eta$ cannot produce singularities strong enough to give nontrivial contributions to integrals such as \eqref{xp-bracket} and \eqref{bracket-phixi}.) The Lie bracket is then explicitly computed as
\be \label{b-genX} \{\CO_1,\CO_2\} = \sum_{i,j} \pm \{g_{1,i}(\phi,\xi),g_{2,j}(\phi,\xi)\} h_{1,i}(\bar\phi,\eta)h_{2,j}(\bar\phi,\eta)\,,
\ee
with signs determined by fermion numbers. The term $\{g_{1,i}(\phi,\xi),g_{2,i}(\phi,\xi)\}$ is the same free-field bracket computed in Section \ref{sec:b-poly}, agreeing up to a sign with the SN bracket. Formula \eqref{b-genX} may be loosely summarized by saying that the SN bracket of polyvector fields controls the secondary bracket on the entire Dolbeault cohomology \eqref{A-polyvec} (at least if one considers polynomial or analytic local operators).

Alternatively, and somewhat more geometrically, we can describe Dolbeault cohomology \eqref{A-polyvec} as the \v{C}ech cohomology of holomorphic polyvector fields, 
\be \CA \simeq H_{\ol \pd}^\bullet\big( \Omega^{0,\bullet}\CX\otimes \Lambda^\bullet T^{1,0}\CX \big)
\simeq H_{\text{\v{C}ech}}^\bullet( \Lambda^\bullet T_{hol}\CX) \,. \ee
This carries a bracket canonically induced by the SN bracket on local holomorphic polyvector fields, which we expect to agree with the secondary product on the entire topological operator algebra, including higher Dolbeault cohomology.

\section{Example: Rozansky-Witten twists of 3d \texorpdfstring{$\cN=4$}{N=4}}
\label{sec:3dn=4}

A novel application of the constructions outlined in this paper is to use topological descent to define a Poisson bracket on the algebra of local operators in three-dimensional $\CN=4$ theories. In three dimensions, the secondary product has even degree, $1-d=-2$, so it maps pairs of bosonic operators to bosonic operators. Indeed, the secondary product turns out to induce an ordinary Poisson bracket in the (bosonic) chiral rings of a three-dimensional $\CN=4$ theory, which sit inside topological algebras $\CA$ of local operators.

We will mainly focus on 3d $\CN=4$ sigma-models, which may also be thought of as the IR limits of gauge theories.
Recall that having 8 supercharges (as in 3d $\CN=4$) requires the target $\CX$ of a sigma-model to be a hyperk\"ahler manifold~\cite{Hitchin:1986ea}. This means that $\CX$ has a $\mathbb{CP}^1$ worth of complex structures; and in each complex structure $\zeta\in \mathbb{CP}^1$, $\CX_\zeta$ is a K\"ahler manifold with a nondegenerate holomorphic symplectic form $\Omega_\zeta$. The existence of the holomorphic symplectic structure turns the ring of holomorphic functions $\C[\CX_\zeta]$ on $\CX_\zeta$ into a Poisson algebra, by the usual formula
\be \{f,g\} := \Omega_\zeta^{-1}(\pd f,\pd g)\,. \label{g-Poisson} \ee

Physically, a 3d $\CN=4$ sigma model admits a $\mathbb{CP}^1$ worth of topological twists $Q^{(\zeta)}$, identified by Blau and Thompson \cite{Blau:1996bx} and then studied by Rozansky and Witten \cite{Rozansky:1996bq}. The local operators in the cohomology of a particular supercharge $Q^{(\zeta)}$ may be identified as Dolbeault cohomology classes
\be \CA_\zeta = H_{\bar\pd}^{0,\bullet}(\CX_\zeta)\simeq H^\bullet(\CX_\zeta, \CO_{\CX_\zeta})\,,\ee
or (by the Dolbeault theorem) as the sheaf cohomology of the structure sheaf of holomorphic functions on $\CX_\zeta$.
Sitting inside this topological algebra are the holomorphic functions
\be \C[\CX_\zeta] = H_{\bar\pd}^{0,0}(\CX_\zeta)\;\subset \; \CA_\zeta\,, \ee
which correspond physically to a half-BPS chiral ring. We will show in Section~\ref{sec:3d-sigma}, by direct calculation, that the secondary product on $\CA_\zeta$ defined by topological descent recovers the natural geometric Poisson bracket on $\C[\CX_\zeta]$. Moreover, the secondary product on all of $\CA$ is controlled (working locally on the target) by the Poisson bracket on holomorphic functions alone.

It may be useful to note that if $\CX_\zeta$ is an affine algebraic variety, all the higher cohomology groups of $\CO_{\CX_\zeta}$ vanish, so that the algebra $\CA_\zeta$ is actually equivalent to the chiral ring $\C[\CX_\zeta]$. For example, the Higgs and Coulomb branches of 3d $\CN=4$ gauge theories with linear matter are (conjecturally) always affine, or admit affine deformations.

In the opposite regime, one could consider compact targets $\CX_\zeta$, as in the original work of Rozansky and Witten. In this case, the chiral ring $\C[\CX_\zeta]$ is trivial (as the only holomorphic functions on compact $\CX_\zeta$ are constants), so the secondary product vanishes tautologically on it.
 In fact we demonstrate that the secondary product vanishes on higher cohomology as well, \emph{i.e.}, on the entire topological algebra $\CA$. This is analogous to the corresponding B-model statement, that the Gerstenhaber bracket vanishes on the Dolbeault cohomology of polyvector fields on compact Calabi-Yau manifolds.

We explore some further applications of the secondary product in Sections \ref{sec:3d-sym}--\ref{sec:nonren}.
We begin by considering some special features of the secondary product in theories with flavor symmetry, where the descendants of moment-map operators are controlled by the structure of current multiplets. We illustrate some of these features in gauge theories, showing how the secondary product can be used to measure magnetic charge of monopole operators. Finally, we emphasize an important physical consequence of the topological nature of the secondary product in 3d $\CN=4$ theories, namely the non-renormalization of holomorphic symplectic structures.

\subsection{Basics}
\label{sec:3d-basics}

The 3d $\CN=4$ SUSY algebra is generated by eight supercharges, transforming as spinors $Q_\alpha^{a\dot a}$ of $SU(2)_E\times SU(2)_H\times SU(2)_C$\,, where $SU(2)_E$ is the Euclidean Lorentz group (acting on the $\alpha=-,+$ index) and $SU(2)_{H,C}$ are R-symmetries (acting on $a$ and $\dot a$ indices).
The supercharges obey
\be [Q_\alpha^{a\dot a},Q_\beta^{b\dot b}] = \epsilon^{ab}\epsilon^{\dot a\dot b}\sigma^\mu_{\alpha\beta}P_\mu \,,\ee
where $(\sigma^1)_\alpha{}^\beta=\bsp 0&1\\1&0\esp\,,\; (\sigma^2)_\alpha{}^\beta = \bsp 0&-i \\ i&0\esp\,,\; (\sigma^3)_\alpha{}^\beta = \bsp 1&0\\0&-1\esp$ are the Pauli matrices, and indices are raised and lowered with antisymmetric tensors $\epsilon^{12}=\epsilon_{21}=1$.

Two $\mathbb{CP}^1$ families of topological twists are available. One family contains the Rozansky-Witten supercharge
\be  Q   = \delta_{\dot a}{}^\alpha Q_\alpha^{1\dot a} = Q_-^{1\dot 1} + Q_+^{1\dot 2}\,, \label{Q0-3d}\ee
as well as its rotations by $SU(2)_H$, which look like $Q^{(\zeta)} = \frac{1}{\sqrt{1+|\zeta|^2}}\delta_{\dot a}{}^\alpha\big(Q_\alpha^{1\dot a}+\zeta Q_\alpha^{2\dot a}\big)$, indexed by an affine parameter $\zeta\in \mathbb{CP}^1$.
Every $Q^{(\zeta)}$ is a scalar under an improved Lorentz group, defined as the diagonal of $SU(2)_E\times SU(2)_C$. Moreover, it is easy to check that every $Q^{(\zeta)}$ obeys $(Q^{(\zeta)})^2 =0$.

To keep things simple, we will just work with $Q = Q^{(\zeta=0)}$ as in \eqref{Q0-3d}. Then the vector supercharge 
\be Q_\mu := -\tfrac i2 (\sigma^\mu)_{\dot a}{}^\alpha Q_\alpha^{2\dot a} \ee
obeys the desired relation
\be [Q,Q_\mu] = iP_\mu\,.\ee

The second family of topological supercharges is related to the first by swapping the roles of $SU(2)_C$ and $SU(2)_H$, \emph{i.e.} by applying 3d mirror symmetry. It contains the topological supercharge
\be \widetilde Q = \delta_a{}^\alpha Q_\alpha^{a1} = Q_-^{1\dot 1}+Q_+^{2\dot 1} \label{mirror-twist}\ee
and all its $SU(2)_C$ rotations. The corresponding vector supercharge is $\widetilde Q_\mu = -\tfrac i2 (\sigma^\mu)_{ a}{}^\alpha Q_\alpha^{a2}$, again obeying $[\widetilde Q,\widetilde Q_\mu]=i P_\mu$. This second family of topological supercharges will be relevant for gauge theory in Section~\ref{sec:3d-gauge}.

\subsection{Sigma model}
\label{sec:3d-sigma}

We now consider a 3d $\CN=4$ sigma-model with smooth hyperk\"ahler target $\CX$. We use the Rozansky-Witten twist with $Q=Q^{(\zeta=0)}$ as the topological supercharge, which amounts to choosing a particular complex structure $\zeta=0$ on the target, and viewing $\CX = \CX_{\zeta=0}$ as a complex symplectic manifold. The ring of topological local operators will contain holomorphic functions on $\CX$.

Much as in the case of the 2d B-model, the analysis of the secondary product reduces to a local computation on $\CX$. This is because
\begin{itemize}
\item The path integral of the RW-twisted 3d $\CN=4$ sigma-model localizes to constant (bosonic) maps \cite{Rozansky:1996bq, Kapustin:2008sc}. Moreover, correlation functions of Q-closed operators can be evaluated in the presence of any fixed vacuum $\phi_0\in \CX$ at spacetime infinity, in which case the path integral only depends on a neighborhood of $\phi_0$. Thus all topological correlators have consistent local specializations.
\item Deformations of the metric on $\CX$ that preserve the complex symplectic structure are $Q$-exact; and as a complex symplectic manifold any local neighborhood in $\CX$ is isomorphic (by Darboux's theorem) to $T^\ast\C^N\simeq \C^{2N}$ with constant symplectic form.
\end{itemize}
Therefore, it suffices to consider a target $\CX=\C^{2N}$ with local complex coordinates $\{X^i\}_{i=1}^{2N}$ and a constant symplectic form $\Omega = \frac12\Omega_{AB} dX^A dX^B$. We could further fix $\Omega_{AB} = \left(\begin{smallmatrix} 0 & I \\ -I & 0\end{smallmatrix}\right)$, but it is more illustrative to leave $\Omega_{AB}$ undetermined.

The 3d $\CN=4$ sigma-model to $\CX=\C^{2N}$ is a theory of free hypermultiplets. Its bosonic fields are conveniently described as $2N$ doublets $\{\phi^{aA}\}_{a=1,2}^{A=1,...,2N}$ of the $SU(2)_H$ R-symmetry (acting on the $a$ index), subject to a reality condition%
\be (\phi^{aA})^\dagger = \epsilon_{ab}\Omega_{AB} \phi^{bB}\,.\ee
We may thus identify the $a=1$ components of $\phi^{aA}$ as holomorphic target-space coordinates, and the $a=2$ components as their complex conjugates
\be \qquad   \phi^{1A} = X^A\,, \qquad \phi^{2A} = -\Omega^{AB}\ol X_B \qquad ((X^A)^\dagger = \ol X_A)\,.  \ee
For example, the bosonic fields of a single free hypermultiplet sit in the $2\times 2$ matrix
\be \phi^{aA} = \begin{pmatrix} X^1 & X^2 \\ \ol X_2 & -\ol X_1\end{pmatrix}\,. \ee

The fermionic fields consist of $2N$ spinors
$\psi_\alpha^{\dot aA}$ of the Lorentz group $SU(2)_E$ and the second R-symmetry $SU(2)_C$. The supercharges act as
\be Q_\alpha^{a\dot a} (\phi^{bA}) = \epsilon^{ab}\psi_\alpha^{\dot aA}\,,\qquad Q_{\alpha}^{a\dot a}(\psi_\beta^{\dot bA}) =  -i\epsilon^{\dot a\dot b} \sigma^\mu_{\alpha\beta}\pd_\mu \phi^{aA}\,,\ee
and preserve the Euclidean action
\be  \label{3d-action}S = \int d^3x \big[ \tfrac12 \epsilon_{ab}\Omega_{AB} \pd_\mu \phi^{aA} \pd^\mu \phi^{bB} + \tfrac i2 \,\epsilon_{\dot a\dot b}\Omega_{AB} \psi^{\dot a A}_\alpha (\sigma^\mu)^{\alpha\beta} \pd_\mu \psi^{\dot b B}_\beta\big]\,. \ee

It is convenient to regroup the fermions into representations of the improved Lorentz group. Following~\cite{Rozansky:1996bq}, we define spacetime scalars
$\eta_A = - \Omega_{AB} \delta_{\dot a}{}^\alpha \psi_\alpha^{\dot a A}$
and 1-forms $\chi_\mu^A = \tfrac i2 (\sigma_\mu)_{\dot a}{}^\alpha \psi_\alpha^{\dot a A}$. Conversely, $\psi_\alpha^{\dot aA} = -\tfrac12\Omega^{AB}\delta_\alpha{}^{\dot a}\eta_B -i(\sigma^\mu)_\alpha{}^{\dot a} \chi_\mu^A\,.$
Then the action reduces to%
\be \label{3d-action-chi} S = \int \big[  dX^A \wedge *d \ol X_A +  \Omega_{AB} \chi^A\wedge d\chi^B - \eta_A d*\chi^A\big]\,.\ee
Setting $\mathbb Q = Q_\mu dx^\mu$, the SUSY transformations relevant for descent are
\be \label{SUSY3d} \begin{array}{c} Q (X^A) = 0\,,\quad Q(\ol X^A) = \eta_A\,,\quad Q(\eta_A) = 0\,,\quad Q\chi^A = d X^A\\[.2cm]
    \mathbb Q (X^A)  = \chi^A\,,\quad \mathbb Q(\ol X_A)=0\,,\quad \mathbb Q(\eta_A) = d \ol X_A\,,\quad
    \mathbb Q(\chi^A) = \Omega^{AB}*d \ol X_B\,.
\end{array} \ee

The identification of the algebra $\CA$ of (polynomial) local operators with Dolbeault cohomology comes about by identifying $\eta_A$ with anti-holomorphic one-forms on the target
\be \eta_A \quad\leftrightarrow \quad \mathbf{d}\ol X_A\,. \ee
The algebra may be constructed from polynomials in the $X^A, \ol X^A$ and $\eta_A$, thought of as $(0,q)$ forms
\be  \omega = \omega^{A_1...A_q}(X,\ol X)\eta_{A_1}...\,\eta_{A_q}\quad \leftrightarrow \quad  \omega^{A_1...A_q}(X,\ol X)\mathbf d \ol X_{A_1}...\, \mathbf d\ol X_{A_q}\,, \ee
with $Q \;\leftrightarrow\;\bar\pd$ acting as the Dolbeault operator.

\subsubsection{Secondary product in the chiral ring}

In the theory with target $\C^{2N}$, the chiral ring is%
\footnote{More generally, one could consider analytic functions of $X$. The analysis here remains unchanged.}
\be \C[\CX] \simeq \{\text{polynomials in the local operators $X^A$}\} = H_{\bar\pd}^{0,0}(\C^{2N})\,. \ee
The primary product is just ordinary multiplication of polynomials.
We would like to show that the secondary product agrees with the geometric Poisson bracket on the generators $X^A$. In particular, we expect
\be \{ X^A,X^B \} = \Omega^{AB}\,. \label{bracket-XAXB}\ee

Since we are in $d=3$ dimensions, we compute the secondary bracket of $X^A$ and $X^B$ by finding the second descendant of the operator $X^A(x)$, and integrating it around~$X^B$. The SUSY transformations \eqref{SUSY3d} yield
\be (X^A)^{(1)} = \mathbb Q(X^A) = \chi^A\,,\qquad (X^A)^{(2)} = \mathbb Q(\chi^A) = \Omega^{AB} *d\ol X_B\,. \ee
Taking another exterior derivative, we find an equation of motion, much like in the B-model:
\be d(X^A)^{(2)} = 
\Omega^{AB} d*d(\ol X^B) =   \Omega^{AB}\frac{\delta S}{\delta X^B}\,.\ee
Since there is a delta-function singularity in the correlation function $\frac{\delta S}{\delta X^B}(x) X^A(y)\sim \delta_B{}^A\delta^3(x-y)$, the secondary product becomes
\begin{align} \{X^A,X^B\}  &= \llbracket X^A(S^2_y) * X^B(y) \rrbracket \notag \\
   &= \Big\llbracket \oint_{S^2_y} (X^A)^{(2)} X^B(y) \Big\rrbracket \notag \\
   &= \Big\llbracket \int_{D^3_y} d(X^A)^{(2)} X^B(y) \Big\rrbracket \notag \\
   &=  \Omega^{AC}\delta_C{}^B \Big\llbracket \int_{D^3_y} \delta^3(x-y) \Big\rrbracket \notag \\
   &= \Omega^{AB}\,. \label{XX-corr}
\end{align}

Note that the derivation property of the secondary product now implies that for arbitrary holomorphic functions $f,g\in \C[\CX]$ we will now have
\be \{f,g\} = \Omega^{AB} \pd_A f \,\pd_B g = \Omega^{-1}(\pd f,\pd g)\,, \label{fg-Omega}\ee
reproducing the familiar definition of the geometric Poisson bracket. The standard properties of the Poisson bracket of functions, such as anti-symmetry $\{f,g\}=-\{g,f\}$ and the Jacobi identity, follow from the general properties of the secondary product in $d=3$ dimensions.

We also recall that, although we computed the bracket by using a cycle $S^2\times p$ in the configuration space $\mathcal C_2(\R^3)$, we could have used any other cycle in the same homology class.
In particular, in $d=3$ dimensions there is a more symmetric choice: we can take a cycle $S_x^1\times S_y^1$ that topologically looks like the configuration space of points on the two strands of the Hopf link in $\R^3$\,, with linking number $1$ --- as in Figure \ref{fig:Hopf} on page \pageref{fig:Hopf}.

It is amusing to do this computation explicitly. Consider the local operators $X^A(x)$ and $X^B(y)$. We know from \eqref{SUSY3d} that the first descendants are
\be (X^A)^{(1)} =  \chi^A\,,\qquad (X^B)^{(1)} = \chi^B\,.\ee
The secondary product between $X^A$ and $X^B$ now comes from the correlation function
\be \{X^A,X^B\}  = \Big\llbracket \oint_{S^1_x}  \oint_{S^1_y} \chi^A(x) \chi^B(y)  \Big\rrbracket \ee
We can evaluate this by choosing a disc $D^2_x$ whose boundary is $S^1_x$, and rewriting the first integral as
 $\oint_{S^1_x}\chi^A(x) = \oint_{D^2_x} d\chi^A(x)$. The two-point function of $d\chi^A(x)$ and $\chi^B(y)$ acquires a  singularity due to the $\Omega_{AB}\chi^A\wedge d\chi^B$ term in the action \eqref{3d-action-chi},
\be d\chi^A(x) \chi^B(y) \sim \Omega^{AB} \delta^{(3)}(x-y)\,.\ee
Since the disc $D^2_x$ intersects the second circle $S^1_y$ at precisely one point, we recover
\be \{X^A,X^B\} = \Big\llbracket \oint_{S^1_x}  \oint_{S^1_y} \chi^A(x) \chi^B(y)  \Big\rrbracket  =  \Big\llbracket \int_{D^2_x\times S^1_y} d\chi^A(x) \chi^B(y)  \Big\rrbracket = \Omega^{AB}\,.\ee

\subsubsection{Global considerations and higher cohomology}

For general target $\CX$, the topological algebra of local operators $\CA = H_{\bar\pd}^{0,\bullet}(\CX)$ is identified as Dolbeault cohomology, with the primary product given by the usual wedge product~\cite{Rozansky:1996bq}. By working locally on $\CX$, we find that the secondary product of any functions $f,g\in H_{\bar\pd}^{0,0}(\CX)$ must be given by \eqref{fg-Omega}, namely
\be \{f,g\} = \Omega^{-1}(\pd f,\pd g)\,. \ee
Technically, this reasoning is valid if $f$ and $g$ are analytic locally on $\CX$, so that the computation leading to \eqref{fg-Omega} makes sense.

An analogous local computation shows that the secondary product of any $\bar\pd$-closed forms $\omega\in \Omega^{0,q}(\CX)$, $\lambda \in \Omega^{0,r}(\CX)$ representing higher cohomology classes in $\CA$ is given by essentially the same formula,
\be \qquad \{\omega,\lambda\} = \Omega^{-1}(\pd \omega,\pd \lambda) 
\;\; \in \Omega^{0,q+r}(\CX)\,. \label{Omega-forms}\ee
In this more general case, we must consider local operators that depend (analytically) on $\ol X$ and $\chi$, as well as $X$. However, correlation functions involving $\ol X$ and $\chi$ do not have strong enough singularities to give a nonvanishing contribution to integrals such as \eqref{XX-corr}, so these operators become invisible to (are in the kernel of) the secondary product.

When $\CX$ is compact K\"ahler, any class $[\omega]\in \Omega^{0,q}(\CX)$ is represented by a $(0,q)$ form that is both $\bar\pd$- and $\pd$-closed. It follows from \eqref{Omega-forms} that the secondary product vanishes on the entire algebra $\CA$ of local operators.

\subsubsection{Gradings}
\label{sec:Zgrading}

Rozansky-Witten theory with complex symplectic target $\CX$ always has a $\Z/2$ grading by fermion number, such that all bosonic fields $\phi^{aA}$ that locally parametrize the target are even, and all fermions $\psi_\alpha^{\dot a A}$ (or $\eta_A,\chi^A$) are odd. Similarly, $Q$ and $Q_\mu$ are odd. The secondary product then becomes even, precisely as one would expect for the Poisson bracket of functions.

Given extra structure on $\CX$, the $\Z/2$ grading can be lifted to a $\Z$ grading, under which the secondary product has degree $-2$. Physically, the $\Z$ grading comes from an unbroken $U(1)_H\subset SU(2)_H$ R-symmetry, which acts on the holomorphic symplectic target $\CX$ as a complex isometry under which the holomorphic symplectic form has degree $+2$.%
\footnote{Viewing $\CX$ as a hyperkahler manifold, $U(1)_H$ is a metric isometry that rotates the twistor $\mathbb{CP}^1$ of complex structures about a fixed axis, leaving fixed the chosen complex structure `$\zeta$' that we use to define the Rozansky-Witten twist.} %
Both Higgs and Coulomb branches of 3d $\CN=4$ (linear) gauge theories have this property, as do Coulomb branches of 4d $\CN=2$ theories compactified on a circle. The latter notably include Hitchin systems.
In some cases, the $U(1)_H$ R-symmetry may need to be mixed with a flavor symmetry (a holomorphic Hamiltonian isometry of $\CX$) to ensure that bosonic fields all have even degree.

For example, the free hypermultiplet parametrizing $\CX=\C^2$ has a naive $U(1)_H$ R-symmetry (corresponding to the superconformal R-symmetry) under which the holomorphic scalars $X^1,X^2$ both have charge $+1$ and the fermions all have charge zero. The holomorphic symplectic form $\Omega = \mathbf{d}X^1\wedge \mathbf{d}X^2$ has charge $+2$ as desired, but it not suitable to have odd-charged bosons and even-charged fermions. In this case, there is an additional $U(1)_f$ flavor symmetry that can be used to define an improved $U(1)_H' = \text{diag}(U(1)_H\times U(1)_f)$, for which bosons are even and fermions are odd:
\be \begin{array}{c|cccc|c}
 &X^1&X^2& \eta_1,\chi^1 & \eta_2,\chi^2 & \Omega  \\\hline
U(1)_H & 1 & 1 & 0 & 0 & 2 \\
U(1)_f & 1 & -1 & 1 & -1 & 0 \\
U(1)_H' & 2 & 0 & 1 & -1 & 2 \end{array}
\ee

\subsection{Flavor symmetry}
\label{sec:3d-sym}

In a 3d $\CN=4$ theory with flavor symmetries, the secondary product is closely related to the structure of current multiplets. We outline the basic relation, beginning with the case of a sigma model.

Recall that a flavor symmetry is a global symmetry that commutes with supersymmetry. 
In a sigma model with hyperk\"ahler target $\CX$, a group $F$ of flavor symmetries corresponds geometrically to tri-Hamiltonian isometries of $\CX$. 
If we view $\CX$ as a holomorphic symplectic manifold in a fixed complex structure, flavor symmetries may be extended (complexified) to complex isometries of $\CX$ that preserve the holomorphic symplectic form $\Omega$. They are generated by holomorphic vector fields $\Omega^{-1}\pd\mu$,
where the complex moment map $\mu$ is a holomorphic function on $\CX$ valued in the complexified dual Lie algebra of $F$
\be \mu\,: \CX \to \mathfrak f_\C^*\,.\ee
In particular, acting on the ring of holomorphic functions $\C[\CX]$, the complexified symmetry is generated by taking Poisson bracket with $\mu$.

Physically, $\mu$ is a local operator (a matrix of local operators) in the topological algebra $\CA$ for a particular Rozansky-Witten twist. We found in Section~\ref{sec:3d-sigma} that the secondary product in $\CA$ coincides with the geometric Poisson bracket. Therefore, we expect the secondary product with $\mu$ to generate the action of flavor symmetries on $\CA$. In the case of an abelian group $F$, one would more commonly say that the bracket $\{\mu,-\}$ should measure flavor charge.

In QFT, there is a canonical $\mathfrak f^*$-valued operator that generates global $F$ symmetries: the current $J=J_\mu dx^\mu$. The infinitesimal action on any local operator $\CO$ is given by an integral of $*J$ on a two-sphere $S^2_y$ surrounding $\CO(y)$,
\be  \int_{S^2_y} *J(x) \CO(y)\,. \label{J-S2} \ee
Comparing this to the definition of the secondary product suggests that the second descendant of the complex moment map should be $\mu^{(2)} = *J$; then $\{\mu,\CO\}$ would coincide with \eqref{J-S2}. Being more careful, we actually find that the bosonic part of $\mu^{(2)}$ is
\be \mu^{(2)} = *\tfrac12(J +  d\mu_\R)\,, \label{mu-J} \ee
where $\mu_\R:\CX\to \mathfrak f^*$ is the real moment map associated to the K\"ahler form on $\CX$ (as opposed to the holomorphic symplectic form).
The complexified current operator $J + d\mu_\R$ generates the complexified $F_\C$ action on the complex symplectic manifold $\CX$. Note that $J + d\mu_\R$ is not conserved, since only $F$ and not $F_\C$ is an exact symmetry of the 3d $\CN=4$ theory (\emph{i.e.} a metric isometry).

We may illustrate this in the theory of a free hypermultiplet, with complex bosonic fields $X^1,X^2$ that have charges $+1,-1$ under a flavor symmetry $F=U(1)\subset USp(1)$. The complexified symmetry is $F_\C=\C^*$, which preserves $\CX=\C^2$ with its holomorphic symplectic form $\Omega=dX^1\wedge dX^2$. The complex moment map is
\be \mu = X^1X^2\,.\ee
The bosonic current, in a convenient normalization, is
\be J = (X^1d\ol X_1-\ol X_1 dX^1) - (X^2 d\ol X_2-\ol X_2 d X^2)\,, \ee
whereas
\be \mu^{(2)} = *(X^1 d\ol X_1-X^2 d\ol X_2) +2 \chi^1\chi^2 = *\tfrac12(J+ d\mu_\R) + 2\chi^1\chi^2\,, \ee
where $\mu_R = |X^1|^2 - |X^2|^2$ is the real moment map. The extra fermionic term $2\chi^1\chi^2$ does not contribute to the secondary bracket of chiral-ring operators.

The relation \eqref{mu-J} between the complex moment map operator and the current is not unique to sigma-models. The relation  follows from the universal structure of $\CN=4$ current multiplets --- which contain moment maps as the bottom components. Every 3d $\CN=4$ theory with a flavor symmetry has moment-map operators that are related to the current in the same way.

\subsubsection{Gauge theories and monopole charge}
\label{sec:3d-gauge}

We can also illustrate the relation between secondary products and flavor symmetries in gauge theories. 
Recall that in 3d $\CN=4$ gauge theory with gauge group $G$, there is a ``topological'' flavor symmetry with the same rank as the center of $G$. This symmetry acts on monopole operators and ``measures'' monopole charge. Its moment maps are traces of adjoint scalars in the gauge multiplet, and its conserved current is (an appropriate trace of) the Hodge-dual of the $G$ field strength. We will verify that the moment maps and current are related as expected by topological descent.

For simplicity, we will consider pure $G=U(N)$ gauge theory. Then the topological symmetry is $U(1)$, with conserved current
\be J = *\text{Tr}(F)\,, \ee
where $F$ is the field strength. The monopole operators that the topological symmetry acts on are detected by the  ``mirror'' $\widetilde Q$ topological twist discussed in \eqref{mirror-twist}. The cohomology of $\widetilde Q$ contains half-BPS disorder operators $V_\lambda$ defined by specifying a local singularity both in the field strength and in one of the three vectormultiplet scalars `$\sigma$' of the form
\be V_\lambda(x)\,:\quad  \sigma \sim  \frac{1}{2r}\text{diag}(\lambda_1,...,\lambda_N)\,,\qquad F \sim *d\frac{1}{2r}\text{diag}(\lambda_1,...,\lambda_N)\,, \label{Vlambda} \ee
where $r_x$ is the distance from the insertion point $x$, and $\lambda = (\lambda_1,...,\lambda_N)\in \mathbb Z^N/S_N$ is a cocharacter of $U(N)$, defined modulo the permutation action of the Weyl group. The topological charge of $V_\lambda$ is $\sum_i\lambda_i$; it is easy to see that this is the integral of the topological current around any $S^2$ surrounding a singularity of the form \eqref{Vlambda},
\be \int_{S^2_x}  \tfrac{1}{2\pi} \text{Tr}F\, V_\lambda(x) =  \Big(\sum_i \lambda_i\Big)V_\lambda(x)\,. \label{FV}\ee
(Mathematically, the integral measures the first Chern class of the gauge bundle in the presence of the $V_\lambda$ singularity.) Note that we keep the operator $V_\lambda(x)$ on the RHS of \eqref{FV}, since the singularity is still present.

The scalar $\sigma\in \mathfrak g$ also plays the role of the real (K\"ahler) moment map for the topological symmetry,
\be \mu_\R = \text{Tr}(\sigma)\,.\ee
The remaining two vectormultiplet scalars form a complex combination $\varphi\in \mathfrak g_\C$, which supplies the complex moment map
\be \mu = \text{Tr}(\varphi) \ee
Just like the $V_\lambda$'s, $\mu=\text{Tr}(\varphi)$ is an element of the topological ring of local operators
\be V_\lambda, \text{Tr}(\varphi)\in \C[\mathcal M_C] \subset \CA\,.\ee
Since $\text{Tr}(\varphi)$ is the complex moment map, we \emph{expect} that its secondary bracket with a monopole operator is
\be \{\tfrac{1}{2\pi}\text{Tr}(\varphi),V_\lambda\} = \Big(\sum_i \lambda_i\Big)V_\lambda\,. \ee

The key to this identity lies, as usual, in identifying the second descendant. A straightforward computation gives
\be \text{Tr}(\varphi)^{(2)} = \mathbb Q(\mathbb Q(\text{Tr}(\varphi))) = \tfrac12\text{Tr}(F+ *D\sigma)\,. \ee
Note that, just as in \eqref{mu-J}, we do not find the flavor current on the nose, but rather a holomorphic modification thereof. The integral of $\text{Tr}(*D\sigma)$ around the $\sigma$ singularity in \eqref{Vlambda} (required for a $\widetilde Q$-closed monopole) produces exactly the same contribution as the integral of $\text{Tr}(F)$ around the singularity in the field strength. With a suitable normalization, the two contributions combine to give
\be \{\tfrac{1}{2\pi}\text{Tr}(\varphi),V_\lambda\}  =\tfrac{1}{2\pi} \int_{S^2} \text{Tr}(\varphi)^{(2)} V_\lambda(x) = \Big(\sum_i \lambda_i\Big)V_\lambda \ee
as desired.

\subsection{Non-renormalization of Poisson brackets}
\label{sec:nonren}

In the works \cite{Gaiotto:2008cd,Bullimore2015} (closely related to the mathematical works \cite{Nakajima:2015txa,Braverman:2016wma})
the Coulomb branches $\cM$ of 3d $\cN=4$ supersymmetric field theories are studied,
as holomorphic symplectic spaces.
In both cases the key step is a reduction to 
the IR description in terms of abelian gauge theory, 
which turns out to give enough information to
completely describe $\cM$ as a holomorphic symplectic manifold. (One then goes
on to describe its \hk structure, by considering it as a holomorphic symplectic
manifold in all of its complex structures simultaneously.)

One subtle point in this analysis has never been quite clear:
why can one compute the holomorphic symplectic form on $\cM$ \ti{exactly}
using only the IR description of the theory?
Our discussion in this section suggests an answer:
the Poisson bracket in the Coulomb-branch chiral ring 
of a 3d $\CN=4$ gauge theory is defined in a purely topological
way, and thus it can be computed exactly either in the UV or the IR.

\section{Deformation quantization in the \texorpdfstring{$\Omega$}{Omega}-background}
\label{sec:omega}

In this section we explain an application of the discussion in Section~\ref{sec:3dn=4}: 
we give a topological derivation of the statement that,
when a 3d $\cN=4$ theory is placed in $\Omega$-background, the
chiral ring undergoes deformation quantization. Under some optimistic assumptions about the properties of the physical $\Omega$-background we derive this result as an aspect of equivariant localization in the context of disc algebras. Namely, we will explain a new result on disc operads: the Poisson bracket underlying an oriented 3-disc algebra has a canonical ``deformation quantization"\footnote{In fact we naturally get the structure of algebra over a graded form of the $BD_1$ operad controlling deformation quantizations.}
 over a graded affine line --- \emph{i.e.}, an associative algebra over $\C[\epsilon]$ (where $F(\epsilon)=2$), recovering the Poisson bracket from the commutator to first order in $\epsilon$ (though without any a priori guarantee of flatness).

 More generally, given a $d$-dimensional TQFT, one can in principle turn on $\Omega$-backgrounds corresponding to rotations in any collection of $n$ independent planes, $n \leq \lfloor d/2\rfloor$, thereby deforming the structure of the operator algebra
\be \text{$\Omega$ in $n$ planes\,:}\quad E_d \leadsto E_{d-2n}\,. \ee 
This is studied further in forthcoming work~\cite{BZNeitzke}.

\subsection{Properties of the \texorpdfstring{$\Omega$}{Omega}-background}

The $\Omega$-background in 3d $\CN=4$ theories may be thought of as a deformation of a topological supercharge $Q$. For example, this may be a Rozansky-Witten supercharge \eqref{Q0-3d} or its mirror \eqref{mirror-twist}; the discussion here is general, and applies equally well to either one. We fix an axis in flat three-dimensional Euclidean spacetime $\R^3$, and let $U(1)_E$ denote rotations about this axis. We also assume there is an unbroken $U(1)_R$ symmetry such that the topological supercharge $Q$ is invariant under diagonal $U(1)'\subset U(1)_E\times U(1)_R$ rotations.
(When there exists an improved Lorentz group $SU(2)_E'$, we can just take $U(1)' \subset SU(2)_E'$ to be the subgroup fixing an axis.)
Let
\be X \in \mathfrak{u}(1)' \ee
be a generator of the $U(1)'$ symmetry, normalized so that $\exp(X)=1\!\!1$.

$\Omega$-backgrounds in 3d $\N=4$
theories have been considered before in \cite{Yagi2014}, generalizing their 4d $\CN=2$ cousins \cite{Nekrasov:2002qd, Nekrasov:2009rc}; they also played a major role in recent constructions of the quantized algebra of functions on the Coulomb branches of gauge theories \cite{Nakajima:2015txa, Braverman:2016wma, Bullimore2015, Bullimore:2016hdc}.
We will not commit ourselves to a specific construction of the $\Omega$-background;
we just assume that it is a $1$-parameter deformation of the theory, with the following
properties:
\begin{itemize} 
	\item Considered as a vector space,
	the space $\ops_\delta$ is independent of $\eps$, and thus 
	the operators $Q$ and $Q_\mu$ acting on $\ops_\delta$ 
	continue to make sense in the
	deformed theory --- though they need no longer be symmetries.
	\item The deformed theory is invariant
under a deformed supercharge
\begin{equation} \label{eq:Q-eps}
	Q_\eps = Q + \eps Q_X\,,
\end{equation}
where $Q_X$ obeys
\begin{equation}
	[Q, Q_X] = X\,, \qquad [Q_\mu, Q_X] = 0\,, \qquad [Q_X, Q_X] = 0\,.
\end{equation}
\end{itemize}

It follows that the operator $Q_\eps$ has $Q_\eps^2 = \eps X$. In particular, 
acting on $U(1)'$-invariant operators we have $Q_\eps^2 = 0$.
Also note that $[Q_\eps, Q_\mu] = i P_\mu$, so all translations
are $Q_\eps$-exact, despite the fact that translations that
do not commute with the $U(1)'$ action are not 
symmetries of the $\Omega$-deformed theory.\footnote{This might at
first seem puzzling to readers used to the idea that being $Q$-exact 
is even \ti{stronger}
than being a symmetry. The point is that the strong consequences
of $Q_\eps$-exactness only hold for $U(1)$-invariant operators,
since only on these do we have $Q_\eps^2 = 0$;
a translation in one of the broken directions will break
the $U(1)$-invariance.}

One succinct way to define the $\Omega$-background, at least formally,
comes from considering our 3d $\cN=4$ theory as a 
1d $\cN=4$ theory.\footnote{of a somewhat unconventional
sort: in a Lagrangian description it would have infinitely many fields, corresponding to the infinitely many
modes of the field in the two suppressed dimensions.} 
From the 1d point of view, the $U(1)'$ symmetry generated by $X$
is just a global symmetry, and the $\Omega$-background can be viewed
as a complex twisted-mass deformation associated to this global symmetry. 
In particular, even after the deformation, we still have a conventional
1d $\cN=4$ theory.
For more on this perspective, see \emph{e.g.} \cite{Bullimore:2016hdc}.

\subsection{Deformation quantization}

Now let us restrict to the 
subspace $\ops_\delta^X \subset \ops_\delta$ consisting of $X$-invariant operators. Acting on 
$\ops_\delta^X$, we have $Q_\eps^2 = 0$. 
Let $\cA_\eps$ denote the cohomology of $Q_\eps$.

Since $\cA_\eps$ is the $Q_\eps$-cohomology in the 1d theory, it carries the usual structure we have in a 
1d theory, namely a not-necessarily-commutative product $\pprod_\eps$ 
as discussed in Section~\ref{sec:products-1d}.
It has been proposed in \cite{Yagi2014} that as $\eps \to 0$ 
the commutator in $\cA_\eps$ is controlled by the Poisson bracket.
To formulate this precisely consider $Q$-closed, $X$-invariant operators
$\phi_1, \phi_2$ in the 1d theory, and
assume that they admit deformations to $Q_\eps$-closed operators $\phi_{1,\eps}$, $\phi_{2,\eps}$.
Then, the proposal is that
\begin{equation} \label{eq:deformation-quantization-prelim}
	\lim_{\eps \to 0} \frac{ \cclass{\phi_{1,\eps}} \pprod_\eps \cclass{\phi_{2,\eps}} - \cclass{\phi_{2,\eps}} \pprod_\eps \cclass{\phi_{1,\eps}} }{\epsilon} =   \sprod{\cclass{\phi_1}}{\cclass{\phi_2}} \,.
\end{equation}
Said otherwise, $\cA_\eps$ is a \ti{deformation quantization} of the Poisson algebra $\cA$.

A tautological reformulation of \eqref{eq:deformation-quantization-prelim} is to say that there
exists an operation $\sprod{\cdot}{\cdot}_\eps$ that as $\eps \to 0$ reduces to $\sprod{\cdot}{\cdot}$,
and obeys the exact relation
\begin{equation} \label{eq:deformation-quantization}
\cclass{\phi_{1,\eps}} \pprod_\eps \cclass{\phi_{2,\eps}} - \cclass{\phi_{2,\eps}} \pprod_\eps \cclass{\phi_{1,\eps}} = \eps \sprod{\cclass{\phi_{1,\eps}}}{\cclass{\phi_{2,\eps}}}_\eps\,.
\end{equation}
This is the version of the deformation quantization that we will derive from topological arguments
below.

The desired \eqref{eq:deformation-quantization} 
is a relation between binary operations on $\cA_\eps$.
When $\eps = 0$, we have seen in Section~\ref{sec:algebras} 
that these operations originate
from homology classes on $\cC_2(B) \simeq S^2$. Namely,
$\pprod$ comes from a degree $0$ class while $\sprod{\cdot}{\cdot}$
comes from a degree $2$ class. 
Thus \eqref{eq:deformation-quantization} looks a bit bizarre:
it says that after we set $\eps \neq 0$ there
is a relation between homology classes of different degrees!
Where could such a relation come from?

The key is that, when $\eps \neq 0$, binary operations on $\cA_\eps$ arise
from classes in \ti{$U(1)$-equivariant} homology $H_\bullet^{\eps}(\cC_2(B))$ rather
than ordinary homology. Once this is understood, \eqref{eq:deformation-quantization} follows directly. In the rest of this 
section we develop this story.

\subsection{Equivariant homology}
We quickly recall some background on equivariant homology.
Let $M$ be any space with $U(1)$ action.
A convenient model for $H_\bullet^{\eps}(M)$ is the homology of the complex 
of singular chains $S_\bullet(M)$, with a deformed differential
\begin{equation}
	\partial_\eps = \partial + \eps J\,.
\end{equation}
Here $\partial: S_k(M) \to S_{k-1}(M)$ is the usual boundary operator,
and $J: S_k(M) \to S_{k+1}(M)$ is the operator of ``sweeping out'' by
the $U(1)$ action.  It can be defined as
\be J(C) = \rho_*([U(1)]\times C)\ee
where $\rho: U(1) \times M \to M$ is the group action, $[U(1)]$ is (some simplicial representative of) the fundamental class of $U(1)$, and $\times: S_\bullet(U(1))\otimes S_\bullet(M)\to S_\bullet(U(1)\times M)$ is (some simplicial approximation to) the cross-product map. 
Note that if $C$ is a $U(1)$-invariant chain then $J(C) = 0$.
So, if $C$ has no boundary and is $U(1)$-invariant,
then $\partial_\eps C = 0$ and we get a class 
$\cclass{C} \in H_\bullet^\eps(M)$.

There is an equivariant analogue of the usual Stokes theorem.
To formulate it, we define the equivariant differential
$\de_\eps$ on  $\Omega^\bullet(M)$, by
\begin{equation}
	\de_\eps = \de + \eps \iota_X\,.
\end{equation}
Then, for any $U(1)$-invariant form $\alpha \in \Omega^*(M)$, we have
\begin{equation} \label{eq:equivariant-stokes}
	\int_C \de_\eps \alpha = \int_{\partial_\eps C} \alpha\,.
\end{equation}
In particular, there is a well defined pairing between
$\de_\eps$-cohomology classes and $\partial_\eps$-homology classes.

\subsection{Equivariant homology of \texorpdfstring{$S^2$}{S2}} \label{sec:equivariant-S2}

Our basic example is $M = S^2$ with $U(1)$ acting by rotations.
See \autoref{fig:equivariant-relation}.
The $0$-chains associated to the fixed points $a, b \in S^2$ have corresponding
classes $\cclass{a}, \cclass{b} \in H_\bullet^\eps(S^2)$.
We also choose a $1$-chain $\gamma$ running from $a$ to $b$.
In ordinary homology, we have $\partial \gamma = b - a$
and thus $\cclass{b} = \cclass{a}$.
In contrast, in equivariant homology, there is a correction
coming from the fact that $\gamma$ is not $U(1)$-equivariant:
since $\gamma$ sweeps out to $J(\gamma)=S^2$, we have
\begin{equation} \label{eq:basic-relation}
	\partial_\eps \gamma = b - a + {\eps} \,S^2,
\end{equation}
and thus\footnote{The relation \eqref{eq:equivariant-relation} says that 
as far as equivariant homology classes go,
we can replace the whole $S^2$ by $\frac{1}{\eps}$ times
the difference of the $U(1)$-fixed points. This might sound familiar
to readers familiar with equivariant localization. Indeed, 
if $\de_\eps \alpha = 0$,
pairing $\alpha$ with 
\eqref{eq:equivariant-relation} gives the Atiyah-Bott-Duistermaat-Heckman
localization formula for $M = S^2$,
\begin{equation}
	\int_{S^2} \alpha = \frac{1}{\eps} (\alpha(a) - \alpha(b))\,.
\end{equation}
Thus we can interpret \eqref{eq:equivariant-relation} as a homology
version of the familiar localization in equivariant cohomology.}
\begin{equation} \label{eq:equivariant-relation}
	\cclass{a} - \cclass{b} = {\eps} \cclass{S^2}\,.
\end{equation}

\insfigscaled{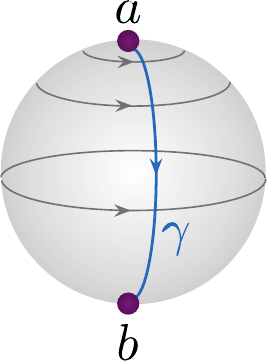}{0.9}{Chains on $S^2$ entering the basic relation \eqref{eq:basic-relation}.}

\subsection{Equivariant descent}

In the $\Omega$-deformed theory we still have a version of topological
descent. Indeed, consider a $Q$-closed, $U(1)$-invariant operator
$\phi$ in the 1d theory, and
assume as above that $\phi$ admits a deformation to a $Q_\eps$-closed operator $\phi_\eps$.

Now we can build the total descendant $\phi_\eps^*$, following much the same
strategy as we reviewed in Section~\ref{sec:descent}.
Despite the fact that we view the theory as a 1d theory, we can still define a ``position-dependent'' operator 
$\phi_\eps(x)$ for $x \in \R^3$, just
by exponentiating the action of $P_\mu$: \emph{i.e.} we define $\phi_\eps(x)$ to match
the 1d $\phi_\eps(x)$ when $x$ is on the axis of rotation,
and in general require it to satisfy 
$\partial_{x^\mu} \phi_\eps(x) = i P_\mu \phi_\eps(x)$.
Then we build up the higher form operators by successively applying the $Q_\mu$\,:
\begin{equation}
	\phi_\eps^*(x) = \sum_{k=0}^3 \frac{1}{k!}(Q_{\mu_1} \cdots Q_{\mu_k} \phi_\eps(x)) \, \de x^{\mu_1} \wedge \cdots \wedge \de x^{\mu_k}\,.
\end{equation}
We claim that we have the key relation
\begin{equation} \label{eq:Qeps-deps-total}
	Q_\eps \phi_\eps^*(x) = \de_\eps \phi_\eps^*(x)\,.
\end{equation}

To prove \eqref{eq:Qeps-deps-total}, first note that it holds when $x$ is on the axis, using the fact that
$[Q_\eps, Q_\mu] = i P_\mu$, and on the axis $\de_\eps = \de$, $Q_\eps \phi_\eps(x) = 0$. 
Next, we observe that the difference between the LHS and RHS is covariant with respect to translations:
\begin{align}
(\partial_{x^\mu} - i P_\mu) (\de_\eps - Q_\eps) \phi_\eps^*(x) &= ([\partial_{x^\mu}, \de_\eps] + [i P_\mu,Q_\eps] + (\de_\eps - Q_\eps)(\partial_{x^\mu}-i P_\mu)) \phi_\eps^*(x) \notag \\ &=  \eps(\partial_\mu X^\nu) (\iota_{\partial_{x^\nu}} - Q_\nu) \phi_\eps^*(x) \\
&= 0\,. \notag
\end{align}
Thus $(\de_\eps - Q_\eps) \phi^*_\eps(x)$ obeys a first-order linear ODE in $x$ and vanishes at a point,
implying that it must vanish everywhere, as desired.

We
can also define equivariant descent on configuration space,
following what we did in Section~\ref{sec:descent-configuration-space}: 
given two $Q_\eps$-closed operators $\phi_{1,\eps}$ and $\phi_{2,\eps}$
we construct a form $(\phi_{1,\eps} \boxtimes \phi_{2,\eps})^*$ on
$\cC_2(B)$,\footnote{Here it is important that we assume that the product $\phi_{1,\eps} (x_1) \phi_{2,\eps}(x_2)$ is well defined
when $x_1 \neq x_2$, even in $\Omega$-background.}
 obeying
\begin{equation} \label{eq:Qeps-deps-total-confspace}
	Q_\eps (\phi_{1,\eps} \boxtimes \phi_{2,\eps})^{*} = \de_\eps (\phi_{1,\eps} \boxtimes \phi_{2,\eps})^{*}\,.
\end{equation}
The equation \eqref{eq:Qeps-deps-total-confspace} plays the same role in the
equivariant story as \eqref{eq:Q-d-total} in the ordinary one:
using the pairing between $\de_\eps$-cohomology and $\partial_\eps$-homology,
it allows us to construct binary operations on
$\cA_\eps$ from classes $\cclass{\Gamma} \in H^\eps_\bullet(\cC_2(B))$,
by
\begin{equation}
	\cclass{\phi_{1,\eps}} \star_\Gamma \cclass {\phi_{2,\eps}} = \int_\Gamma (\phi_{1,\eps} \boxtimes \phi_{2,\eps})^*\,.
\end{equation}

\subsection{Deriving the quantization}

We are now in a position to derive the deformation quantization
statement \eqref{eq:deformation-quantization}.
We consider the the $U(1)$-equivariant homology $H^\eps_\bullet(\cC_2(B))$.
Since we can $U(1)$-equivariantly retract $\cC_2(B)$ to $S^2$,
we may as well consider $H^\eps_\bullet(S^2)$: any class $\Gamma \in H^\eps_\bullet(S^2)$
gives rise to a binary operation on $\cA_\eps$.

In Section~\ref{sec:equivariant-S2} we considered three such classes:
point classes $\cclass{a}$, $\cclass{b}$ associated to the $U(1)$-fixed 
points on $S^2$, and the fundamental class $\cclass{S^2}$.
Now, $\cclass{a}$ and $\cclass{b}$ correspond to the two primary products
$\cclass{\phi_1} \pprod_\eps \cclass{\phi_2}$ and $\cclass{\phi_2} \pprod_\eps \cclass{\phi_1}$
on $\cA_\eps$. The fundamental relation \eqref{eq:equivariant-relation}
says that the difference of these two products is ${\eps}$ times the secondary product
associated to $\cclass{S^2}$. But by construction, as $\eps \to 0$, this product limits to the
secondary product associated to $\cclass{S^2}$ in the non-equivariant setup, \emph{i.e.}
to the Poisson bracket. This is the desired \eqref{eq:deformation-quantization}.

 \section{Secondary operations for extended operators}
 \label{sec:defects}
 
So far in this paper we have discussed and illustrated a secondary product between local operators.
This secondary product has a natural generalization to \eo s (a.k.a. topological defects) of arbitrary dimensions, which we briefly outline in this section before illustrating it concretely in the next. 

In addition to considering the usual product, in which \eo s aligned in parallel are brought together in the transverse dimensions, we can use descent to define a secondary operation, in which a descendent of one \eo\ is integrated over a small sphere linking the other. More abstractly, there are operations on $m$-tuples of $k$-dimensional \o s given by the topology of configuration spaces of $m$ points (or little discs) in $\R^{d-k}$.

Before describing these, it's useful to mention a simple way to think of $E_d$ algebras in the setting of homotopical algebra, which goes under the name of Dunn additivity (\emph{cf.}~\cite{HA}).
Namely, there's a precise sense in which a $d$-disc algebra structure consists simply of the data of $d$ compatible associative multiplications.
A toy example is the fact that if we're given two associative multiplications on a set which commute with each other then the two are forced to be equal and further to be commutative.
Geometrically, we think of topological local operators in $\R^n$ equipped with the associative (primary) product along the $d$ coordinate axes.
Compatibility, when carefully formulated, expresses the fact that these $d$ multiplications come from a locally $Q$-trivial family of multiplications around the $(d-1)$-sphere of possible directions, and thus encodes the secondary product as well.

In the mathematical language of extended topological field theory, $k$-dimensional \eo s in an $d$-dimensional TQFT have the structure of a $k$-category, in which objects are given by the \eo s themselves, morphisms are given by $(k-1)$-dimensional topological interfaces between \eo s, 2-morphisms by interfaces between interfaces and so on (\emph{cf.} ~\cite{Kapustin:2010} for a physical explanation of this mathematical structure). For example, line \o s form a category: the vector space of topological interfaces between two line \o s $\cL,\cM$ is interpreted as the space of morphisms $\text{Hom}(\cL,\cM)$ in the category. Moreover, topological interfaces enjoy an associative product: an interface from $\cL$ to $\cM$ and one from $\cM$ to $\cN$ can be brought together to define an interface from $\cL$ to $\cN$. This defines the associative composition of morphisms in the category.  

Note that we can recover lower-dimensional \o s from categories of higher-dim\-en\-sional ones.
As an important special case, we can think of local operators as self-interfaces of a trivial line \o, stretching along one axis in $\R^d$. Thus we think asymmetrically of the primary product along this line and the product in the $d-1$ other directions, which together make up the disc structure of local operators. The product of local operators in the directions transverse to the line can be interpreted as a primary product of line \o s, in which we bring two parallel trivial lines together in a transverse direction. 

More abstractly, there are operations on line \o s given by the topology of the space of configurations of points, or little $(d-1)$-discs, inside a large $(d-1)$-disc: line \o s form an {\it $(d-1)$-disc category}. This means in particular that we have a full $S^{d-2}$ of directions in which to take the product of two line operators, with the result depending in a locally constant way on the direction. For example, for line operators $\cL,\cM$ in 3d we have a locally constant family of tensor products $\{\cL\ot\cM\}_\theta$, $\theta\in S^1$. Equivalently, this may be described as a single tensor product at $\theta=0$ together with a monodromy operator $R_{\cL\ot\cM}\in \text{End}(\cL\ot \cM)$ (Figure \ref{fig:braid}). These monodromy operators are the R-matrices for the familiar braided tensor structure --- a.k.a. 2-disc category structure --- on line operators in 3d, such as the braiding of Wilson lines in Chern-Simons theory.

\insfigscaled{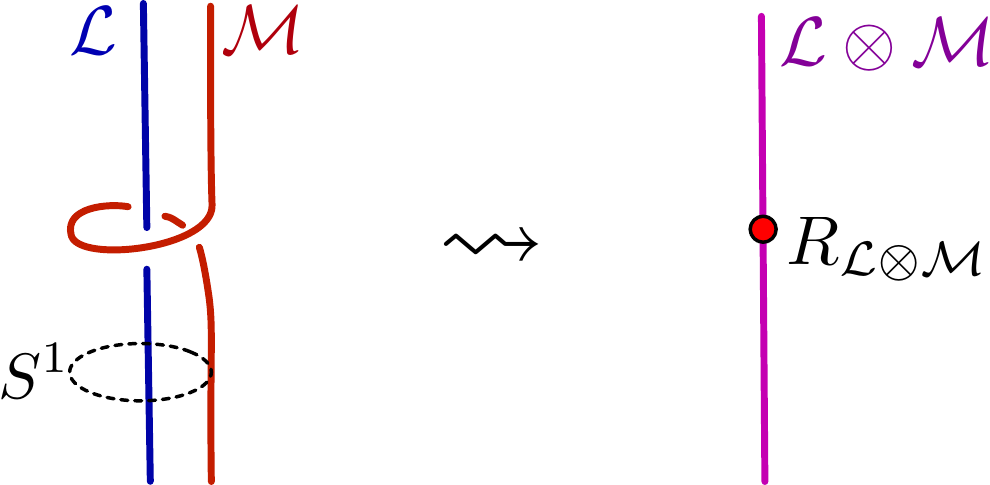}{0.7}{Monodromy in the $S^1$ family of tensor products $\{\cL\otimes\cM\}_\theta$ interpreted as a braiding interface.}

Likewise, $k$-dimensional \eo s form a $(d-k)$-disc $k$-category: we trade $k$ directions of the product of local operators into the structure of a $k$-category, while the transverse $d-k$ directions define multiplication operations among extended operators.

We would like to spell out a concrete aspect of this abstract general structure, in the form of new bracket operations between topological interfaces. In the following sections we provide physical manifestations of these structures in 3d and 4d theories.

\subsection{Secondary OPE of local and line operators}
\label{sec:local-line}

We can fix a $k$-dimensional \eo\ in space and consider configurations of local operators placed at points in the transverse $d-k$ directions. Integrating the descendant of a local operator on a small $S^{d-k-1}$ linking the \eo\ results in a secondary product. 
We illustrate this operation and some of its topological properties in the simplest case of line \o s,  $k=1$.

Let us assume that the spacetime dimension satisfies $d\geq 3$. Given a local operator $\cO$ and a line \o\ $\cL$, we can construct two self-interfaces of $\cL$: a primary interface $\CO * \cL$ given by simply colliding $\cO$ with $\cL$, and a secondary interface
\be \CO_\cL := 
 \int_{S^{d-2}} \CO^{(d-2)} \; \cL \label{def-OL} \ee
defined by integrating the $(d-2)$-form descendent of $\cO$ on a small $S^{d-2}$ that links the line (Figure \ref{fig:local-line}).%
\footnote{As in Section~\ref{sec:top-2}, we work with cohomology classes rather than actual operators; but we will omit the $\cclass{...}$ to simplify the notation.} %

\insfigscaled{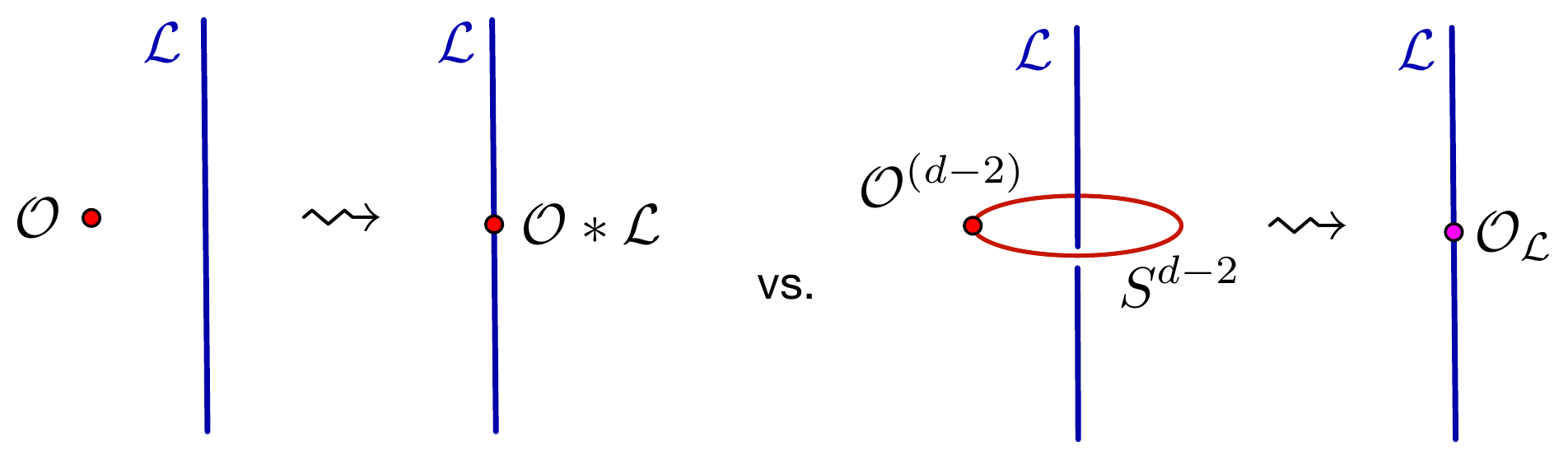}{0.7}{The primary and secondary products of a local operator with a line}

Both the primary and secondary self-interfaces are special, in that they commute with all other interfaces among line operators. This is well known for the primary product. Heuristically, given any two line \o s $\cL_1,\cL_2$, any interface $\cP\in \text{Hom}(\cL_1,\cL_2)$, and a local operator $\CO$, we can continuously move $\CO$ through the ``bulk'' to collide with either $\cL_1$ or $\cL_2$, and then with $\cP$. Topological invariance (specifically, the $Q$-exactness of this motion) ensures that the result will be the same:
\be \cP * (\CO * \cL_1) = (\CO * \cL_2) * \cP\,. \label{primary-int} \ee
A more topological argument for \eqref{primary-int} follows from the fact that the configuration space of two points in $\R^{d\geq 3}$, with one of the points constrained to a fixed line, is simply connected. This configuration space is homotopic to the linking sphere $S^{d-2}$.

Commutativity for the secondary interface follows from a similar argument. Consider the same setup, involving an interface $\cP$ between $\cL_1$ and $\cL_2$.
Heuristically, we can can continuously slide a small sphere $S^{d-2}$ linking $\cL_1$ ``above'' $\cP$ to a small sphere linking $\cL_2$ ``below'' $\cP$. Topological invariance of the descent procedure then ensures that
\be  \cP *\CO_{\cL_1} = \CO_{\cL_2} * \cP\,. \ee

\insfigscaled{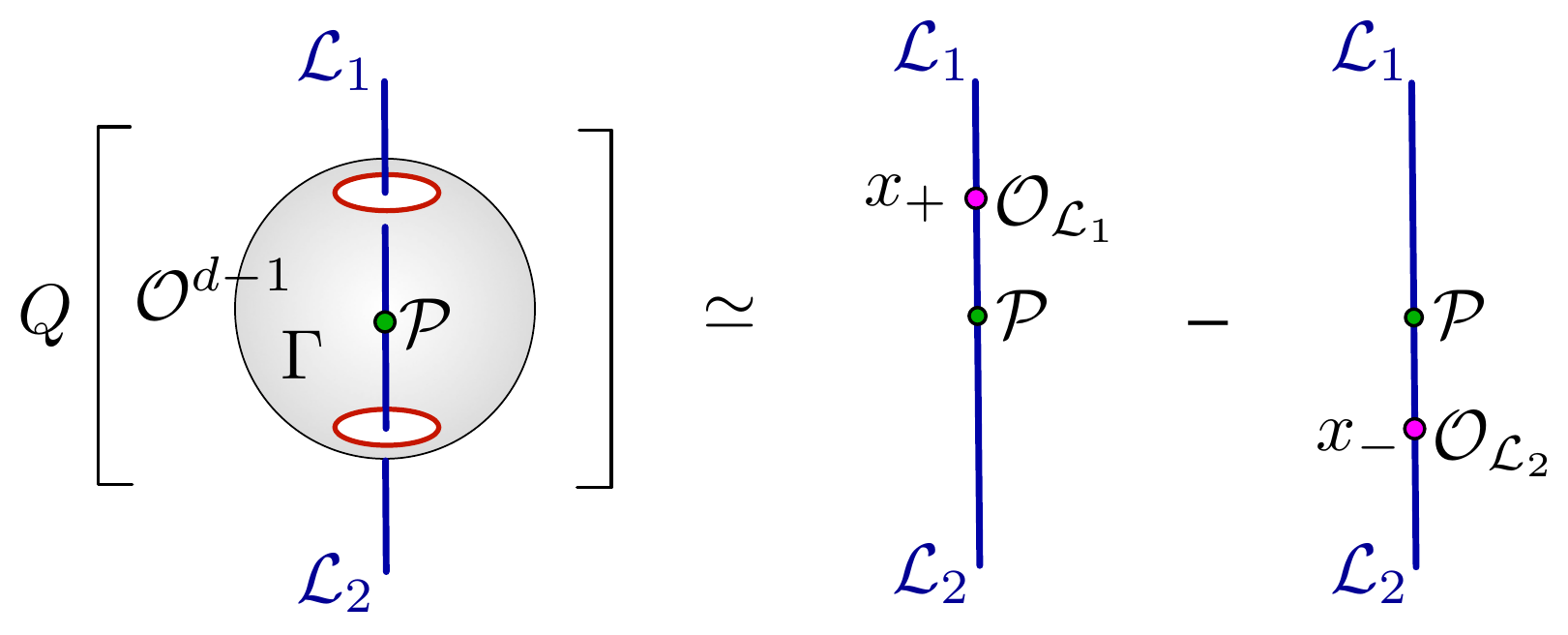}{0.7}{Topological argument for commutativity of secondary products with all other interfaces.}

A more explicit way to see this commutativity is the following. Consider a large $S^{d-1}$ sphere linking the interface $\cP$, as in Figure \ref{fig:slide}. This sphere intersects the support of the line \o s in two points $x_+,x_-$. Let us remove small discs surrounding these two points from $S^{d-1}$, obtaining $\Gamma = S^{d-1}\backslash(D^{d-2}_+\sqcup D^{d-2}_-)$. Consider the integral of the $(d-1)$-th descendant of $\CO$ along $\Gamma$, in the presence of $\cP$. Since $\Gamma$ has a boundary, we find
\begin{align} Q \bigg[\Big(\int_\Gamma \CO^{(d-1)}\Big) \cP\bigg] &= \Big(\int_{\pd \Gamma} \CO^{(d-2)}\Big)\cP \notag \\ &= \Big(\int_{S^{d-2}_+}\CO^{(d-2)}\Big)\cP-\Big(\int_{S^{d-2}_-}\CO^{(d-2)}\Big)\cP \notag \\ &= \cP*\CO_{\cL_1} - \CO_{\cL_2}*\cP\,,\end{align}
which demonstrates the $Q$-exactness of the commutator of secondary interfaces with $\cP$.

One might be tempted to directly compute a secondary product of $\CO$ and the interface~$\cP$ by integrating the $(d-1)$-th descendant of $\CO$ around a full $S^{d-1}$ that links $\cP$. We note, however, that this is not an allowed operation. The points $x_{\pm}$ where the $S^{d-1}$ intersects the lines can produce genuine, non-$Q$-exact, singularities in correlation functions. The only sensible (topologically-invariant) way to define a secondary product of $\CO$ and $\cP$ is via computing $\cP*\CO_{\cL_1}$, or equivalently $\CO_{\cL_2}*\cP$, as above.

When the spacetime dimension is less than three, some of the statements above must be modified, in fairly obvious ways. For $d=1$, line \o s are space-filling, so there are no local operators separated in transverse dimensions. For $d=2$, there are two distinct primary products $\CO * \CL$ and $\CL *\CO$, coming from placing $\CO$ to the ``left'' or ``right'' of a line. Moreover, the secondary product is just the difference of the two primary operations, $\CO_\CL = \CO * \CL - \CL *\CO$.

\subsection{Mathematical formulation}

We may connect the structures just described with a more abstract mathematical characterization, in the following way.

The endomorphisms of the unit object (trivial line \o) in the $(d-1)$-disc category of line \o s carries a natural $d$-disc structure and is identified as such with the 
disc algebra of local operators. Now in any monoidal category $\cC$, the endomorphisms $\cA=\text{End}(1_\cC)=\Omega_{1_\cC}\cC$ of the unit object give endomorphisms of any object $M\simeq 1_\cC\otimes M$, via the action on the first factor. In other words, we may upgrade $M$ to the status of $\cA$-module in $\cC$. In fact this comes from a homomorphism $\text{End}(1_\cC)\to \text{End}(Id_\cC)$ to the center of the category $\cC$ --- \emph{i.e.}, the induced homomorphisms of objects commute with all maps in $\cC$. This gives the usual (primary) product of local operators on interfaces between line \o s.

In a $(d-1)$-disc category, this structure gets enhanced: any object $M$ becomes an {\em $(d-1)$-disc module} in $\cC$ for $\cA$, meaning we have operations of $\cA$ on $M$ labelled by configurations of little $(d-1)$-discs in a large $(d-1)$-disc with $M$ placed at the origin. This structure is captured, on the level of homology, by the two binary operations, primary and secondary products, coming from the two homology groups of the space of configurations of points in $\R^{d-1}\setminus 0\sim S^{d-2}$. More abstractly, these actions by endomorphisms of any object come from a central action, making the identity functor $Id_\cC$ an $(d-1)$-disc module for $\cA$ in $\text{End}(\cC)$. 

Likewise for higher-dimensional \o s, we can recover the $d$-disc algebra $\cA$ of local operators from the $(d-k)$-disc $k$-category $\cC$ as its ``k-fold based loops" $\cA=\Omega^k_{1_\cC}\cC$ (endomorphisms of the unit endomorphism of the unit endomorphism ... of the unit). The identification of tensoring with the unit with the identity functor gives rise to an analogous higher structure, an {\em $E_{k\subset d}$-structure} on the pair $(\cA,\Omega^k_{Id_\cC}(\text{End} \cC))$.
The theory of $E_{k\subset n}$ algebras was introduced in~\cite{AyalaFrancisTanaka} --- for example an $E_d$ (or $d$-disc) module $M$ for an $E_d$ algebra $\cA$ is equivalent to the data of a $E_{0\subset d}$ algebra.
These structures perfectly capture the algebraic structure involving products of operators of different dimensions.

\subsection{Towards a secondary product of line operators}
\label{further OPEs}

So far we have discussed the secondary structures on local operators as well as those pairing local and line operators. These probe, but do not fully capture, the product structure of line operators. In particular in situations in which there are very few local operators at all, such as Rozansky-Witten theory on a compact target, we certainly need to delve further to find interesting higher structures for line operators. Here we briefly comment on the higher product that exists between two line operators. A more complete analysis of secondary operations amongst extended operators and their implementation in standard examples is beyond the scope of the present paper, though we intend to return to it in future work.

Given two line operators $\cL$ and $\cM$, one should be able to define a secondary product $\{\cL,\cM\}$ in a manner analogous to the construction for local operators. In particular, these line operators only need be topological at the level of $Q$-cohomology, so infinitesimal variations of the configuration of a line in spacetime should be $Q$-exact. By performing an analogue of the descent procedure for local operators, one can produce line operators that are differential forms on the configuration space of lines with the property that their $Q$-image is closed (these will be integrals of descendants of the displacement operator on the line over the line). In the path integral, such an object can be inserted when integrated over homology classes in the configuration space to define physical observables in the twisted theory. As with local operators, one can then use this construction to define a secondary composite operator by restricting to the configuration space of parallel lines and integrating the $(d-2)$-form descendent of $\cL$ over a linking $(d-2)$-sphere around~$\cM$.

At a more formal level, this structure can be described as follows. The $(d-1)$-disc structure on line operators produces a line operator $(\cL\ast\cM)_c$ for every pair of embedded $(d-1)$ discs in a large $(d-1)$ disc. Topological invariance then endows this family of operators with a flat connection. In other words, there is a local system $(\cL\ast \cM)_{S^{d-2}}$ valued in the category of line operators over the (homotopy type of the) configuration space $\cC_2(\R^{d-1})\sim S^{d-2}$. (The fiber of this local system at any point on $S^{d-2}$ is equivalent to the primary product $\cL\ast \cM$.) We now define the secondary product as the integral,\footnote{Formally the integral is defined as the homotopy colimit of the local system, considered as a diagram valued in the $\infty$-category of line operators.} or total cohomology, of this local system over the configuration space:
\begin{equation}
\{\cL,\cM\}:=\int_{S^{d-2}} (\cL\ast \cM)_{S^{d-2}} = R\Gamma(S^{d-2}, (\cL\ast \cM)_{S^{d-2}})~.
\end{equation}
Functoriality of the construction implies that there is a map from self-interfaces of $\cL$ to self-interfaces of the secondary product, which generalizes the secondary product between local and line operators described previously.

The secondary product of line operators accesses the topology of the $d-2$ sphere only via its homology (or chains). This idea is made precise in To\"en's notion of a {\em unipotent} disc algebra structure~\cite{Toenbranes} (we thank Pavel Safronov for teaching us about unipotent disc structures~\cite{pavel}). For $d>3$, the sphere $S^{d-2}$ is simply connected and thus we expect the disc structure to be unipotent, so that the primary and secondary products
\begin{equation}
\cL,\cM\mapsto \cL\ast \cM, \{\cL,\cM\}
\end{equation}
capture the full disc structure in a suitable sense. For $d=3$ we have a local system $(\cL\ast \cM)_{S^1}$ on the circle, which is equivalent to the data of the primary product $\cL\ast\cM$ and its braiding automorphism (R-matrix) giving the monodromy of the local system. In this case the secondary product $\{\cL,\cM\}$ can be identified with the (derived) invariants of the braiding automorphism, which is only sensitive to the (generalized) 1-eigenspace of the R-matrix. However in the case of Rozansky-Witten theory the braiding (and disc structure) is in fact unipotent -- this follows from the description of the 2-disc category of line operators (locally) as modules for the 3-disc algebra of local operators. Hence again in this case one can expect the secondary product $\{\cL,\cM\}$ to play a central role.

\section{Extended operators and Hamiltonian flow in RW theory}
\label{sec:extended RW example}

The secondary product of extended operators is particularly useful when there aren't enough local operators to adequately capture features of a theory, such as the full structure of its moduli space. For example, in Rozansky-Witten theory, local operators (corresponding to holomorphic functions) can only distinguish all points of the target $\CX$ if $\CX$ is affine. Otherwise there simply aren't enough holomorphic functions; looking at higher Dolbeault cohomology does not help the situation. In full generality one must utilize line operators --- given by holomorphic vector bundles and more general complexes of vector bundles or coherent sheaves on $\CX$ --- along with their 2-disc structure.  

Our goal in this section is to describe the primary and secondary products between a local operator, $\CO$, and a line operator, $\cL$, in Rozansky-Witten theory with complex symplectic target $\CX$. To this end, we first recall the geometric description of line operators as coherent sheaves on $\CX$, following \cite{Rozansky:1996bq, Kapustin:2008sc}. We then compute the primary and secondary products by identifying \emph{both} as primary products between local operators and a boundary condition in a two-dimensional B-model. (A computation of secondary products directly in the 3d theory appears in Section~\ref{sec:1d-fermi}.)

The main geometric result is that the secondary product between a holomorphic function $\CO = f \in \C[\CX]$ and a sheaf $\CL$ is the fermionic endomorphism $\CO_\CL\in \text{Ext}^1(\CL,\CL)$ corresponding to an infinitesimal Hamiltonian flow 
\be \CO_\CL \sim \Omega^{-1} \pd f\,. \ee
We will obtain this in several steps, learning along the way how dimensional reduction interacts with higher products.
The result is a concrete measurement of the nontriviality of the braided tensor structure on line operators in RW theory.

We assume throughout this section that we are working with $\Z$-graded Rozansky-Witten theories, as discussed in Section~\ref{sec:Zgrading}. This ensures that the category of line operators will be $\Z$-graded as well, and can be identified with the ordinary derived category of coherent sheaves in \eqref{line-Coh} below.

\subsection{The category of line operators}
\label{sec:line-cat}

Consider a 3d $\CN=4$ sigma-model on $\R^3$ spacetime, with a line operator supported along a straight line~$\ell$. We assume that the line operator preserves the Rozansky-Witten supercharge $Q$; physically, it should be a quarter-BPS operator that preserves at least a 1d $\CN=2$ subalgebra%
\footnote{More precisely, one might call this a 1d $\CN=(0,2)$ subalgebra, with the supercharge $Q$ corresponding to a B-type (Dolbeault-type) twist. If the 3d bulk theory were empty, the line would support matter in chiral and fermi multiplets.} %
of 3d $\CN=4$ SUSY.

A convenient way to identify the category of line operators is by reduction on a circle linking $\ell$.
In the complement of a small neighborhood the line $\ell$, the spacetime geometry looks like $S^1\times \R_+\times \R$, where the $S^1$ circle links $\ell$. Geometrically, this $S^1$ is fibered over $\R_+$ --- its radius increases the further one gets from the line $\ell$. However, up to Q-exact terms, we may deform the metric to an honest product. It was further argued in \cite{Kapustin:2008sc} that the topological theory on $S^1\times \R_+\times \R$ is equivalent to a \emph{purely two-dimensional} B-model on $\R_+\times \R$  with the same target $\CX$. Physically, one would have to be careful to include all the Kaluza-Klein modes of fields on $S^1$. However, at least for the bosons, all but the zero-modes are $Q$-exact, and may be neglected. (The story for fermions is slightly more interesting; it will be discussed below.)

In the course of this dimensional reduction 
any line operator supported on $\ell$ becomes identified with a boundary condition for the 2d B-model. The category of boundary conditions in a 2d B-model with target $\CX$ is the derived category of coherent sheaves,%
\be \cC = D^b\text{Coh}(\CX)\,. \label{line-Coh} \ee
(equivalently, since $\CX$ is smooth, objects of $\cC$ are described simply as complexes of vector bundles).

The simplest example of a coherent sheaf is a holomorphic vector bundle $V$ on $\CX$. In this case, there is an easy physical description of the corresponding line operator in RW theory. It may be realized as a quarter-BPS ``Wilson line'' in 3d $\CN=4$ theory, defined by pulling back the bundle $V$ to spacetime, and computing the holonomy of its complexified (anti-holomorphic) connection along a line $\ell$.

In order to construct line operators corresponding to more general sheaves (and complexes of sheaves) in 3d $\CN=4$ theory, one may introduce additional 1d $\CN=2$ supersymmetric matter along $\ell$. For example, a skyscraper sheaf supported at a point on the target $\CX$ comes from coupling the bulk theory to 1d fermi multiplets (Section~\ref{sec:1d-fermi}).

The category $\cC$ carries the natural commutative tensor product operation, making it a symmetric monoidal category. 
On the other hand, the interpretation as a category of line operators operators endows $\cC$ with a 2-disc ($E_2$ or braided) 
monoidal structure (as in Section~\ref{sec:defects}). It was explained in \cite{Rozansky:1996bq, RobertsWillerton, Kapustin:2008sc} that the OPE of lines may be identified with the tensor product of coherent sheaves, but carries a nontrivial braiding governed (to leading order) by the holomorphic symplectic form. The braided structure was rigorously but somewhat implicitly constructed on the cohomology level in~\cite{RobertsWillerton} using weight systems and associators. The braided structure can be described precisely on the chain level, locally on the target, using the disc structure of local operators, \emph{i.e.}, the holomorphic Poisson structure of functions on $X$ (up to even degree shifts, which we suppress). Namely, locally on $X$ the derived category\footnote{For these constructions it is essential to be working ``on the chain level", \emph{i.e.} with dg categories or $\infty$-categories, rather than with the standard derived category.} can be written as the derived category of modules for the ring of holomorphic functions, \emph{i.e.}, as modules for the endomorphism ring of the structure sheaf $\CO_X$, which is the unit for the tensor structure. Finally, a general construction (see \emph{e.g.} Section 6.3.5 in~\cite{HA}) produces a (d-1)-disc structure on the category of modules for any d-disc algebra.

Our goal is to identify precisely and explicitly a global aspect of the braided (2-disc) product structure on line operators in RW theory, namely the secondary product between local operators and line operators.

\subsection{Collision with boundaries in 2d}
\label{sec:bdy-2d}

Another useful piece of information is the geometric description of primary products of local operators and boundaries in the 2d B-model. We collect and motivate relevant results here; see \emph{e.g.} \cite{Aspinwall:2009isa} for a review.

We saw above that objects $\CL\in \cC$ (viewed as boundary conditions in the B-model to~$\CX$) are coherent sheaves on $\CX$ or complexes thereof. It suffices for us to understand the primary product between local operators and single coherent sheaves. The product extends in a straightforward way to complexes --- using the fact that the primary product commutes with all other morphisms, as in \eqref{primary-int}.

\insfigscaled{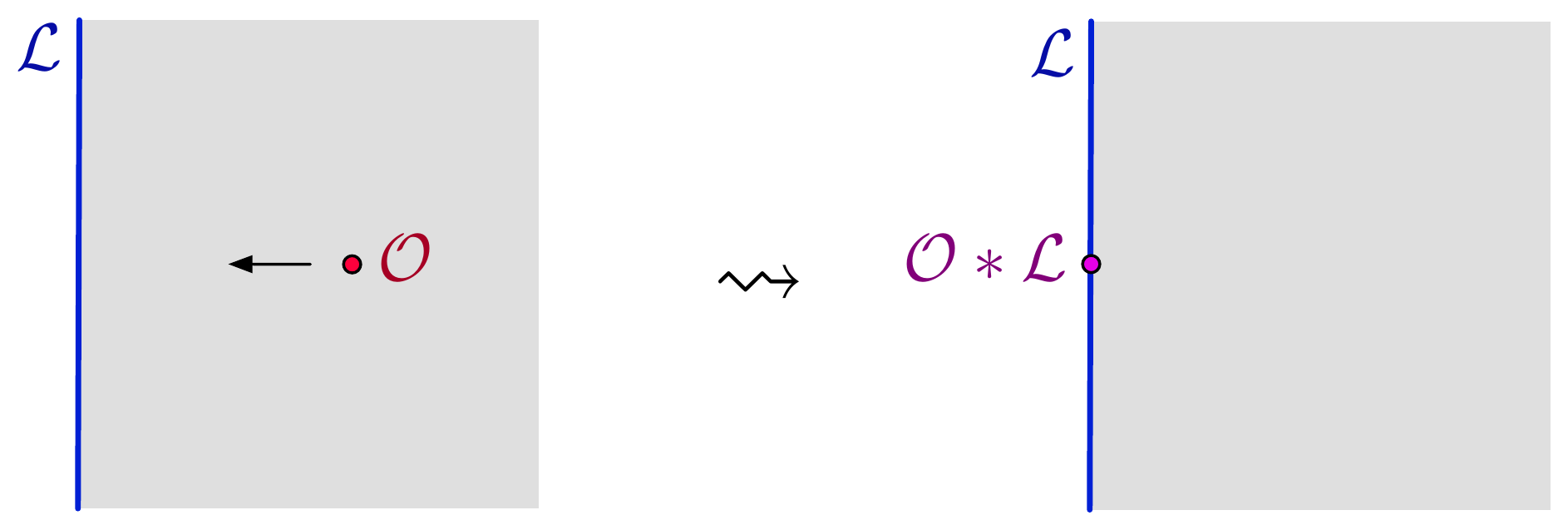}{0.6}{In $d=2$, there exists only a primary product between local operators and boundaries.}

Let us start with the simple case that $\CL=\CO_\CY$ is the structure sheaf (the trivial holomorphic line bundle) of a holomorphic submanifold $\CY\subset \CX$. Mathematically, its derived endomorphism algebra is
\be \text{End}(\CO_\CY) = \text{Ext}^\bullet(\CO_\CY,\CO_\CY) \simeq H_{\bar\pd}^\bullet\big(\Lambda^\bullet(N^{(1,0)}\CY)\otimes \Omega^{0,\bullet}(\CY)\big)\,,\ee
where $N^{(1,0)}\CY = T^{(1,0)}\CX/T^{(1,0)}\CY$ denotes the holomorphic normal bundle of $\CY$.
In particular, $\text{End}(\CO_\CY)$ contains functions on $\CY$ (elements of $H_{\bar\pd}^0\big(\Lambda^0(N^{(1,0)}\CY)\otimes \Omega^{0,\bullet}(\CY)\big)$), which act via multiplications on the sections of $\CO_\CY$. It also contains odd normal vector fields (elements of $H_{\bar\pd}^0\big(\Lambda^1(N^{(1,0)}\CY)\otimes \Omega^{0,\bullet}(\CY)\big)$), which represent infinitesimal deformations of $\CY$ itself. An important limiting case is the skyscraper sheaf $\CO_p$ supported at a point $p\in \CX$; its endomorphisms are a finite-dimensional exterior algebra, entirely generated by the odd tangent vectors at $p$
\be \text{End}(\CO_p) = \Lambda^\bullet(T_p^{(1,0)}\CX)\,. \ee

As reviewed Section~\ref{sec:b-model}, the topological operators in the B-model with target $\CX$ are polyvector fields,
\be \CA_{2d} \simeq H_{\bar\pd}^\bullet \big(\Lambda^\bullet (T^{(1,0)}\CX) \otimes \Omega^{0,\bullet}(\CX)\big)\,.\ee
The primary product between a local operator $\CO\in \CA_{2d}$ and a structure sheaf $\CL=\CO_\CY$ has a natural geometric description: it is the image of $\CO$ under a combination of pull-back from $\CX$ to $\CY$ and projection from the full tangent bundle to the normal bundle of $\CY$:
\be \Lambda^\bullet (T^{(1,0)}\CX) \otimes \Omega^{0,\bullet}(\CX) \;\overset{\iota^*}\to\; \Lambda^\bullet (T^{(1,0)}\CX\big|_\CY) \otimes \Omega^{0,\bullet}(\CY) \;\overset{q}{\to}\; \Lambda^\bullet(N^{(1,0)}\CY)\otimes \Omega^{0,\bullet}(\CY)\,, \ee
\be \CO * \CL = q\circ \iota^*(\CO) \,\in\, \text{End}(\CL)\,. \label{Otobdy}\ee
For example, if $\CL=\CO_p$ is a skyscraper sheaf at $p$ and $\CO\in \Lambda^r(T^{(1,0)}\CX) \otimes \Omega^{0,s}(\CX)$ represents a $Q$-cohomology class of local operators, the primary product is obtained by evaluating the $0$-form part of $\CO$ at the point $p$
\be \CO * \CL = \delta_{s,0}\, \CO\big|_p \in  \Lambda^r(T^{(1,0)}_p\CX)\,.\ee

We can also offer a more explicit physical description of the primary product. If we work locally on the target $\CX$, a local operator $\CO$ is represented as some polynomial in the  B-model scalars $\phi^i,\bar\phi_{i}$ and fermions $\eta_i,\xi_i$, discussed in Section~\ref{sec:freechiral}. At a boundary labelled by $\CL$, these fields all satisfy some relations --- the physical boundary conditions. The primary product of $\CO$ and $\CL$ is simply the result of imposing the boundary conditions on the fields that make up $\CO$. (Technically, this is only a semi-classical description of the product. It is a feature of the B-model that there are no further quantum corrections.) For example, a boundary condition $\CL=\CO_p$ corresponding to the skyscraper sheaf at $p$ sets $\phi^i,\bar\phi_i\to \phi^i(p),\bar\phi_i(p)$ (the coordinates of $p$), sets $\eta_i=0$, and leaves $\xi_i$ unconstrained. The primary product of $\CO$ and the skyscraper boundary leaves behind a polynomial in the $\xi_i$, \emph{i.e.} an element of $\Lambda^\bullet (T^{(1,0)}_p\CX)$.

\subsection{Reduction of operators to 2d}
\label{sec:3d2d-ops}

A final result we will require is the relation between local operators \emph{and their descendants} in 3d Rozansky-Witten theory to $\CX$, and local operators in the 2d B-model to $\CX$ obtained by placing Rozansky-Witten theory on $\R^2\times S^1$.

Certainly local operators in the 3d theory become local operators in the 2d theory. Local operators in 3d are elements of
\be \CA = H_{\bar\pd}^\bullet(\Omega^{0,\bullet}\CX)\,. \label{A3d-red} \ee
In the notation of Section~\ref{sec:3d-sigma}, the local operators are represented locally on the target as polynomials in the complex coordinates $X^A,\ol X_A$ and the fermions $\eta_A$. These are directly identified with local operators in the B-model, with a simple change of notation
\be \phi^A = X^A \ee
to match the conventions in Section~\ref{sec:freechiral}. (The symplectic `$A$' index is reinterpreted as the unitary `$i$' index of the B-model.) 

The 3d algebra of local operators \eqref{A3d-red} does not account for all the expected B-model local operators: it is missing holomorphic polyvector fields. However, we can recover (ordinary, Hamiltonian) vector fields from integrating the 1-form descendants of 3d local operators around the compactification circle!

Let's derive this explicitly, working locally on the target and assuming a flat metric. The first descendant of the 3d local operator $X^A$ is the 1-form fermion $(X^A)^{(1)} = \chi^A$. If we identify
\be \xi_A = -2\Omega_{AB} \oint_{S^1} \chi^B \label{xi-red} \ee
as the zero-mode of the component of $\chi^A_\mu$ parallel to the compactification circle, and also identify a two-dimensional 1-form fermion $\chi^{A}_{(2d)}$ with the remaining components of $\chi^A_\mu$ along $\R^2$, we find that the 3d action and SUSY transformations \eqref{3d-action-chi}-\eqref{SUSY3d} reduce precisely to the 2d action and SUSY transformations \eqref{b-xiaction}-\eqref{b-SUSY}. (Physically, the reduction requires taking a zero-radius limit, \emph{i.e.} keeping only the zero-modes of all the fields.)

The full algebra of local operators in the local B-model is thus generated by 3d local operators and their secondary reductions  \eqref{xi-red}. Crucially, the relation between the usual holomorphic vector fields of the B-model and the 3d one-form fermions $\chi^A$ involves the holomorphic symplectic form $\Omega$.

Inverting \eqref{xi-red}, we discover that the compactified descendant $\oint_{S^1}(X^A)^{(1)}$ is the Hamiltonian vector field generated by the function $X^A$. More generally, taking $\CO=f(X)$ to be any holomorphic function on the target $\CX$, we find $\CO^{(1)}=\pd_A f\, \chi^A$, and identify the compactified descendant
\be \oint_{S^1_\ell} \CO^{(1)} = -\tfrac12 \Omega^{AB} \pd_A f\, \xi_B = \tfrac12 \Omega^{-1}(\pd f)\,,  \label{Ham-end} \ee
with the Hamiltonian vector field generated by $f$.
A similar description holds for 3d operators $\CO=\omega \in \Omega^{0,\bullet}(\CX)$ represented by higher forms; a straightforward local calculation produces $\oint_{S^1_\ell} \omega^{(1)} =  \tfrac12\Omega^{-1}(\pd \omega)$.

We can also describe the 2d local operators obtained by this ``secondary reduction" procedure by testing them against operators obtained by ordinary reduction, \emph{i.e.}, holomorphic functions. Thus let $\cO$ and $\cO'$ be 3d local operators, and let $\overline{\CO} = \oint_{S^1}\CO^{(1)}$ denote the B-model operator obtained as a compactified descendant of $\CO$. 
We can consider $\CO'$ as a B-model local-operator as well, obtained by straightforward reduction.
We wish to calculate the 2d secondary product (Gerstenhaber/SN bracket) of $\overline{\cO}$ with $\CO'$.
This is achieved by integrating the 1-form descendent of $\CO'$ along a circle linking the insertion point $x$ of $\overline{\cO}$. However, rewriting from the 3d point of view, we are integrating the descendants of both operators along two simply linked circles! As we saw in Section~\ref{sec:symmetry} (Figure \ref{fig:Hopf}), the result is the ordinary 3d bracket of $\cO$ and $\cO'$, \emph{i.e.} the holomorphic Poisson bracket.
The 2d and 3d computations agree precisely if $\overline{\cO}$ is the Hamiltonian vector field generated by~$\CO$. 
This property can be used to uniquely characterize $\overline{\CO}$.

\subsection{Primary and secondary products}
\label{sec:RW12}

Now consider a line operator $\cL\in D^b\text{Coh}(\CX) $ supported along a line $\ell$, and a local operator $\CO \in \CA = H_{\bar\pd}^\bullet(\Omega^{0,\bullet}\CX)$.

The primary product $\CO *\CL \in \text{End}(\CL)$ is easy to interpret in the language of sheaves by reducing to a 2d B-model along a circle $S^1_\ell$ that links $\ell$. As above, $\CL$ becomes a boundary condition and $\CO$ remains a local operator, interpreted as an element of
\be \CO \in H_{\bar\pd}^\bullet(\Omega^{0,\bullet}\CX) \;\subset\; \CA_{2d} = H_{\bar\pd}^\bullet \big(\Lambda^\bullet (T^{(1,0)}\CX) \otimes \Omega^{0,\bullet}(\CX)\big)\,.\ee
The primary product $\CO *\CL$ in 3d is equivalent to a primary product (collision) of $\CO$ and the boundary condition $\CL$ in 2d. We saw in Section~\ref{sec:bdy-2d} precisely what this means. In particular, if $\CO = f \in \C[\CX]$ is a holomorphic function on $\CX$, then $\CO *\CL$ is the central endomorphism of the sheaf $\CL$ that multiples its sections by $f$.

What about the secondary product $\CO_\cL$? It is defined by integrating the first descendant of $\CO$ along a circle $S^1_\ell$ linking the line $\ell$,
\be \CO_{\cL}  = \int_{S^1_\ell} \CO^{(1)} * \cL \quad \in\; \text{End}(\cL)\,. \ee
If we reduce this configuration along $S^1_\ell$ to a 2d B-model, we find as usual that $\CL$ becomes a boundary condition. In addition, $\int_{S^1_\ell} \CO^{(1)}$ becomes an ordinary local operator.
Following Section~\ref{sec:3d2d-ops}, it must be an element of
\be \int_{S^1_\ell} \CO^{(1)} \,\in\, H_{\bar\pd}^\bullet\big( \Lambda^1(T^{1,0}\CX)\otimes \Omega^{0,\bullet}\CX\big) \subset \CA_{2d}\,, \ee
\emph{i.e.} a holomorphic vector field on $\CX$. Then the secondary product from 3d is reinterpreted as an ordinary primary product in the B-model,
\be \CO_{\CL} = \Big( \int_{S^1_\ell} \CO^{(1)}\Big) *_{2d} \CL\,, \ee
and corresponds to an infinitesimal deformation of $\cL$ (Figure \ref{fig:3d2d}).

\insfigscaled{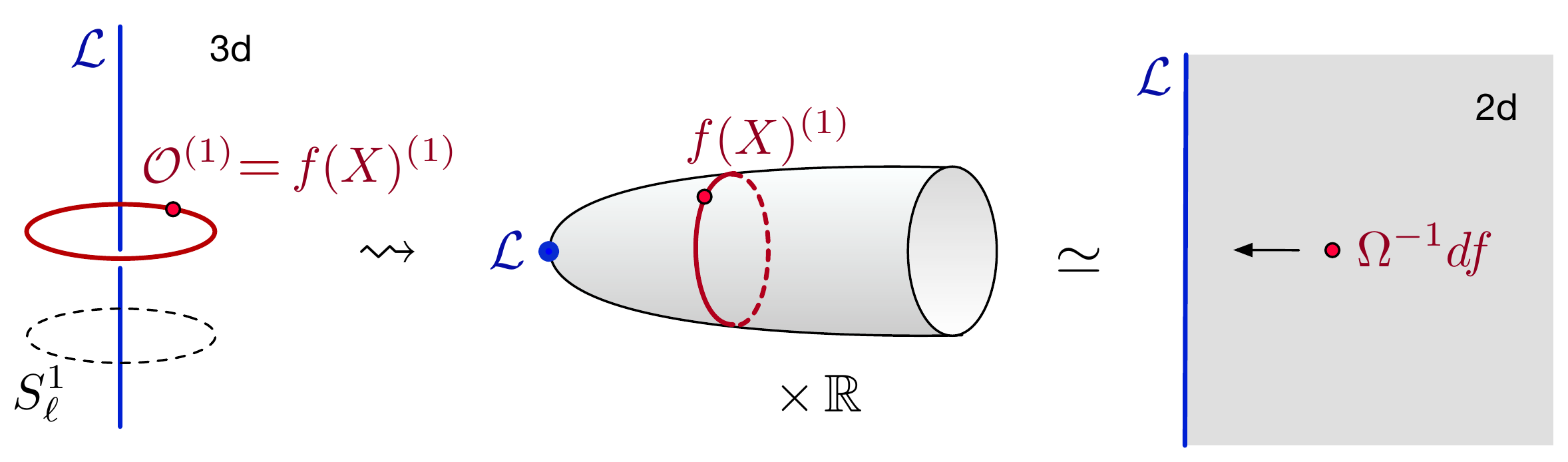}{0.6}{Deriving the secondary product by dimensional reduction.}

More concretely, we saw in Section \ref{sec:3d2d-ops} that $\oint_{S^1}\CO^{(1)}$ is identified as the Hamiltonian vector field generated by $\CO$. Then, if $\CO=f(X) \in \C[\CX]$ is a holomorphic function,  $\CO_\CL \in \text{Ext}^1(\CL,\CL)$ is the endomorphism of $\CL$ that corresponds to an infinitesimal Hamiltonian flow. Similarly, if $\CO = \omega \in \Omega^{0,q}(\CX)$ is a higher form, then $\CO_\CL\in \text{Ext}^{1+q}(\CL,\CL)$ is the corresponding derived Hamiltonian flow.

\subsection{Skyscraper sheaf from a 3d perspective}
\label{sec:1d-fermi}

It is also possible to compute secondary products of local and line operators in Rozansky-Witten theory \emph{without} reducing to a 2d B-model. We provide an example of this in the case that $\CL=\CO_p$ is a skyscraper sheaf. 

Working locally on the target, we may assume that $\CX=\C^{2N}$. We can construct a skyscraper sheaf at the origin $p=0$ by coupling the bulk 3d $\CN=4$ theory to a collection of 1d $\CN=2$ fermi multiplets $\rho_A$, $A=1,...,2N$. (These are matter multiplets for ``$\CN=(0,2)$'' SUSY in one dimension. In this case they can be paired up into $N$ fermi multiplets for $\CN=(0,4)$ SUSY, reflecting the fact that the skyscraper sheaf is hyperholomorphic, and actually preserves four rather than two supercharges of the 3d $\CN=4$ algebra.) The bulk hypermultiplets are coupled to the 1d fermis via J-term superpotentials $J^A = X^A\big|_\ell$. The additional contribution to the action is 
\be \begin{aligned} \label{S-skyscraper}
S_\CL 
& = \int_\ell \Big( \bar\rho^A\partial_\tau \rho_A  + J^A\bar J_A + \chi^A|_\ell \frac{\partial J^B}{\partial X^A|_\ell}\rho_B + \bar\rho^A \frac{\partial \bar J_B}{\partial \bar X_B|_\ell}  \eta_B|_\ell \Big) \\
& = \int_{\R^3} \Big( \bar\rho^A d \rho_A  + X^A\ol X_A + \chi^A \rho_A + \bar\rho^A   \eta_A  \Big) \delta^{(2)}_\ell\,,
\end{aligned} \ee
where $\tau$ is a coordinate along the line $\ell$, and $\delta^{(2)}_\ell$ is a delta-function 2-form supported on $\ell$.

The endomorphism algebra $\text{End}(\CL) = \text{Ext}^\bullet(\CL,\CL)$ is the $Q$-cohomology of local operators bound to the line. In this case, these are polynomials in the $2N$ fermions $\rho_A$. This corresponds to the expected result for the skyscraper sheaf
\be \text{End}(\CL) = \Lambda^\bullet (T^{(1,0)}_0 \CX) = \C[\rho_1,...,\rho_{2N}]\,. \ee
Note that the restriction of bulk local operators consisting of polynomials in $X^A|_\ell$ are not in the cohomology of $Q$ because the $J$-term contribution to the supercharge makes them exact.

Let us now compute the secondary product of the bulk local operator $\CO=X^A$ and the line operator $\CL$. The first descendent of $X^A$ is the 1-form $\chi^A$ so we should consider the correlation function of the line operator with the insertion
\be \oint_{S^1_\ell} \chi^A =  \int_{D_\ell} d \chi^A  \ee
where $S^1_\ell$ links the line $\ell$, and $D_\ell$ is disc with boundary $S^1_\ell$, which is pierced by $\ell$. In the absence of the line operator, the bulk action \eqref{3d-action-chi} relates $d \chi^A$ to an equation of motion  $d\chi^A = \Omega^{AB} \frac{\delta S}{\delta \chi^B}$, which would causes correlation functions involving $\int_{D_\ell} d \chi^A$ to vanish (because they are total derivatives). However, in the presence of the line operator we now have
\be \frac{\delta S}{\delta \chi^A} = \frac{\delta}{\delta \chi^A}(S_{\rm bulk}+S_\CL) = \Omega_{AB} d\chi^B +\rho_A\delta^{(2)}_\ell\,,\ee
whence $d\chi^A = \Omega^{AB} \frac{\delta S}{\delta \chi^B} - \Omega^{AB}\rho_B \delta^{(2)}_\ell$. Therefore,
\be \CO_\CL= \oint_{S^1_\ell}(X^A)^{(1)} = \int_{D_\ell} d\chi^A = -\Omega^{AB}\int_{D_\ell}\rho_B\delta_\ell^{(2)} = -\Omega^{AB}\rho_B\,. \label{rho-bracket} \ee
In other words, the insertion of $\int_{D_\ell} d\chi^A$ in any correlation function is equivalent to an insertion of the local operator $-\Omega^{AB}\rho_B$ on the line.

This agrees with the general prediction \eqref{Ham-end}: up to a numerical factor (which can be absorbed in the normalization of $\rho$), $\Omega^{AB}\rho_B$ is precisely the element of $\text{End}(\CL)$ coming from evaluation of the Hamiltonian vector field $\Omega^{-1}(\partial X^A)$ at the support of the skyscraper sheaf.
The secondary product with arbitrary functions $\CO\in \C[\CX]$ can be obtained from \eqref{rho-bracket} using a derivation property --- and reproduces more general Hamiltonian flows.

\subsection{Non-degeneracy}

An attractive feature of the secondary product with lines is that it may be nontrivial even when the secondary product of local operators vanishes.

A familiar example arises in compactifications of 4d $\CN=2$ theories. The Coulomb branch $\CX$ of a 4d $\CN=2$ theory on $\R^3\times S^1$ is a complex integrable system
\be \pi:\CX\to \cB \ee
with compact fibers over an affine base $\cB$. For example, in a 4d theory of class S, $\CX$ is the Hitchin integrable system \cite{Klemm:1996bj, Witten:1997sc, Gaiotto:2009hg}. 
A supply of topological local operators $\cO\in \cA$ is given by pullbacks of holomorphic functions on $\cB$. Formally, there is a map of algebras $\C[\cB]\to \CA$. However, the secondary product among all functions on the base necessarily vanishes
\be \{\CO,\CO'\} = 0\,,\qquad \CO,\CO' \in \C[\cB]\,.\ee
(This is precisely because $\cB$ is the base of an integrable system: functions on $\cB$ are Poisson-commuting Hamiltonians.)
Fortunately, there is also a large supply of topological line operators $\CL$ given by coherent sheaves on $\CX$. For any $\CL$, the secondary product gives an odd map
\be \C[\cB] \to \text{End}(\CL)\,. \label{BtoL}\ee
The functions on $\cB$ generate Hamiltonian flows along the fibers. Therefore, the image of the map \eqref{BtoL} will be nontrivial as long as (say) the support of $\CL$ is localized in the fiber directions of the integrable system.

\section{Descent structures in \texorpdfstring{$\cN=4$}{N=4} SYM}
\label{sec:4d}

In this section we describe a manifestation of the secondary product in the context of four-dimensional gauge theory, and in particular we refine and reinterpret a construction of Witten~\cite{Witten:2009mh} in the context of $\cN=4$ SYM, which in turn interprets a result of Ginzburg~\cite{Ginzburg1995} in the geometric Langlands program. Below we review some general features of local and line operators in the Geometric Langlands twist of the $\cN=4$ theory, after which we address some of the algebraic structures that arise amongst them. But first let us make some preliminary comments about our expectations for four-dimensional topological field theories.

\subsection{General considerations in four-dimensional TQFT}
\label{subsec:general_4d}

On completely general grounds we expect local operators in a four-dimensional TQFT to carry an $4$-disc structure. In particular, the primary product endows local operators with a commutative ring structure, and the secondary product should define a Poisson bracket of cohomological degree $-3$ that acts as a derivation of the ring structure. An example of what we might expect to see is the ring of holomorphic functions on a thrice-shifted cotangent bundle $T^\ast[3]\C$. This is in analogy to the two-dimensional B-model of a free chiral multiplet where we saw functions $T^\ast[1]\C$, and the Rozansky-Witten theory of a free hypermultiplet where we saw $T^\ast[2]\C$. A free four-dimensional $\cN=2$ hypermultiplet does contain bosons and fermions that could look like functions on $T^\ast[3]\C$; however, there is no twist for which these become topological local operators.

One immediately observes that the odd degree of the bracket implies that it should send pairs of bosonic operators to fermionic operators, just as was the case in the B-model. Unfortunately, in all standard twists of four-dimensional gauge theories --- the Donaldson twist \cite{Witten:1988ze} of $\CN=2$ gauge theories with linear matter, and the Vafa-Witten \cite{Vafa:1994tf} and Langlands \cite{Marcus:1995mq, Kapustin:2006pk} twists of $\CN=4$ Super Yang-Mills --- there are no fermionic operators at all in the topological algebra. Therefore, the Poisson bracket one local operators will vanish for trivial degree reasons in all of these cases.

Despite this, our suggestion is that the structure of higher products in topologically twisted theories is in fact typically nondegenerate --- one must, however, look sufficiently deep into the theory to observe the nondegeneracy. Namely, we must expand our view to include higher-dimensional extended operators. For the GL twists of $\cN=4$ SYM, the construction of~\cite{Witten:2009mh} is a manifestation of a secondary product involving line operators and local operators. In the remainder of this section we will recall this result and place it into a more general context.

\subsection{Local operators}
\label{subsec:GL_local_operators}

We recall the GL twist of $\cN=4$ super-Yang-Mills theory with gauge group $G$, at the value $\Psi=0$ of the canonical parameter. This four-dimensional TQFT was called the $\wh{A}$-model in~\cite{Witten:2009mh}; we shall denote it by $\Ahat$. Its admits an $S$-dual description as the $\Psi=\infty$ twist of $\cN=4$ SYM with Langlands-dual gauge group $G^\vee$, called the $\wh B$-model; we shall denote this dual description as $\Bhatv$.

In $\Ahat$, the topological local operators are gauge-invariant polynomials of a complex, adjoint-valued scalar field $\sigma$, which has degree ($R$-charge) $R[\sigma]=+2$. Algebraically, we have for the topological operator algebra
\begin{equation}
\label{4d-locA}
\CA \simeq (\text{Sym}\, \mathfrak g^*[-2])^{G_\C}\simeq \C[\mathfrak g[2]]^{G_\C} \simeq \C[\mathfrak h[2]]^W~,
\end{equation}
where $\mathfrak h$ is the complexification of the Cartan subalgebra of $G$, $W$ is its Weyl group, and the shift by two keeps track of the degree (with the standard but counterintuitive convention that $V[n]$ denotes the vector space $V$ placed in cohomological degree $-n$). In fact, local operators comprise such gauge-invariant polynomials in any of the GL twists (as well as in the Vafa-Witten twist and the Donaldson twist of $\cN=2$ gauge theory). One may think of $\CA$ as polynomial functions on the Coulomb branch $\mathfrak g/\text{Ad}(G_\C)\simeq \mathfrak h/W$ of the four-dimensional theory.

It follows from the above comment that the topological algebra of local operators in $\Bhatv$ should just be 
\begin{equation}
\label{4d-locB}
\CA^\vee \simeq 
\C[\mathfrak g^\vee[2]]^{G^\vee_\C}~.
\end{equation}
As $\Ahat$ and $\Bhatv$ are $S$-dual descriptions of the same theory, their algebras of local operators must be isomorphic, $\CA\simeq \CA^\vee$. As has been discussed in \cite[Sec 2.10]{Witten:2009mh}, this isomorphism depends on the invariant quadratic form that is used to define the kinetic terms of the underlying physical theories. For a simple gauge group $G$, it is the Cartan-Killing form, normalized by the physical gauge coupling (which can be chosen independently of $\Psi$). 

Naively, this seems to imply that the equivalence between $\Ahat$ and $\Bhatv$ is non-canonical in the topological theory, but this turns out not to be the case. In the algebraic description of the B-model $\Bhatv$ described in~\cite{Ben-Zvi:2016mrh,Elliott:2015rja,Elliott:2017ynt}, the local operators naturally appear as invariant polynomials of a complex {\em coadjoint} scalar,
\begin{equation}
(\text{Sym}\, \mathfrak g^{\vee}[-2])^{G^\vee_\C}\simeq \C[(\mathfrak g^{\vee})^\ast[2]]^{G^\vee_\C} \simeq \C[\mathfrak h[2]]^W~,
\end{equation}
giving a match with the topological algebra in $\Ahat$ that is independent of the choice of invariant form.

In any case, it is clear that the secondary product on $\CA$ must vanish: all elements of $\CA$ have even degree, while the secondary product is odd by construction. To see some hint of the secondary structure, we will need to look beyond local operators and include line operators in our discussion.

\subsection{Line operators}
\label{sec:GL_Line_operators}

Both theories $\Ahat$ and $\Bhatv$ allow for topological line operators. There are 't Hooft lines in $\Ahat$ and Wilson lines in $\Bhatv$, both labelled by representations of $G^\vee$, that are $S$-dual to each other \cite{Kapustin:2006pk}. Both theories include the topological algebra $\CA$ of local operators, which must appear as endomorphisms of the trivial line operator. This means that the categories of topological line operators are richer objects than just the category of representations of $G^\vee$. Below we will give a complete (and mathematically involved) description of these categories that includes this richer structure. We note that the main calculation we wish to describe is that of the secondary product of local operators with Wilson (or 't Hooft) lines, and for this purpose the full description of the category is not required.

The general derivation of~\cite{Kapustin:2006pk} gives us the category of line operators in $\Ahat$ as the equivariant derived category of $\cD$-modules (or perverse sheaves) on the affine Grassmannian
\begin{equation}
\cC = D^b_{L_+G_\C}(\cD{\rm -}mod(\text{Gr}_{G}))\simeq D^b_{L_+G_\C}(Perv(\text{Gr}_{G}))~,
\end{equation}
also known as the spherical (or derived) Satake category -- the geometric form of the spherical (or unramified) Hecke algebra. It is a mathematical avatar of the category of A-model boundary conditions on the moduli space of $G$-Higgs bundles on the two-sphere (the link of a line operator in four dimensions). In this description, local operators appear naturally in the form of the $L_+G$-equivariant cohomology ring of a point, which is equivalent to the previous description of $\CA$. 

On the $S$-dual side, the category of line operators in $\Bhatv$ has been described in \cite{Ben-Zvi:2016mrh,Elliott:2017ynt}, interpreting work of~\cite{BezFink,ArinkinGaitsgory}. This category is identified with the category of B-model boundary conditions (\emph{i.e.}, the derived category of coherent sheaves) on the moduli space of $G^\vee$ flat connections on $S^2$. Naively this moduli space is a point (the trivial flat connection), but it is corrected first to a stacky point $pt/G^{\vee}$ by taking into account the automorphisms of the trivial bundle, and then to a super- or derived stacky point $\fg^\vee[-1]/G^\vee_\C$ by taking into account ghosts measuring the nontransversality of the defining equations (\emph{e.g.}, as Hamiltonian reduction at a nonregular value of the moment map). The resulting category is given by,
\begin{equation}
\label{cat-W}
\cC^\vee = D^b\text{Coh}_{G_\C^\vee}(\mathfrak g^{\vee}[-1])\simeq D^b\text{Coh}_{G_\C^\vee}((\mathfrak g^\vee)^\ast[2])~,
\end{equation}
where in the second equality Koszul duality gives an equivalence with $G^\vee_\C$-equivariant coherent sheaves on the graded coadjoint representation $(\mathfrak g^{\vee})^\ast[2]$. (See \emph{e.g.} Section 11 of~\cite{ArinkinGaitsgory}.)

In these terms, a Wilson line in representation $R$ is associated to the trivial $R$-bundle on $\mathfrak g^{\vee,\ast }$, treated equivariantly with respect to the simultaneous $G_\C^\vee$ action on $R$ and on the coadjoint representation. In other words, a Wilson line makes sense at any point of the Coulomb branch.
Note that $\cC^\vee$ admits the structure of a symmetric monoidal category with respect to the tensor product of sheaves. However, this is {\em not} the natural structure that arises when considering line operators. Instead, the latter gives a nontrivial $3$-disc deformation of this category, and it is this structure that we aim to measure. The $3$-disc structure on the spherical category was first explained to the second-named author by Lurie in 2005. It can be constructed using the formalism of~\cite{BFN}, and it is a motivating example of To\"en's ``brane operations" construction~\cite{Toenbranes} (though the compatibility of the two constructions is not currently documented). The factorization homology of this $3$-disc structure (\emph{i.e.}, the structure of ``line operator Ward identities" for the geometric Langlands program) is calculated in~\cite{darioTFT}. See also~\cite{LurieICM,ArinkinGaitsgory}.

Amusingly, in the formalism of derived stacks, there is an isomorphism
\begin{equation}
(\fg^{\vee})^\ast[2]/G_\C^\vee\simeq T^\ast[3] (pt)/G_\C^\vee
\end{equation}
between the equivariant coadjoint representation and the thrice-shifted cotangent bundle of a $G_\C^\vee$-equivariant point. In other words, the category of line operators in $\Bhatv$ is equivalent to the derived category of a thrice-shifted cotangent bundle, which is a shifted symplectic (in particular $P_3$) space.  This makes manifest the sort of structure we previously identified as what one would naively expect to see on the moduli space of a four-dimensional TQFT. The holomorphic functions on $T^\ast[3](pt)/G_\C^\vee$ (\emph{i.e.} local operators) do not detect the shifted Poisson structure, since they only see the space through its map to $\mathfrak h[2]/W$, but sheaves (line operators) do: they inherit a $3$-disc structure from the general quantization formalism of~\cite{CPTVV,PantevVezzosi}.

The equivalence $\cC\simeq\cC^\vee$ of monoidal categories, \emph{i.e.}, the $S$-duality of categories of line operators respecting OPE, is the content of the derived geometric Satake theorem of Bezrukavnikov and Finkelberg \cite{BezFink} (see also~\cite{ArinkinGaitsgory}). It can be upgraded to an equivalence of $3$-disc categories following along the lines of~\cite{ArinkinGaitsgory}.

\subsection{Primary products}
\label{subsec:GL_line_local_primary}

Now let us consider more concretely the algebraic interactions of local and line operators. The first thing is the primary product of local between local and line operators, which is simplest to describe in the $\Bhatv$ description. Since local operators $\CO\in \CA^\vee$ are just holomorphic functions on $(\mathfrak g^{\vee})^\ast[2]/G^\vee_\C$, they act naturally on coherent sheaves $\CL\in \cC^\vee$ via ordinary pointwise multiplication. In other words,
\begin{equation}
\label{4d-Acat}
\CO *\CL \in \text{End}(\CL)~, 
\end{equation}
is the bosonic endomorphism that multiplies sections of $\CL$ by the function $\CO$. This is directly analogous to the B-model discussion from Section~\ref{sec:bdy-2d}. 

In the $S$-dual description $\Ahat$, the endomorphism algebras of perverse sheaves $\CL\in \cC$ are described in terms of $G_\C$-equivariant cohomology. Then the primary product $\CA \to \text{End}(\CL)$ identifies polynomials in $\sigma$ with polynomials in the $G_\C$-equivariant parameters.

Physically speaking, we recall the interpretation of the ring $\CA$ of local operators as functions on the Coulomb branch $\mathfrak h/W$, The action on line operators is then given by specifying where on the Coulomb branch we set when considering a given line operator, and then replacing the local operator by its expectation value at that point.

\subsection{Secondary products: formal description} 
\label{subsec:GL_secondary_formal}

How should we understand the secondary structure in this case? We will first describe the formal structure implied by the general construction of secondary products when applied to the case of the GL twisted $\CN=4$ theory. We will then apply a result of Witten, which calculates a particular specialization of this structure, to deduce the nontriviality of the construction.

The secondary product defines an action of a local operator on a line operator that is the four-dimensional version of the operation we previously met in three dimensions, namely that of deforming vector bundles by the Hamiltonian flow defined by a function. In four dimensions, however, this action is bosonic, \emph{i.e.}, the secondary actions by bosonic local operators give bosonic self-interfaces of lines. Rather than interpreting the Hamiltonian flow as a deformation (an $Ext^1$ class, \emph{i.e.}, an endomorphism of degree $1$), we interpret it as an even (derived) endomorphism. For simplicitly, we will mostly suppress the (always even) cohomological/R-charge grading in the discussion below.

To illustrate how this action will look, let us first consider the classical situation of a family of Poisson commuting Hamiltonians on a Poisson manifold $X$, formulated as the data of a Poisson map $H:X\to B$ where the base $B$ carries the zero Poisson bracket (typically $B$ is a vector space, and after identifying $B$ with $\R^k$ the data of $H$ is equivalent to $k$ commuting Hamiltonians $H_1,\dots, H_k$). In this case we can describe the family of Poisson commuting Hamiltonian flows as a (fiberwise) action on $X$ of the vector bundle $T^\ast B$, considered as a family over $B$ of commutative Lie algebras (or as a trivial Lie algebroid).

We will use the same kind of picture to understand the secondary bracket of local operators on line operators --- again note that in contrast to the bracket for local operators themselves, this operation is a bosonic Poisson bracket (of degree $-2$). In our setting, the base $B$ is the Coulomb branch $(\fg^{\vee})^\ast/\Gv_\C\simeq \fh/W$, and the polynomial functions on $B$ correspond to local operators. The space $X$ in our Hamiltonian analogy is slightly more abstract --- it's the stack quotient
\begin{equation}
(\fg^{\vee})^\ast[2]/G_\C^\vee\simeq T^\ast[3](pt)/G_\C^\vee~,
\end{equation}
which maps to $B$ by the characteristic polynomial map -- or concretely, by virtue of the fact that the functions on this stack are the same as the topological algebra $\CA=\C[B]$. Recall that $X$ is designed so that line operators in $\Bhatv$ are coherent sheaves on it. Combining all the data together, we have the following abstract description of the secondary action on line operators:
\begin{proposition}\label{infi Ngo}
Any line operator $\CL$ in $\Bhatv\simeq \Ahat$ carries an action by central self-interfaces of the family of abelian Lie algebras $T^\ast \fh/W$ over the Coulomb branch $B\simeq \fh/W$. 
\end{proposition}
\noindent In other words, fixing a vacuum $\chi\in B$, there is an action of the abelian Lie algebra $T^\ast \fh/W$  by (even) self-interfaces of $\CL$, which moreover commutes with all interfaces of line operators. 

\subsection{Secondary products: concrete realization}
\label{Witten section}

We now relate our construction with a result in Section 2 of~\cite{Witten:2009mh}, showing in particular that the action described in Proposition~\ref{infi Ngo} is faithful and recovers a well-known construction of Ginzburg.

\bigskip

We would like to ``measure" our line operator by embedding it in a physical configuration that will produce an ordinary vector space, and we will then understand the secondary action on this vector space. Physically, Witten considers the Hilbert space of theories $\Ahat$ and $\Bhatv$ on the three-dimensional space $S^2\times I$, in the presence of 't Hooft and Wilson lines, respectively. The line operators sit at a point in $S^2\times I$, and are extended in Euclidean time. He chooses pairs of boundary conditions for the endpoints of the interval $I$ that greatly simplify the Hilbert space, effectively trivializing the contribution of bulk fields. In the $\Ahat$-model, he places Dirichlet boundary conditions on one end and Neumann on the other, while in the $\Bhatv$-model he places their more subtle $S$-dual boundary conditions: the ``universal kernel" on one end and the regular Nahm pole on the other. This setup is designed so that in the $\Bhatv$ theory with Wilson line $\CL_R^\vee$ in representation $R$, the Hilbert space simply becomes the finite-dimensional representation space $R$. In the $\Ahat$ theory with an $S$-dual 't Hooft line $\CL_R$, the Hilbert space is naturally identified as the intersection cohomology of a finite-dimensional orbit closure $\ol{\text{Gr}_G^R} \subset \text{Gr}_G$ in the affine Grassmannian for $G$. $S$-duality then reduces to the statement that
\begin{equation}
\label{Hilb-4d}
H^\bullet\Big(\ol{\text{Gr}_G^R}\Big) \simeq R~.
\end{equation}

Note that ordinary, rather than equivariant, cohomology appears here (despite the categorical equivariance in \eqref{4d-Acat}) due to the choice of boundary conditions, which restrict both theories $\Ahat$ and $\Bhatv$ to the origin of the Coulomb branche $B\simeq \fh/W$. In particular, the boundary conditions set to zero the bulk field $\sigma$ that plays the role of equivariant parameter in $\Ahat$. Correspondingly, the primary action of local operators on \eqref{Hilb-4d} is trivial. One can modify this setup so as to introduce dependence on the equivariant (\emph{i.e.}, Coulomb branch) parameters. 

The mathematical counterpart to this construction and its match across $S$-duality is a key compatibility of the equivalence of the derived geometric Satake theorem $\cC\simeq \cC^\vee$~\cite{BezFink} -- it respects natural functors to vector spaces. On the A-side, the natural measurement of an equivariant sheaf on the Grassmannian is its equivariant cohomology, which is a module for the equivariant cohomology ring, \emph{i.e.}, the topological algebra $\CA$. In other words it defines a vector space for each choice of point (vacuum) $\chi\in \fh/W$ on the Coulomb branch. On the B-side, we can measure an equivariant coherent sheaf on the coadjoint representation $\CL\in D^b\text{Coh}_{G_\C^\vee}(\mathfrak g^{\vee,\ast }[2])$ by restricting to the {\em Kostant slice} (the principal Slodowy slice). Namely, we consider a principal ${\mathfrak sl}_2$ triple $(e,h,f)$ in $\fg^\vee$ (the same data that appears in the description of the regular Nahm pole). Here $e$ denotes the image of the raising operator of $\mathfrak {sl}_2$. We let $(\fg^\vee)^e$ denote the centralizer of $e$ in $\mathfrak g^\vee$ -- an abelian subalgebra of dimension $\text{rank}(\fg)=\text{rank}(\fg^\vee)$. The Kostant slice is the embedding,
\begin{equation}
Kos:\fh/W\simeq \{e + (\fg^\vee)^f\}\hookrightarrow \fg^\vee~,
\end{equation}
of the Coulomb branch into the coadjoint representation. Thus given a line operator $\CL$ as an object in $\Bhatv$, we can restrict it to the Kostant slice, obtaining a module over $\CA$, or family of vector spaces over the Coulomb branch, and~\cite{BezFink} prove this matches with equivariant cohomology under $S$-duality. The physical construction described above corresponds to the further restriction of these families of vector spaces to the origin of the Coulomb branch.

\medskip


Witten considered the action on the vector space $R$ arising by integrating two-form descendants of local operators $\CO=f(\sigma^\vee)\in \CA^\vee$ on a two-sphere linking the Wilson line $\CL_R^\vee$. In other words, this is the restriction to the measurement $R$ of the secondary action,
\begin{equation}
\CO_{\CL_R^\vee} \in \text{End}(\CL_R^\vee)~.
\end{equation}
Witten demonstrated that this action agrees with that of the regular nilpotent centralizer $(\fg^\vee)^e$ of a principal $\mathfrak{sl}_2$ embedding in $\mathfrak g^\vee$. 
This provides a physical interpretation of a result of Ginzburg \cite{Ginzburg1995}, who showed that the cohomology ring of the entire affine Grassmannian is identified with the enveloping algebra of the regular nilpotent centralizer in a way manner that is compatible with the geometric Satake equivalence,
\begin{equation}
H^\ast(\text{Gr}_G)\simeq U(\fg^\vee)^e~.
\end{equation}
This identification relates representations of $\Gv$ with the cohomology of corresponding perverse sheaves on the Grassmannian \eqref{Hilb-4d}.

We can now combine Witten's calculation with Proposition~\ref{infi Ngo}. To do so we will need an explicit description of the cotangent bundle $T^\ast \fh/W$ in Lie algebraic terms that was explained in~\cite{BFM} Sections 2.2 and 2.4 (as part of what may be interpreted as the derivation of the Coulomb branch of pure 3d $\CN=4$ gauge theory). The description uses a principal ${\mathfrak sl}_2$ triple $(e,h,f)$ in $\fg^\vee$. First the cotangent fiber at the origin of the Coulomb branch can be identified with the centralizer of the principal nilpotent element $e\in \fg^\vee$,
\begin{equation}
T^\ast_0 B \simeq (\fg^{\vee})^e~.
\end{equation}
More generally, given a point $\chi\in B\simeq (\fg^{\vee})^\ast/\Gv_\C$, there is an isomorphism with the $\text{rank}(\fg)$-dimensional abelian Lie algebra,
\begin{equation}
T^\ast_\chi B\simeq (\fg^\vee)^{Kos(\chi)}~,
\end{equation}
which is the centralizer of the image of $\chi$ under the Kostant slice. It is also straightforward to see from the derivation of this isomorphism from Hamiltonian reduction of $T^\ast \fg^\vee$ under $\Gv$ that it agrees with Witten's identification of the principal nilpotent centralizer with $T^\ast_0 \fh/W$ in Section 2.11 (in particular after equation (2.17)) of~\cite{Witten:2009mh}. Putting all the pieces together we find the following result:

\begin{theorem}\label{infi Ngo redux}
The secondary action of local operators on a line operator $\CL$ in $\Ahat\simeq\Bhatv$, specialized at the origin $0\in \fh/W$ of the Coulomb branch, defines an action of the abelian Lie subalgebra $(\fg^\vee)^e\subset \fg^\vee$ by central self-interfaces of $\CL$, lifting its action on the underlying $\Gv$-representation space of $\Gv$ for $\CL$ a Wilson or 't Hooft line.
\end{theorem}

More generally, an extension of Witten's calculation of the descent bracket along the entire Coulomb branch is expected to show the following:

\begin{claim}
The secondary action of local operators on a line operator $\CL$ in $\Ahat\simeq\Bhatv$ defines an action of the family of abelian Lie algebras over $\fh/W$ given by regular centralizers
$$\chi \mapsto T^\ast_\chi \fh/W\simeq (\fg^\vee)^{Kos(\chi)}\subset \fg^\vee$$ by central self-interfaces of $\CL$, lifting its action on the underlying $\Gv$-representation space of $\Gv$ for $\CL$ a Wilson or 't Hooft line\footnote{For general line operators $\CL$, the action lifts the natural action of the regular centralizer $(\fg^\vee)^{Kos(\chi)}$ on the Kostant-Whittaker reduction of $\CL$ as in~\cite{BezFink}.}
\end{claim}

Theorem~\ref{infi Ngo redux} shows in a very concrete way the nontriviality of the $E_3$ structure on the category of line operators in GL-twisted $\CN=4$ SYM. Namely, the $E_3$ structure is measured through the secondary action of local operators on line operators, and in the case of Wilson lines the action is a lift (to the level of central self-interfaces of lines) of the action of a particular $\text{rank}(\fg)$-dimensional subalgebra of $\fg^\vee$ (depending on the chosen point $\chi\in B$ on the Coulomb branch) on the corresponding representation. This nontriviality is a measurement of the shifted symplectic nature of the space $T^\ast[3](pt)/\Gv$ on which line operators are realized as coherent sheaves. 

In fact a stronger result (though without the relation to the $E_3$ structure) appears from a closely related perspective in~\cite{BZG}: the Lie algebra action above is the derivative of the {\em Ng\^o action}~\cite{Knop, Ngo}. The Ng\^o action is a central action of the family of abelian groups of centralizers in the group $\Gv$ of Kostant slice elements,
\begin{equation}
\chi\in B\mapsto J_{\Gv}(\chi)=Z_{\Gv}(Kos(\chi))~,
\end{equation}
on the spherical category, \emph{i.e.}, the category of line operators. The total space of the family $J_\Gv$ is familiar physically as the Coulomb branch of pure three-dimensional $\CN=4$ $G$-gauge theory, \emph{i.e.}, the (partially completed) Toda integrable system~\cite{BFM}, though its group structure is more natural from the 4d $\CN=4$ perspective. Physically  the Ng\^o action can be interpreted as an action of the family of groups $J_\Gv$ over the Coulomb branch by one-form symmetries of the topologically twisted theory $\Ahat\simeq\Bhatv$. We plan to explain this interpretation in detail in a future publication.

\subsection{Donaldson theory and surface operators}
\label{Donaldson section}

Finally, we briefly comment on the possibility of higher operations in Donaldson theory, \emph{i.e.}, $\cN=2$ SYM in the Donaldson-Witten twist. In this theory there are no topological line operators, but we may still expect to find interesting higher products involving surface operators.
Indeed, we claim that there should be nontrivial secondary products of local operators in Donaldson theory with suitable (though somewhat nonstandard) surface operators.

Recall \cite{Witten:1988ze} that local operators in Donaldson theory with gauge group $G$ consist of gauge-invariant polynomials in the $\mathfrak g_\C$-valued complex scalar field, often denoted $\phi$. This is the same algebra \eqref{4d-locA} that appeared in the GL twist. 

A useful perspective for analyzing surface operators in a four-dimensional TQFT is to identify them with boundary conditions in a circle compactification of the theory. (This is analogous to the identification of line operators in a 3d TQFT with boundary conditions for its circle compactification, \emph{cf.} Section \ref{sec:line-cat}.) In the case of the Donaldson twist of a 4d $\CN=2$ gauge theory, the circle compactification may roughly be identified with 3d Rozansky-Witten theory on the Seiberg-Witten integrable system $M_{SW}$. 
Then, thanks to the description of boundary conditions in Rozansky-Witten theory~\cite{Kapustin:2008sc} we expect topological surface operators corresponding to arbitrary holomorphic Lagrangians on the Seiberg-Witten integrable system (as well as more general operators coming roughly from sheaves of categories over holomorphic Lagrangians).

In this compactified perspective, secondary products of a 4d local operator $\CO$ and a surface operator should translate to primary products between a 3d local operator $\oint_{S^1}\CO^{(1)}$ and a boundary condition.
(This is analogous to the 3d/2d setup in Section \ref{sec:RW12}.) If $\CO = p(\phi)$ is a bosonic guage-invariant polynomial, then $\oint_{S^1}\CO^{(1)}$ is a fermionic local operator in RW theory on $M_{SW}$. The relevant fermionic local operators are given by classes in $H^{0,1}(M_{SW})=H^1(M_{SW},\cO)=TPic(M_{SW})$, \emph{i.e.}  tangent vectors to the Picard group of line bundles on $M_{SW}$. The secondary action of these operators on a boundary condition, realized by a sheaf of categories on a holomorphic Lagrangian, is tangent to the natural action of the Picard group by automorphisms of a sheaf of categories --- \emph{i.e.}, to tensoring by line bundles.

For the familiar surface operators in Donaldson theory (namely, generalizations of Gukov-Witten surface operators \cite{Gukov:2006jk}) this action is trivial: the corresponding Lagrangians wrap (multi-)sections of the integrable system \cite{Gaiotto:2011tf}, to which all line bundles restrict trivially. However, since the line bundles are nontrivial along fibers of the integrable system they will act nontrivially as endomorphisms of topological boundary condition corresponding to a holomorphic Lagrangian \emph{wrapping a fiber}. The surface operators corresponding to such holomorphic Lagrangians give our sought-for examples of nontrivial secondary products.

\bibliographystyle{utphys}
\bibliography{descent-short-paper}

\providecommand{\href}[2]{#2}\begingroup\raggedright\begin{thebibliography}{100}

\bibitem{Atiyah:1989vu}
M.~Atiyah, ``{Topological quantum field theories},''
\href{http://dx.doi.org/10.1007/BF02698547}{{\em Inst. Hautes Etudes Sci. Publ.
  Math.} {\bfseries 68} (1989) 175--186}.

\bibitem{Segal:1987sk}
G.~B. Segal, ``{The Definition of Conformal Field Theory},'' in {\em {In *COMO
  1987 Proceedings, Differential Geometric Methods in Theoretical Physics}},
  pp.~165--171.
\newblock
1987.
\newblock

\bibitem{Witten:1988ze}
E.~Witten, ``Topological quantum field theory,''
{\em Commun. Math. Phys.} {\bfseries 117} (1988) 353.

\bibitem{Witten:1988xj}
E.~Witten, ``{Topological Sigma Models},''
\href{http://dx.doi.org/10.1007/BF01466725}{{\em Commun. Math. Phys.}
  {\bfseries 118} (1988) 411}.

\bibitem{Getzler:1994yd}
E.~Getzler, ``{Batalin-Vilkovisky algebras and two-dimensional topological
  field theories},'' \href{http://dx.doi.org/10.1007/BF02102639}{{\em Commun.
  Math. Phys.} {\bfseries 159} (1994) 265--285},
\href{http://arxiv.org/abs/hep-th/9212043}{{\ttfamily arXiv:hep-th/9212043
  [hep-th]}}.

\bibitem{Segal:99}
G.~Segal, ``Topological field theory.'' http://www.cgtp.duke.edu/ITP99/segal/,
  Notes of lectures at Stanford university., 1999.

\bibitem{Costello:2004ei}
K.~Costello, ``{Topological conformal field theories and Calabi-Yau
  categories},'' \href{http://dx.doi.org/10.1016/j.aim.2006.06.004}{{\em Adv.
  Math.} {\bfseries 210} (2007) 165--214},
\href{http://arxiv.org/abs/math/0412149}{{\ttfamily arXiv:math/0412149
  [math-qa]}}.

\bibitem{Baez:1995xq}
J.~C. Baez and J.~Dolan, ``{Higher dimensional algebra and topological quantum
  field theory},'' \href{http://dx.doi.org/10.1063/1.531236}{{\em J. Math.
  Phys.} {\bfseries 36} (1995) 6073--6105},
\href{http://arxiv.org/abs/q-alg/9503002}{{\ttfamily arXiv:q-alg/9503002
  [q-alg]}}.

\bibitem{Lurie:2009keu}
J.~Lurie, ``{On the Classification of Topological Field Theories},''
\href{http://arxiv.org/abs/0905.0465}{{\ttfamily arXiv:0905.0465 [math.CT]}}.

\bibitem{Ayala:2017wcr}
D.~Ayala and J.~Francis, ``{The cobordism hypothesis},''
\href{http://arxiv.org/abs/1705.02240}{{\ttfamily arXiv:1705.02240 [math.AT]}}.

\bibitem{Freed:2012hx}
D.~S. Freed, ``{The cobordism hypothesis},''
\href{http://arxiv.org/abs/1210.5100}{{\ttfamily arXiv:1210.5100 [math.AT]}}.

\bibitem{Moore:1997pc}
G.~W. Moore and E.~Witten, ``Integration over the {$u$}-plane in {D}onaldson
  theory,'' {\em Adv. Theor. Math. Phys.} {\bfseries 1} (1998) 298,
\href{http://arxiv.org/abs/hep-th/9709193}{{\ttfamily hep-th/9709193}}.

\bibitem{May:1972}
J.~P. May, {\em The geometry of iterated loop spaces}.
\newblock Springer-Verlag, Berlin-New York, 1972.
\newblock Lectures Notes in Mathematics, Vol. 271.

\bibitem{Cohen:1976}
F.~Cohen, T.~Lada, and P.~May, {\em The homology of iterated loop spaces}.
\newblock Springer-Verlag, Berlin,
  http://www.math.uchicago.edu/~may/BOOKS/homo\_iter.pdf, 1976.

\bibitem{Sinha}
D.~P. Sinha, ``The (non-equivariant) homology of the little disks operad,'' in
  {\em O{PERADS} 2009}, vol.~26 of {\em S\'emin. Congr.}, pp.~253--279.
\newblock Soc. Math. France, Paris, 2013.

\bibitem{HA}
J.~Lurie, ``Higher algebra.''. \texttt{www.math.harvard.edu/$\sim$lurie/}.

\bibitem{ayalafrancis}
D.~Ayala and J.~Francis, ``Factorization homology of topological manifolds,''
  {\em J. of Topology} {\bfseries 8} no.~4, (2015) 1045--1084,
  \href{http://arxiv.org/abs/arXiv:1206.5522}{{\ttfamily arXiv:1206.5522}}.

\bibitem{ayalafrancisPoincare}
D.~Ayala and J.~Francis, ``Poincar\'e/koszul duality.'' arXiv:1409.2478, 2018.

\bibitem{Cattaneo:2006}
A.~Cattaneo, D.~Fiorenza, and R.~Longoni, ``Graded poisson algebras,'' in {\em
  Encyclopedia of Mathematical Physics}, N.~Fran{\c c}oise and Tsou, eds.,
  vol.~2, pp.~560--567.
\newblock Oxford: Elsevier,
  https://www.math.uzh.ch/fileadmin/math/preprints/15-05.pdf, 2006.

\bibitem{WahlSalvatore}
P.~Salvatore and N.~Wahl, ``Framed discs operads and batalin-vilkovisky
  algebras.,'' {\em Quarterly Journal of Mathematics} {\bfseries 54} no.~2,
  (2003) 213--231.

\bibitem{Lada:1992wc}
T.~Lada and J.~Stasheff, ``Introduction to sh lie algebras for physicists,''
  {\em Int. J. Theor. Phys.} {\bfseries 32} (1993) 1087--1104,
  \href{http://arxiv.org/abs/arXiv:hep-th/9209099}{{\ttfamily
  arXiv:hep-th/9209099}}.

\bibitem{Kontsevich-notes}
M.~Kontsevich, ``Notes on deformation theory.'' Berkeley course notes,
  available at \texttt{www.math.uchicago.edu/$\sim$mitya/langlands.html}.

\bibitem{Gaiotto:2015aoa}
D.~Gaiotto, G.~W. Moore, and E.~Witten, ``{Algebra of the Infrared: String
  Field Theoretic Structures in Massive ${\cal N}=(2,2)$ Field Theory In Two
  Dimensions},''
\href{http://arxiv.org/abs/1506.04087}{{\ttfamily arXiv:1506.04087 [hep-th]}}.

\bibitem{Gaiotto:2015zna}
D.~Gaiotto, G.~W. Moore, and E.~Witten, ``{An Introduction To The Web-Based
  Formalism},''
\href{http://arxiv.org/abs/1506.04086}{{\ttfamily arXiv:1506.04086 [hep-th]}}.

\bibitem{PTVV}
T.~Pantev, B.~To\"en, M.~Vaqui\'e, and G.~Vezzosi, ``Shifted symplectic
  structures.,'' {\em Publ. Math. Inst. Hautes \'Etudes Sci.} {\bfseries 117}
  (2013) 271--328, \href{http://arxiv.org/abs/arXiv:1111.3209}{{\ttfamily
  arXiv:1111.3209}}.

\bibitem{CPTVV}
D.~Calaque, T.~Pantev, B.~To\"en, M.~Vaqui\'e, and G.~Vezzosi, ``Shifted
  poisson structures and deformation quantization,'' {\em J. Topology}
  {\bfseries 10} no.~2, (2017) 483--584,
  \href{http://arxiv.org/abs/arXiv:1506.03699}{{\ttfamily arXiv:1506.03699}}.

\bibitem{PantevVezzosi}
T.~Pantev and G.~Vezzosi, ``Symplectic and poisson derived geometry and
  deformation quantization.'' arXiv:1603.02753, 2016.

\bibitem{Safronov}
P.~Safronov, ``Lectures on shifted poisson geometry.'' arXiv:1709.07698, 2017.

\bibitem{AKSZ}
M.~Alexandrov, M.~Kontsevich, A.~Schwarz, and O.~Zaboronsky, ``The geometry of
  the master equation and topological quantum field theory,'' {\em
  Int.J.Mod.Phys.} {\bfseries A12} (1997) 1405--1430,
  \href{http://arxiv.org/abs/arXiv:hep-th/9502010}{{\ttfamily
  arXiv:hep-th/9502010}}.

\bibitem{Kotov:2010wr}
A.~Kotov and T.~Strobl, ``{Generalizing Geometry - Algebroids and Sigma
  Models},'' \href{http://arxiv.org/abs/1004.0632}{{\ttfamily arXiv:1004.0632
  [hep-th]}}.

\bibitem{Fiorenza:2011jr}
D.~Fiorenza, C.~L. Rogers, and U.~Schreiber, ``{A Higher Chern-Weil derivation
  of AKSZ $\sigma$-models},''
  \href{http://dx.doi.org/10.1142/S0219887812500788}{{\em Int. J. Geom. Meth.
  Mod. Phys.} {\bfseries 10} (2013) 1250078},
\href{http://arxiv.org/abs/1108.4378}{{\ttfamily arXiv:1108.4378 [math-ph]}}.

\bibitem{nlab:AKSZ}
``nlab: Aksz sigma-model.''. \texttt{ncatlab.org/nlab/show/AKSZ+sigma-model}.

\bibitem{CG}
K.~Costello and O.~Gwilliam, {\em Factorization Algebras in Quantum Field
  Theory}.
\newblock New Mathematical Monographs 31. Cambridge University Press, 2017.

\bibitem{BD}
A.~Beilinson and V.~Drinfeld, {\em Chiral Algebras}.
\newblock No.~51 in American Mathematical Society Colloquium Publications.
  American Mathematical Society, 2004.

\bibitem{CostelloScheimbauer}
K.~Costello and C.~Scheimbauer, ``Lectures on mathematical aspects of (twisted)
  supersymmetric gauge theories,'' in {\em Mathematical aspects of quantum
  field theories}, Math. Phys. Stud., pp.~57--87.
\newblock Springer, Cham, 2015.

\bibitem{ElliottSafronov}
C.~Elliott and P.~Safronov, ``Topological twists of supersymmetric algebras of
  observables,'' \href{http://arxiv.org/abs/arXiv:1805.10806}{{\ttfamily
  arXiv:1805.10806}}.

\bibitem{Lian:1992mn}
B.~H. Lian and G.~J. Zuckerman, ``{New perspectives on the BRST algebraic
  structure of string theory},''
  \href{http://dx.doi.org/10.1007/BF02102111}{{\em Commun. Math. Phys.}
  {\bfseries 154} (1993) 613--646},
\href{http://arxiv.org/abs/hep-th/9211072}{{\ttfamily arXiv:hep-th/9211072
  [hep-th]}}.

\bibitem{Penkava:1992sh}
M.~Penkava and A.~S. Schwarz, ``{On some algebraic structure arising in string
  theory},''
\href{http://arxiv.org/abs/hep-th/9212072}{{\ttfamily arXiv:hep-th/9212072
  [hep-th]}}.

\bibitem{Kontsevich93}
M.~Kontsevich, ``Formal (non)commutative symplectic geometry,'' in {\em The
  Gel'fand Mathematical Seminars, 1990--1992}, pp.~173--187.
\newblock Birkh\"user Boston, 1993.

\bibitem{Witten:1992yj}
E.~Witten and B.~Zwiebach, ``{Algebraic structures and differential geometry in
  2-D string theory},''
  \href{http://dx.doi.org/10.1016/0550-3213(92)90018-7}{{\em Nucl. Phys.}
  {\bfseries B377} (1992) 55--112},
\href{http://arxiv.org/abs/hep-th/9201056}{{\ttfamily arXiv:hep-th/9201056
  [hep-th]}}.

\bibitem{KKP08}
L.~Katzarkov, M.~Kontsevich, and T.~Pantev,
  \href{http://dx.doi.org/10.1090/pspum/078/2483750}{``Hodge theoretic aspects
  of mirror symmetry,''} in {\em From {H}odge theory to integrability and
  {TQFT} tt*-geometry}, vol.~78 of {\em Proc. Sympos. Pure Math.}, pp.~87--174.
\newblock Amer. Math. Soc., Providence, RI, 2008.
\newblock \url{https://doi.org/10.1090/pspum/078/2483750}.

\bibitem{KKP14}
L.~Katzarkov, M.~Kontsevich, and T.~Pantev, ``Bogomolov-tian-todorov theorems
  for landau-ginzburg models.,'' {\em J. Diff. Geom} {\bfseries 105} no.~1,
  (2017) 55--117.

\bibitem{Seidel}
P.~Seidel, \href{http://dx.doi.org/10.1090/pspum/080.1/2483942}{``Symplectic
  homology as {H}ochschild homology,''} in {\em Algebraic geometry---{S}eattle
  2005. {P}art 1}, vol.~80 of {\em Proc. Sympos. Pure Math.}, pp.~415--434.
\newblock Amer. Math. Soc., Providence, RI, 2009.
\newblock \url{https://doi.org/10.1090/pspum/080.1/2483942}.

\bibitem{Ganatra}
S.~Ganatra, {\em Symplectic {C}ohomology and {D}uality for the {W}rapped
  {F}ukaya {C}ategory}.
\newblock ProQuest LLC, Ann Arbor, MI, 2012.
\newblock
  \url{http://gateway.proquest.com/openurl?url_ver=Z39.88-2004&rft_val_fmt=info:ofi/fmt:kev:mtx:dissertation&res_dat=xri:pqm&rft_dat=xri:pqdiss:0828785}.
\newblock Thesis (Ph.D.)--Massachusetts Institute of Technology.

\bibitem{GPS}
S.~Ganatra, J.~Pardon, and V.~Shende, ``{Covariantly functorial wrapped Floer
  theory on Liouville sectors},''
  \href{http://arxiv.org/abs/arXiv:1706.03152}{{\ttfamily arXiv:1706.03152}}.

\bibitem{Kontsevich:1994dn}
M.~Kontsevich, ``{Homological Algebra of Mirror Symmetry},'' in {\em
  Proceedings of the International Congress of Mathematicians, Zurich, Vol. I},
  pp.~120--139.
\newblock Birkhauser, 1994.
\newblock
\href{http://arxiv.org/abs/alg-geom/9411018}{{\ttfamily arXiv:alg-geom/9411018
  [alg-geom]}}.
\newblock

\bibitem{KontsevichSoibelman:2009}
M.~Kontsevich and Y.~Soibelman, ``Notes on {$A_\infty$}-algebras,
  {$A_\infty$}-categories and non-commutative geometry,'' in {\em Homological
  mirror symmetry}, vol.~757 of {\em Lecture Notes in Phys.}, pp.~153--219.
\newblock Springer, Berlin, 2009.

\bibitem{Kapustin:2004df}
A.~Kapustin and L.~Rozansky, ``{On the relation between open and closed
  topological strings},''
  \href{http://dx.doi.org/10.1007/s00220-004-1227-z}{{\em Commun. Math. Phys.}
  {\bfseries 252} (2004) 393--414},
\href{http://arxiv.org/abs/hep-th/0405232}{{\ttfamily arXiv:hep-th/0405232
  [hep-th]}}.

\bibitem{Gerst}
M.~Gerstenhaber, ``The cohomology structure of an associative ring,'' {\em Ann.
  Math.} {\bfseries 78} (1963) 267--288.

\bibitem{Tamarkin:98}
D.~Tamarkin, ``Another proof of m. kontsevich formality theorem.''
  arXiv:math/9803025, 1998.

\bibitem{McClureSmith:99}
J.~McClure and J.~Smith, ``A solution of deligne's conjecture,'' in {\em Recent
  progress in homotopy theory (Baltimore, MD, 2000)}, vol.~293 of {\em Contemp.
  Math.}, pp.~153--193.
\newblock Amer. Math. Soc., Providence, RI, 2002.
\newblock \href{http://arxiv.org/abs/arXiv:math/9910126}{{\ttfamily
  arXiv:math/9910126}}.

\bibitem{KontsevichSoibelman:2000}
M.~Kontsevich and Y.~Soibelman, ``Deformations of algebras over operads and the
  deligne conjecture.,'' in {\em Conf\'erence Mosh\'e Flato 1999, Vol. I
  (Dijon)}, vol.~21 of {\em Math. Phys. Stud.}, pp.~255--307.
\newblock Kluwer Acad. Publ, 2000.

\bibitem{BergerFresse:2001}
C.~Berger and B.~Fresse, ``Combinatorial operad actions on cochains.,'' {\em
  Math. Proc. Cambridge Philos. Soc.} {\bfseries 137} no.~1, (2004) 135--174.

\bibitem{Rozansky:1996bq}
L.~Rozansky and E.~Witten, ``Hyper-kaehler geometry and invariants of
  three-manifolds,'' {\em Selecta Math.} {\bfseries 3} (1997) 401--458,
\href{http://arxiv.org/abs/hep-th/9612216}{{\ttfamily hep-th/9612216}}.

\bibitem{Moore:1997dj}
G.~W. Moore, N.~Nekrasov, and S.~Shatashvili, ``{Integrating over Higgs
  branches},'' \href{http://dx.doi.org/10.1007/PL00005525}{{\em Commun. Math.
  Phys.} {\bfseries 209} (2000) 97--121},
\href{http://arxiv.org/abs/hep-th/9712241}{{\ttfamily arXiv:hep-th/9712241
  [hep-th]}}.

\bibitem{Moore:1998et}
G.~W. Moore, N.~Nekrasov, and S.~Shatashvili, ``{D particle bound states and
  generalized instantons},''
  \href{http://dx.doi.org/10.1007/s002200050016}{{\em Commun. Math. Phys.}
  {\bfseries 209} (2000) 77--95},
\href{http://arxiv.org/abs/hep-th/9803265}{{\ttfamily arXiv:hep-th/9803265
  [hep-th]}}.

\bibitem{Nekrasov:2002qd}
N.~A. Nekrasov, ``{Seiberg-Witten prepotential from instanton counting},''
  \href{http://dx.doi.org/10.4310/ATMP.2003.v7.n5.a4}{{\em Adv. Theor. Math.
  Phys.} {\bfseries 7} no.~5, (2003) 831--864},
\href{http://arxiv.org/abs/hep-th/0206161}{{\ttfamily arXiv:hep-th/0206161
  [hep-th]}}.

\bibitem{Nekrasov:2003rj}
N.~A. Nekrasov and A.~Okounkov, ``{S}eiberg-{W}itten theory and random
  partitions,''
\href{http://arxiv.org/abs/hep-th/0306238}{{\ttfamily hep-th/0306238}}.

\bibitem{Nekrasov:2009rc}
N.~A. Nekrasov and S.~L. Shatashvili,
  \href{http://dx.doi.org/10.1142/9789814304634_0015}{``{Quantization of
  Integrable Systems and Four Dimensional Gauge Theories},''} in {\em
  {Proceedings, 16th International Congress on Mathematical Physics (ICMP09):
  Prague, Czech Republic, August 3-8, 2009}}, pp.~265--289.
\newblock 2009.
\newblock \href{http://arxiv.org/abs/0908.4052}{{\ttfamily arXiv:0908.4052
  [hep-th]}}.
\newblock
\url{https://inspirehep.net/record/829640/files/arXiv:0908.4052.pdf}.
\newblock

\bibitem{Shadchin:2006yz}
S.~Shadchin, ``{On F-term contribution to effective action},''
  \href{http://dx.doi.org/10.1088/1126-6708/2007/08/052}{{\em JHEP} {\bfseries
  08} (2007) 052},
\href{http://arxiv.org/abs/hep-th/0611278}{{\ttfamily arXiv:hep-th/0611278
  [hep-th]}}.

\bibitem{Dimofte:2010tz}
T.~Dimofte, S.~Gukov, and L.~Hollands, ``{Vortex Counting and Lagrangian
  3-manifolds},'' \href{http://dx.doi.org/10.1007/s11005-011-0531-8}{{\em Lett.
  Math. Phys.} {\bfseries 98} (2011) 225--287},
\href{http://arxiv.org/abs/1006.0977}{{\ttfamily arXiv:1006.0977 [hep-th]}}.

\bibitem{Yagi2014}
J.~Yagi, ``$\omega$-deformation and quantization,''
  \href{http://arxiv.org/abs/1405.6714v3}{{\ttfamily 1405.6714v3}}.

\bibitem{Dimofte:2009bv}
T.~Dimofte and S.~Gukov, ``{Refined, Motivic, and Quantum},''
  \href{http://dx.doi.org/10.1007/s11005-009-0357-9}{{\em Lett. Math. Phys.}
  {\bfseries 91} (2010) 1},
\href{http://arxiv.org/abs/0904.1420}{{\ttfamily arXiv:0904.1420 [hep-th]}}.

\bibitem{Alday:2009fs}
L.~F. Alday, D.~Gaiotto, S.~Gukov, Y.~Tachikawa, and H.~Verlinde, ``{Loop and
  surface operators in N=2 gauge theory and Liouville modular geometry},''
  \href{http://dx.doi.org/10.1007/JHEP01(2010)113}{{\em JHEP} {\bfseries 01}
  (2010) 113},
\href{http://arxiv.org/abs/0909.0945}{{\ttfamily arXiv:0909.0945 [hep-th]}}.

\bibitem{Drukker:2009id}
N.~Drukker, J.~Gomis, T.~Okuda, and J.~Teschner, ``{Gauge Theory Loop Operators
  and Liouville Theory},''
  \href{http://dx.doi.org/10.1007/JHEP02(2010)057}{{\em JHEP} {\bfseries 02}
  (2010) 057},
\href{http://arxiv.org/abs/0909.1105}{{\ttfamily arXiv:0909.1105 [hep-th]}}.

\bibitem{Teschner:2010je}
J.~Teschner, ``{Quantization of the Hitchin moduli spaces, Liouville theory,
  and the geometric Langlands correspondence I},''
  \href{http://dx.doi.org/10.4310/ATMP.2011.v15.n2.a6}{{\em Adv. Theor. Math.
  Phys.} {\bfseries 15} no.~2, (2011) 471--564},
\href{http://arxiv.org/abs/1005.2846}{{\ttfamily arXiv:1005.2846 [hep-th]}}.

\bibitem{Gaiotto:2010be}
D.~Gaiotto, G.~W. Moore, and A.~Neitzke, ``{Framed BPS States},''
  \href{http://dx.doi.org/10.4310/ATMP.2013.v17.n2.a1}{{\em Adv. Theor. Math.
  Phys.} {\bfseries 17} no.~2, (2013) 241--397},
\href{http://arxiv.org/abs/1006.0146}{{\ttfamily arXiv:1006.0146 [hep-th]}}.

\bibitem{Nekrasov:2010ka}
N.~Nekrasov and E.~Witten, ``{The Omega Deformation, Branes, Integrability, and
  Liouville Theory},'' \href{http://dx.doi.org/10.1007/JHEP09(2010)092}{{\em
  JHEP} {\bfseries 09} (2010) 092},
\href{http://arxiv.org/abs/1002.0888}{{\ttfamily arXiv:1002.0888 [hep-th]}}.

\bibitem{Kapustin:2001ij}
A.~Kapustin and D.~Orlov, ``{Remarks on A branes, mirror symmetry, and the
  Fukaya category},''
  \href{http://dx.doi.org/10.1016/S0393-0440(03)00026-3}{{\em J. Geom. Phys.}
  {\bfseries 48} (2003) 84},
\href{http://arxiv.org/abs/hep-th/0109098}{{\ttfamily arXiv:hep-th/0109098
  [hep-th]}}.

\bibitem{Kapustin:2006pk}
A.~Kapustin and E.~Witten, ``{Electric-Magnetic Duality And The Geometric
  Langlands Program},''
  \href{http://dx.doi.org/10.4310/CNTP.2007.v1.n1.a1}{{\em Commun. Num. Theor.
  Phys.} {\bfseries 1} (2007) 1--236},
\href{http://arxiv.org/abs/hep-th/0604151}{{\ttfamily arXiv:hep-th/0604151
  [hep-th]}}.

\bibitem{Gukov:2008ve}
S.~Gukov and E.~Witten, ``{Branes and Quantization},''
  \href{http://dx.doi.org/10.4310/ATMP.2009.v13.n5.a5}{{\em Adv. Theor. Math.
  Phys.} {\bfseries 13} no.~5, (2009) 1445--1518},
\href{http://arxiv.org/abs/0809.0305}{{\ttfamily arXiv:0809.0305 [hep-th]}}.

\bibitem{BZNeitzke}
D.~Ben-Zvi and A.~Neitzke, ``Algebraic geometry of topological field theory.''
  In preparation.

\bibitem{Kapustin:2010ta}
A.~Kapustin, ``{Topological Field Theory, Higher Categories, and Their
  Applications},'' in {\em {International Congress of Mathematicians (ICM 2010)
  Hyderabad, India, August 19-27, 2010}}.
\newblock 2010.
\newblock \href{http://arxiv.org/abs/1004.2307}{{\ttfamily arXiv:1004.2307
  [math.QA]}}.
\newblock
\url{https://inspirehep.net/record/852088/files/arXiv:1004.2307.pdf}.
\newblock

\bibitem{RobertsWillerton}
J.~Roberts and S.~Willerton, ``On the rozansky-witten weight systems,'' {\em
  Algebr. Geom. Topol.} {\bfseries 10} (2010) 1455--1519,
  \href{http://arxiv.org/abs/arXiv:math/0602653}{{\ttfamily
  arXiv:math/0602653}}.

\bibitem{Kapustin:2008sc}
A.~Kapustin, L.~Rozansky, and N.~Saulina, ``Three-dimensional topological field
  theory and symplectic algebraic geometry i,'' {\em Nucl. Phys.} {\bfseries
  B816} (2009) 295--355,
  \href{http://arxiv.org/abs/arXiv:hep-th/0810.5415}{{\ttfamily
  arXiv:hep-th/0810.5415}}.

\bibitem{BFN}
D.~Ben-Zvi, J.~Francis, and D.~Nadler, ``Integral transforms and drinfeld
  centers in derived algebraic geometry,'' {\em J. Amer. Math. Soc.} {\bfseries
  23} (2010) 909--966, \href{http://arxiv.org/abs/arXiv:0805.0157}{{\ttfamily
  arXiv:0805.0157}}.

\bibitem{CostelloFrancis}
K.~Costello and J.~Francis, ``In preparation.''.

\bibitem{Costello:13}
K.~Costello, ``{Supersymmetric gauge theory and the Yangian},''
\href{http://arxiv.org/abs/1303.2632}{{\ttfamily arXiv:1303.2632 [hep-th]}}.

\bibitem{AyalaFrancisTanaka}
D.~Ayala, J.~Francis, and H.~Tanaka, ``Factorization homology of stratified
  spaces.,'' {\em Selecta Math. (N.S.)} {\bfseries 23} no.~1, (2017) 293--362,
  \href{http://arxiv.org/abs/arXiv:1409.0848}{{\ttfamily arXiv:1409.0848}}.

\bibitem{Witten:2009mh}
E.~Witten, ``{Geometric Langlands And The Equations Of Nahm And Bogomolny},''
\href{http://arxiv.org/abs/0905.4795}{{\ttfamily arXiv:0905.4795 [hep-th]}}.

\bibitem{Ginzburg1995}
V.~Ginzburg, ``Perverse sheaves on a loop group and langlands' duality,''
  \href{http://arxiv.org/abs/alg-geom/9511007v4}{{\ttfamily
  alg-geom/9511007v4}}.

\bibitem{LurieICM}
J.~Lurie, ``Moduli problems for ring spectra,'' in {\em Proceedings of the
  International Congress of Mathematicians. Volume II,}, pp.~1099--1125.
\newblock Hindustan Book Agency, New Delhi,
  http://www.math.harvard.edu/~lurie/, 2010.

\bibitem{Toenbranes}
B.~To\"en, ``Operations on derived moduli spaces of branes,''
  \href{http://arxiv.org/abs/arXiv:1307.0405}{{\ttfamily arXiv:1307.0405}}.

\bibitem{ArinkinGaitsgory}
D.~Arinkin and D.~Gaitsgory, ``Singular support of coherent sheaves, and the
  geometric langlands conjecture.,'' {\em Selecta Math. (N.S.)} {\bfseries 21}
  no.~1, (2015) 1--199, \href{http://arxiv.org/abs/arXiv:1201.6343}{{\ttfamily
  arXiv:1201.6343}}.

\bibitem{darioTFT}
D.~Beraldo, ``The topological chiral homology of the spherical category,''
  \href{http://arxiv.org/abs/arXiv:1802.08118}{{\ttfamily arXiv:1802.08118}}.

\bibitem{BZG}
D.~Ben-Zvi and S.~Gunningham, ``{Symmetries of categorical representations and
  the quantum Ng\^o action},''
  \href{http://arxiv.org/abs/arXiv:1712.01963}{{\ttfamily arXiv:1712.01963}}.

\bibitem{KontsevichSoibelmanBook}
M.~Kontsevich and Y.~Soibelman, ``Deformation theory, vol.1.'' Preliminary
  draft at https://www.math.ksu.edu/~soibel/.

\bibitem{Park:2003kn}
J.-S. Park, ``Pursuing the quantum world: Flat family of qfts and quantization
  of d algebras,'' \href{http://arxiv.org/abs/arXiv:hep-th/0308130}{{\ttfamily
  arXiv:hep-th/0308130}}.

\bibitem{SoibelmanCFT}
Y.~Soibelman, ``Collapsing cfts and quantum spaces with non-negative ricci
  curvature.'' Unfinished draft, https://www.math.ksu.edu/~soibel/.

\bibitem{Gaiotto:2014kfa}
D.~Gaiotto, A.~Kapustin, N.~Seiberg, and B.~Willett, ``Generalized global
  symmetries,'' {\em JHEP} {\bfseries 2} (2015) 172,
  \href{http://arxiv.org/abs/arXiv:1412.5148}{{\ttfamily arXiv:1412.5148}}.

\bibitem{Johansen:1994aw}
A.~Johansen, ``{Twisting of $N=1$ SUSY gauge theories and heterotic topological
  theories},'' \href{http://dx.doi.org/10.1142/S0217751X9500200X}{{\em Int. J.
  Mod. Phys.} {\bfseries A10} (1995) 4325--4358},
\href{http://arxiv.org/abs/hep-th/9403017}{{\ttfamily arXiv:hep-th/9403017
  [hep-th]}}.

\bibitem{Kapustin:2006hi}
A.~Kapustin, ``{Holomorphic reduction of N=2 gauge theories, Wilson-'t Hooft
  operators, and S-duality},''
\href{http://arxiv.org/abs/hep-th/0612119}{{\ttfamily arXiv:hep-th/0612119
  [hep-th]}}.

\bibitem{Costello:2011np}
K.~J. Costello, ``{Notes on supersymmetric and holomorphic field theories in
  dimensions 2 and 4},''
\href{http://arxiv.org/abs/1111.4234}{{\ttfamily arXiv:1111.4234 [math.QA]}}.

\bibitem{Elliott:2015rja}
C.~Elliott and P.~Yoo, ``Geometric langlands twists of n = 4 gauge theory from
  derived algebraic geometry,'' {\em arXiv:1507.03048} (2015) .

\bibitem{pavel}
P.~Safronov, ``Unipotent $e_n$ structures.'' Private communication.

\bibitem{Witten:1991zz}
E.~Witten, ``{Mirror manifolds and topological field theory},''
  \href{http://arxiv.org/abs/hep-th/9112056}{{\ttfamily arXiv:hep-th/9112056
  [hep-th]}}.
[AMS/IP Stud. Adv. Math.9,121(1998)].

\bibitem{Hori:2003ic}
K.~Hori, S.~Katz, A.~Klemm, R.~Pandharipande, R.~Thomas, C.~Vafa, R.~Vakil, and
  E.~Zaslow, {\em {Mirror symmetry}}, vol.~1 of {\em Clay mathematics
  monographs}.
\newblock AMS, Providence, USA, 2003.
\newblock
\url{http://www.claymath.org/library/monographs/cmim01.pdf}.
\newblock

\bibitem{Lerche:1989uy}
W.~Lerche, C.~Vafa, and N.~P. Warner, ``{Chiral Rings in N=2 Superconformal
  Theories},''
\href{http://dx.doi.org/10.1016/0550-3213(89)90474-4}{{\em Nucl. Phys.}
  {\bfseries B324} (1989) 427--474}.

\bibitem{Hitchin:1986ea}
N.~J. Hitchin, A.~Karlhede, U.~Lindstrom, and M.~Rocek, ``Hyperkahler metrics
  and supersymmetry,'' {\em Commun. Math. Phys.} {\bfseries 108} (1987) 535.

\bibitem{Blau:1996bx}
M.~Blau and G.~Thompson, ``Aspects of $n(t)\geq 2$ topological gauge theories
  and {D}-branes,'' \href{http://dx.doi.org/10.1016/S0550-3213(97)00161-2}{{\em
  Nucl. Phys.} {\bfseries B492} (1997) 545--590},
\href{http://arxiv.org/abs/hep-th/9612143}{{\ttfamily arXiv:hep-th/9612143
  [hep-th]}}.

\bibitem{Gaiotto:2008cd}
D.~Gaiotto, G.~W. Moore, and A.~Neitzke, ``{Four-dimensional wall-crossing via
  three-dimensional field theory},''
  \href{http://dx.doi.org/10.1007/s00220-010-1071-2}{{\em Commun. Math. Phys.}
  {\bfseries 299} (2010) 163--224},
\href{http://arxiv.org/abs/0807.4723}{{\ttfamily arXiv:0807.4723 [hep-th]}}.

\bibitem{Bullimore2015}
M.~Bullimore, T.~Dimofte, and D.~Gaiotto, ``{The Coulomb Branch of 3d
  $\mathcal{N}=4$ Theories},''
  \href{http://arxiv.org/abs/1503.04817}{{\ttfamily 1503.04817}}.

\bibitem{Nakajima:2015txa}
H.~Nakajima, ``{Towards a mathematical definition of Coulomb branches of
  $3$-dimensional $\mathcal{N}=4$ gauge theories, I},''
  \href{http://dx.doi.org/10.4310/ATMP.2016.v20.n3.a4}{{\em Adv. Theor. Math.
  Phys.} {\bfseries 20} (2016) 595--669},
\href{http://arxiv.org/abs/1503.03676}{{\ttfamily arXiv:1503.03676 [math-ph]}}.

\bibitem{Braverman:2016wma}
A.~Braverman, M.~Finkelberg, and H.~Nakajima, ``{Towards a mathematical
  definition of Coulomb branches of $3$-dimensional $\mathcal N=4$ gauge
  theories, II},''
\href{http://arxiv.org/abs/1601.03586}{{\ttfamily arXiv:1601.03586 [math.RT]}}.

\bibitem{Bullimore:2016hdc}
M.~Bullimore, T.~Dimofte, D.~Gaiotto, J.~Hilburn, and H.-C. Kim, ``{Vortices
  and Vermas},''
\href{http://arxiv.org/abs/1609.04406}{{\ttfamily arXiv:1609.04406 [hep-th]}}.

\bibitem{Kapustin:2010}
A.~Kapustin, ``Topological field theory, higher categories, and their
  applications.,'' in {\em Proceedings of the International Congress of
  Mathematicians.}, vol.~III, pp.~2021--2043.
\newblock Hindustan Book Agency, New Delhi, 2010.
\newblock \href{http://arxiv.org/abs/arXiv:1004.2307}{{\ttfamily
  arXiv:1004.2307}}.

\bibitem{Aspinwall:2009isa}
P.~S. Aspinwall, T.~Bridgeland, A.~Craw, M.~R. Douglas, M.~Gross, A.~Kapustin,
  G.~W. Moore, G.~Segal, B.~Szendr\H~oi, and P.~M.~H. Wilson, {\em Dirichlet
  branes and mirror symmetry}, vol.~4 of {\em Clay Mathematics Monographs}.
\newblock American Mathematical Society, Providence, RI; Clay Mathematics
  Institute, Cambridge, MA, 2009.

\bibitem{Klemm:1996bj}
A.~Klemm, W.~Lerche, P.~Mayr, C.~Vafa, and N.~P. Warner, ``Self-dual strings
  and n=2~{S}upersymmetric field theory,'' {\em Nucl. Phys.} {\bfseries B477}
  (1996) 746--766,
\href{http://arxiv.org/abs/hep-th/9604034}{{\ttfamily hep-th/9604034}}.

\bibitem{Witten:1997sc}
E.~Witten, ``Solutions of four-dimensional field theories via {M}-theory,''
  {\em Nucl. Phys.} {\bfseries B500} (1997) 3--42,
\href{http://arxiv.org/abs/hep-th/9703166}{{\ttfamily hep-th/9703166}}.

\bibitem{Gaiotto:2009hg}
D.~Gaiotto, G.~W. Moore, and A.~Neitzke, ``{Wall-crossing, Hitchin Systems, and
  the WKB Approximation},''
\href{http://arxiv.org/abs/0907.3987}{{\ttfamily arXiv:0907.3987 [hep-th]}}.

\bibitem{Vafa:1994tf}
C.~Vafa and E.~Witten, ``{A Strong coupling test of S duality},''
  \href{http://dx.doi.org/10.1016/0550-3213(94)90097-3}{{\em Nucl. Phys.}
  {\bfseries B431} (1994) 3--77},
\href{http://arxiv.org/abs/hep-th/9408074}{{\ttfamily arXiv:hep-th/9408074
  [hep-th]}}.

\bibitem{Marcus:1995mq}
N.~Marcus, ``{The Other topological twisting of N=4 Yang-Mills},''
  \href{http://dx.doi.org/10.1016/0550-3213(95)00389-A}{{\em Nucl. Phys.}
  {\bfseries B452} (1995) 331--345},
\href{http://arxiv.org/abs/hep-th/9506002}{{\ttfamily arXiv:hep-th/9506002
  [hep-th]}}.

\bibitem{Ben-Zvi:2016mrh}
D.~Ben-Zvi and D.~Nadler, ``Betti geometric langlands,'' {\em Proc. Sympos.
  Pure Math.} {\bfseries 97.2} (2018) ,
  \href{http://arxiv.org/abs/arXiv:1606.08523}{{\ttfamily arXiv:1606.08523}}.

\bibitem{Elliott:2017ynt}
C.~Elliott and P.~Yoo, ``A physical origin for singular support conditions in
  geometric langlands theory,'' {\em arXiv:1707.01292} (2017) .

\bibitem{BezFink}
R.~Bezrukavnikov and M.~Finkelberg, ``Equivariant satake category and
  kostant-whittaker reduction.,'' {\em Mosc. Math. J. 8 (2008), no. 1, 39--72,
  183.} {\bfseries 8} no.~1, (2008) 39--72.

\bibitem{BFM}
R.~Bezrukavnikov, M.~Finkelberg, and I.~Mirkovic, ``Equivariant homology and
  k-theory of affine grassmannians and toda lattices.,'' {\em Compos. Math.}
  {\bfseries 141} no.~3, (2005) 746--768.

\bibitem{Knop}
F.~Knop, ``The asymptotic behavior of invariant collective motion.,'' {\em
  Invent. Math.} {\bfseries 116} no.~1-3, (1994) 309--328.

\bibitem{Ngo}
B.~C. Ng\^o, ``Le lemme fondamental pour les alg\`ebres de {L}ie,''
  \href{http://dx.doi.org/10.1007/s10240-010-0026-7}{{\em Publ. Math. Inst.
  Hautes \'Etudes Sci.} no.~111, (2010) 1--169}.
  \url{https://doi.org/10.1007/s10240-010-0026-7}.

\bibitem{Gukov:2006jk}
S.~Gukov and E.~Witten, ``{Gauge Theory, Ramification, And The Geometric
  Langlands Program},''
\href{http://arxiv.org/abs/hep-th/0612073}{{\ttfamily arXiv:hep-th/0612073
  [hep-th]}}.

\bibitem{Gaiotto:2011tf}
D.~Gaiotto, G.~W. Moore, and A.~Neitzke, ``{Wall-Crossing in Coupled 2d-4d
  Systems},'' \href{http://dx.doi.org/10.1007/JHEP12(2012)082}{{\em JHEP}
  {\bfseries 12} (2012) 082},
\href{http://arxiv.org/abs/1103.2598}{{\ttfamily arXiv:1103.2598 [hep-th]}}.

\end{thebibliography}\endgroup

\end{document}